\documentclass[aps,twocolumn]{revtex4}

\usepackage{amssymb}
\usepackage{amsmath}
\usepackage{bm}
\usepackage[dvips]{epsfig}
\usepackage{graphicx}

\begin{document}

\title{On the analytical properties of the magneto-conductivity in 
the case of presence of stable open electron trajectories on 
a complex Fermi surface.}

\author{A.Ya. Maltsev.}

\affiliation{
\centerline{\it L.D. Landau Institute for Theoretical Physics}
\centerline{\it 142432 Chernogolovka, pr. Ak. Semenova 1A,
maltsev@itp.ac.ru}}

\begin{abstract}
 We consider the electric conductivity in normal metals
in presence of a strong magnetic field. It is assumed here that the
Fermi surface of a metal has rather complicated form such that 
different types of quasiclassical electron trajectories can appear
on the Fermi level for different directions of ${\bf B}$. The effects
we consider are connected with the existence of regular (stable) 
open electron trajectories which arise in general on complicated 
Fermi surfaces. The trajectories of this type have a nice geometric 
description and represent quasiperiodic lines with a fixed mean 
direction in the ${\bf p}$-space. Being stable geometric objects, 
the trajectories of this kind exist for some open regions in
the space of directions of ${\bf B}$, which can be represented by 
``Stability Zones'' on the unit sphere $\mathbb{S}^{2}$. 
The main goal of the paper is to give a description of the analytical 
behavior of conductivity in the Stability Zones, which demonstrates 
in general rather nontrivial properties. 
\end{abstract}

\maketitle

\vspace{5mm}

\section{Introduction.}

 Our considerations here will be connected with the geometry of
the quasiclassical electron trajectories on the Fermi surface in the
presence of a strong magnetic field. The main goal of the paper is
to give a detailed consideration of the contribution of the stable
open trajectories to the conductivity tensor in the limit
$\, \omega_{B} \tau \, \rightarrow \, \infty $. As we will see below,
in spite of rather regular geometric properties of the stable 
(generic) open trajectories, their contribution to magneto-conductivity 
is quite non-trivial from analytical point of view, which is caused by 
nontrivial statistical properties of the trajectories of this kind. 
Here we will try to represent a general picture of the 
magneto-conductivity behavior in the case of presence of such
trajectories on the Fermi surface, including the description
of the dependence of conductivity on both the magnitude and the
direction of $\, {\bf B} $. As we will see, both the structure of a 
``Stability Zone'' on the angle diagram and the conductivity behavior
in strong magnetic fields will demonstrate actually rather non-trivial
properties.

 We will base our considerations on the topological picture for the 
electron dynamics in the space of the quasimomenta, arising for
general dispersion relation $\, \epsilon ({\bf p}) \, $ under the 
presence of a magnetic field. So, according to standard approach
we will assume that the electron states in the conduction band are
parametrized by the values of the quasimomenta ${\bf p}$ which in fact
should be considered as points of the three-dimensional space,
factorized over the vectors of the reciprocal lattice:
\begin{multline*}
\mathbb{T}^{3} \,\,\,\, = \,\,\,\, \left. \mathbb{R}^{3} \, \right/
\, \left\{ n_{1} {\bf a}_{1} \, + \, n_{2} {\bf a}_{2} \, + \,
n_{3} \, {\bf a}_{3} \right\} \quad ,   \\
n_{1} \, , \,\, n_{2} \, , \,\, n_{3} \,\,\, \in \,\,\, \mathbb{Z} 
\end{multline*}

 The basis vectors of the reciprocal lattice are connected with the
basis vectors $\, ({\bf l}_{1}, \, {\bf l}_{2}, \, {\bf l}_{3} ) \, $
of the direct lattice by the formulae

$${\bf a}_{1} \,\,\, = \,\,\, 2 \pi \hbar \,\,
{{\bf l}_{2} \, \times \, {\bf l}_{3} \over 
({\bf l}_{1}, \, {\bf l}_{2}, \, {\bf l}_{3} )} \,\,\, , \quad \quad
{\bf a}_{2} \,\,\, = \,\,\, 2 \pi \hbar \,\,
{{\bf l}_{3} \, \times \, {\bf l}_{1} \over
({\bf l}_{1}, \, {\bf l}_{2}, \, {\bf l}_{3} )} \,\,\, ,  $$
$${\bf a}_{3} \,\,\, = \,\,\, 2 \pi \hbar \,\,
{{\bf l}_{1} \, \times \, {\bf l}_{2} \over
({\bf l}_{1}, \, {\bf l}_{2}, \, {\bf l}_{3} )} $$
and define the fundamental (Brillouen) zone in the  \linebreak
$\, {\bf p}$ - space.

 The opposite faces of the parallelepiped formed by
$\, ({\bf a}_{1}, \, {\bf a}_{2}, \, {\bf a}_{3} ) \, $
should be identified with each other, so the space of the electron 
states (for a given conduction band) represents in fact the
three-dimensional torus $\, \mathbb{T}^{3}$.\footnote{We do not
consider here the spin variables, which will not play an essential
role in our picture.}

\begin{figure}[t]
\begin{center}
\includegraphics[width=0.9\linewidth]{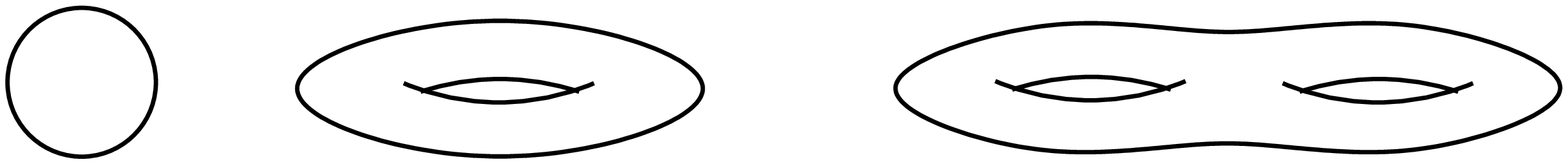}   
\end{center}
\caption{The canonical representation of the surfaces of genus
$\, $ 0, 1, 2 $\, $ (etc ...). }
\label{Genus}
\end{figure}

 The dispersion relation $\, \epsilon ({\bf p}) \, $ can be considered
either as a 3-periodic function in the $\, {\bf p}$ - space or just as 
a smooth function on the compact manifold $\, \mathbb{T}^{3}$, given by
the factorization of the $\, {\bf p}$ - space over the reciprocal 
lattice $\, L $. In the same way, the constant energy levels
$\, \epsilon ({\bf p}) \, = \, {\rm const} \, $ can be considered
either as 3-periodic two-dimensional surfaces in the 
$\, {\bf p}$ - space or as smooth compact surfaces 
$\, S_{\epsilon} \, \subset \, \mathbb{T}^{3} \, $
embedded in the three-dimensional torus $\, \mathbb{T}^{3}$.
Being considered in the last way, every nonsingular energy level
represents a smooth compact orientable surface, which is topologically
equivalent to one of the canonical surfaces defined by genus $g$
(Fig. \ref{Genus}).

\begin{figure}[t]
\begin{center}
\includegraphics[width=0.4\linewidth]{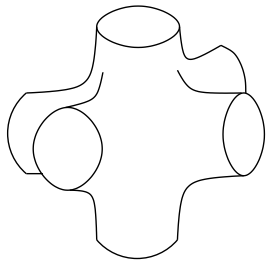}
\end{center}
\caption{The embedding of ``maximal rank'' of the surface of genus 3
in the torus $\, \mathbb{T}^{3}$.}
\label{MaximalRank}
\end{figure}

 The genus $\, g \, $ of the Fermi surface represents an important
topological characteristic of the electron spectrum in metal. Another
important topological characteristic of the electron spectrum is the 
way of embedding of the Fermi surface in $\, \mathbb{T}^{3}$. As an
example, Fig. \ref{MaximalRank} $\, $ represents an embedding of a genus
3 surface in the first Brillouen zone having ``maximal rank''.

 The quasiclassical evolution of the electron states in the presence 
of external homogeneous magnetic field ${\bf B}$ can be described by
the adiabatic equation (see e.g. \cite{Kittel,Ziman}):
\begin{equation}
\label{QuasiclassicalEvolution}
{\dot {\bf p}} \,\,\,\, = \,\,\,\, {e \over c} \,\,
\left[ {\bf v}_{\rm gr} ({\bf p}) \, \times \, {\bf B} \right]
\,\,\,\, = \,\,\,\, {e \over c} \,\, \left[ \nabla \epsilon ({\bf p})
\, \times \, {\bf B} \right]
\end{equation}
in the space of the quasimomenta.

\begin{figure}[t]
\begin{center}
\includegraphics[width=0.9\linewidth]{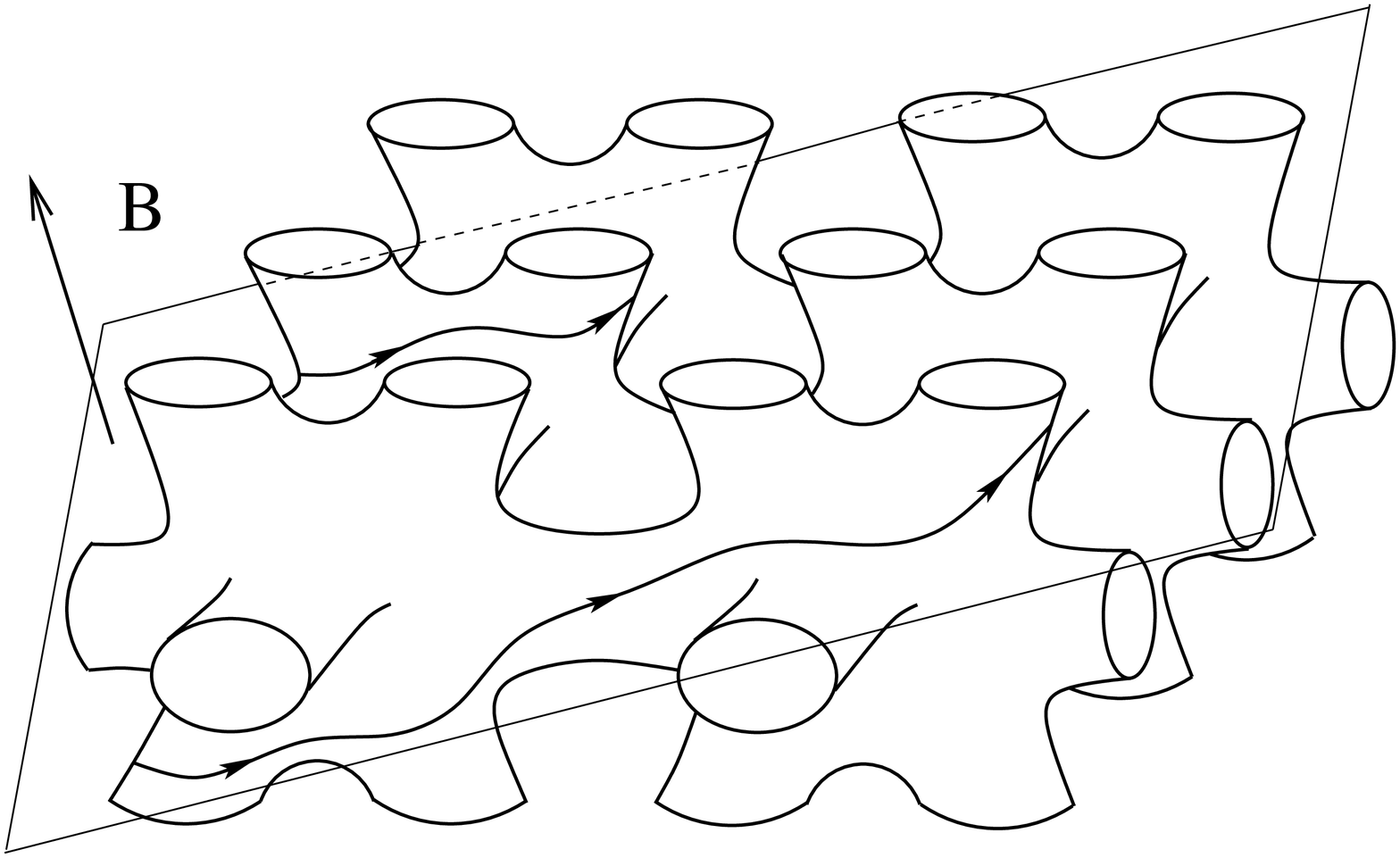}
\end{center}
\caption{The general picture of intersection of a complicated
3-periodic surface with the planes orthogonal to generic direction
of $\, {\bf B} $.}
\label{GeneralSurface}
\end{figure}

 The equation (\ref{QuasiclassicalEvolution}) is analytically
integrable in  \linebreak
$\, {\bf p}$ - space and the quasiclassical trajectories 
are given by the intersections of the constant energy 
surfaces  \linebreak
$\, \epsilon ({\bf p}) \, = \, {\rm const} \, $ with the planes
orthogonal to $\, {\bf B}$. However, despite this analytical property
of system (\ref{QuasiclassicalEvolution}), the global geometry of the
quasiclassical trajectories in  \linebreak
$\, {\bf p}$ - space can be rather
complicated. The reason for this circumstance is that the total
${\bf p}$ - space is not compact while in $\, \mathbb{T}^{3} \, $
the function $\, \epsilon ({\bf p}) \, $ represents the only
one-valued integral of motion of (\ref{QuasiclassicalEvolution}).
In general, we can expect rather complicated picture of intersection
of the 3-periodic surface 
$\, \epsilon ({\bf p}) \, = \, {\rm const} \, $ with the planes 
orthogonal to general (totally irrational) direction of 
$\, {\bf B} \, $ (see e.g. Fig. \ref{GeneralSurface}).

 As it is usual for the case of normal metals, we will assume here that
all the electron states with energies below the Fermi energy
$\, \epsilon_{F} \, $ are occupied by electrons, while all the electron
states with energies above the Fermi energy are empty. The Fermi surface
$\,\, S_{F} : \,\, \epsilon ({\bf p}) \, = \, \epsilon_{F} \, $
represents a two-dimensional surface in $\, \mathbb{T}^{3} \, $
separating occupied electron states from the empty ones. We have naturally
$\, \epsilon_{\rm min} \, < \, \epsilon_{F} \, < \, \epsilon_{\rm max} \,$
where $\, \epsilon_{\rm min} \, $ and $ \, \epsilon_{\rm max} \, $ are
the minimal and maximal possible electron energies in the conductivity
zone.

 System (\ref{QuasiclassicalEvolution}) conserves the energy of a 
particle and also the phase volume element $\, d^{3} p $. As a
consequence, the evolution according to system 
(\ref{QuasiclassicalEvolution}) does not change the Fermi distribution
of the quasiparticles or more general temperature dependent equilibrium
distributions
$$n \left( {\bf p} \right) \,\,\,\, = \,\,\,\, 
{1 \over e^{(\epsilon ({\bf p}) - \mu)/T} \, + \, 1} $$

 However, it appears that the form of the trajectories of
(\ref{QuasiclassicalEvolution}) plays an important role for the response
of the electron system to the small electric fields $\, {\bf E} $,
which defines the magneto-conductivity properties of normal metals.
Due to the high degeneracy of the electron gas in metals, all the
properties of such transport phenomena are defined in fact by the
properties of system (\ref{QuasiclassicalEvolution}) just on the one
energy level $\, \epsilon \, = \, \epsilon_{F} \, , \,\, $
i.e. just on the one (Fermi) energy surface
$\, \epsilon ({\bf p}) \, = \, \epsilon_{F} $.

\vspace{1mm}

\begin{figure}[t]
\begin{center}
\includegraphics[width=0.9\linewidth]{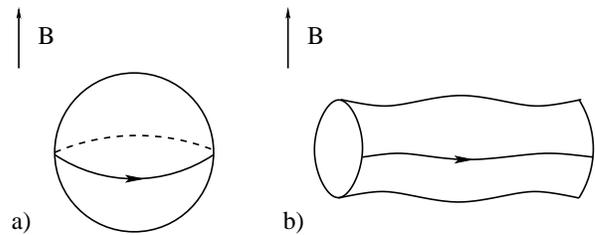}
\end{center}
\caption{Closed (a) and periodic (b) quasiclassical electron 
trajectories on different Fermi surfaces in ${\bf p}$ - space (only
one Brillouen zone is shown).}
\label{ClosedAndPer}
\end{figure}

 The role of the global geometry of the quasiclassical electron 
trajectories in the transport phenomena in strong magnetic fields
was first revealed by the school of I.M. Lifshitz (I.M. Lifshitz, 
M.Ya. Azbel, M.I. Kaganov, V.G. Peschansky) in 1950's
(see \cite{lifazkag,lifpes1,lifpes2,lifkag1,lifkag2,etm}).
Thus, in paper \cite{lifazkag} a crucial difference in contribution
to magneto-conductivity of closed (Fig. \ref{ClosedAndPer}, a) and
open periodic (Fig. \ref{ClosedAndPer}, b) quasiclassical trajectories 
in the limit $\, B \, \rightarrow \, \infty \, $ was pointed out.

 As was shown in \cite{lifazkag}, the contribution to the
magneto-conductivity of the trajectories of the first kind is almost 
isotropic in the plane orthogonal to $\, {\bf B} \, $ and has a
rapidly decreasing character in this plane as
$\, B \, \rightarrow \, \infty $. In contrast, the presence of the
open periodic trajectories gives an additional contribution to the
conductivity in the plane orthogonal to $\, {\bf B} $, which is
strongly anisotropic in the limit $\, B \, \rightarrow \, \infty $.
The asymptotic form of the conductivity tensor in the plane orthogonal 
to $\, {\bf B} \, $ in the described cases can be represented as
\begin{equation}
\label{ClosedTraj}
\sigma^{\alpha\beta} \,\,\, \simeq \,\,\,
{n e^{2} \tau \over m^{*}} \, \left(
\begin{array}{cc}
( \omega_{B} \tau )^{-2}  &  ( \omega_{B} \tau )^{-1}  \cr
( \omega_{B} \tau )^{-1}  &  ( \omega_{B} \tau )^{-2}
\end{array}  \right) \,\, , \quad 
\omega_{B} \tau \, \rightarrow \, \infty 
\end{equation}
(closed trajectories),
\begin{equation}
\label{PeriodicTraj}
\sigma^{\alpha\beta} \,\,\, \simeq \,\,\,
{n e^{2} \tau \over m^{*}} \, \left(
\begin{array}{cc}
( \omega_{B} \tau )^{-2}  &  ( \omega_{B} \tau )^{-1}  \cr
( \omega_{B} \tau )^{-1}  &  *
\end{array}  \right) \,\, , \quad 
\omega_{B} \tau \, \rightarrow \, \infty   
\end{equation}
(open periodic trajectories), after an appropriate choice of 
coordinates. (Here and below the notation $\, * \, $ means just
a constant of the order of unity).

 The formulae (\ref{ClosedTraj}) - (\ref{PeriodicTraj}) represent 
the asymptotic form of the conductivity tensor, so all the equalities 
give actually just the order of the absolute value of the corresponding
quantities.

 The value $\, m^{*} \, $ represents here just some (approximate)
effective mass of electron in crystal and $\, \tau \, $ denotes the
mean free electron time defined by the intensity of the 
scattering processes.

 The value
$$\omega_{B} \,\,\,\, \simeq \,\,\,\, {e B \over m^{*} c} $$
can be considered as an ``approximate value'' of the cyclotron
frequency in the crystal. From geometrical point of view 
$\, \omega_{B} \, $ represents approximately the inverse time
of motion along a closed trajectory (Fig. \ref{ClosedAndPer}, a)
or the inverse time of motion through one Brillouen zone for
periodic open trajectories (Fig. \ref{ClosedAndPer}, b). It becomes
clear from this point of view that the global geometry of the
quasiclassical trajectories reveals exactly in the limit
$\, \omega_{B} \tau \, \rightarrow \, \infty \, $ when the mean 
free electron time exceeds the time of motion inside one Brillouen
zone. Let us say here that for experimental observation of the
``geometric strong magnetic field limit'' rather pure materials
under rather low temperatures ($T \, \sim \, 1 K$) and rather strong
magnetic fields ($B \, \sim \, 10^{4} \, Gs$) should be usually used.

\begin{figure}[t]
\begin{center}
\includegraphics[width=0.9\linewidth]{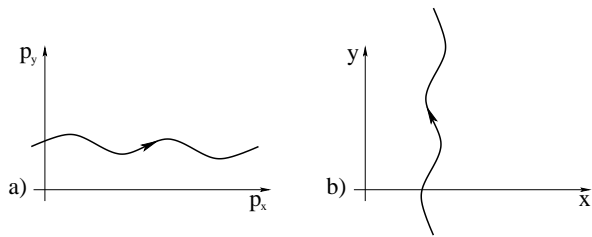}
\end{center}
\caption{The quasiclassical electron trajectory 
in $\, {\bf p}$ - space (a) $\, $ and its projection on the plane,
orthogonal to $\, {\bf B} $, in $\, {\bf x}$ - space (b).}
\label{pandxtraject}
\end{figure}

 In formula (\ref{PeriodicTraj}) the $\, x$ - axis is chosen along the 
mean direction of the periodic trajectories in $\, {\bf p}$ - space.
As can be easily derived from system (\ref{QuasiclassicalEvolution}),
the projection of the corresponding trajectory in coordinate space
onto the plane orthogonal to $\, {\bf B} \, $ is given just by the
rotation of the trajectory in $\, {\bf p}$ - space by $90^{\circ}$
(Fig. \ref{pandxtraject}).

 We can see then that even very strong magnetic field does not block
the electron motion along the $\, y$ - axis in the coordinate space in
the presence of the trajectories shown at Fig. \ref{ClosedAndPer}, b.
As a result, the $\, y$ - component of conductivity in the plane 
orthogonal to $\, {\bf B} \, $ remains finite in the limit
$\, \omega_{B} \tau \, \rightarrow \, \infty $. The strong anisotropy
of the conductivity tensor in this situation gives an experimental
possibility to observe this phenomenon as well as to determine the
mean direction of the open trajectories in $\, {\bf p}$ - space.

 The conductivity along the direction of $\, {\bf B} \, $ remains
finite in the limit $\, \omega_{B} \tau \, \rightarrow \, \infty \, $
in both the cases described above. The asymptotic forms of the full
conductivity tensor in $\, {\bf x}$ - space can be represented as:
\begin{multline}
\label{3DimClosedTr}
\sigma^{kl} \,\,\,\, \simeq \,\,\,\,
{n e^{2} \tau \over m^{*}} \, \left(
\begin{array}{ccc}
( \omega_{B} \tau )^{-2}  &  ( \omega_{B} \tau )^{-1}  &
( \omega_{B} \tau )^{-1}  \cr
( \omega_{B} \tau )^{-1}  &  ( \omega_{B} \tau )^{-2}  &
( \omega_{B} \tau )^{-1}  \cr
( \omega_{B} \tau )^{-1}  &  ( \omega_{B} \tau )^{-1}  &  *
\end{array}  \right) \,\, ,   \\
\omega_{B} \tau \,\, \rightarrow \,\, \infty
\end{multline}
(closed trajectories),
\begin{multline}
\label{3DimPeriodicTr}
\sigma^{kl} \,\,\,\, \simeq \,\,\,\,
{n e^{2} \tau \over m^{*}} \, \left(
\begin{array}{ccc}
( \omega_{B} \tau )^{-2}  &  ( \omega_{B} \tau )^{-1}  &
( \omega_{B} \tau )^{-1}  \cr
( \omega_{B} \tau )^{-1}  &  *  &  *  \cr
( \omega_{B} \tau )^{-1}  &  *  &  *
\end{array}  \right) \,\, ,   \\
\omega_{B} \tau \,\, \rightarrow \,\, \infty
\end{multline}
(open periodic trajectories).

 Let us note here that both the trajectories shown at 
Fig. \ref{ClosedAndPer} $\, $ are actually closed in the torus
$\, \mathbb{T}^{3}$. However, their embeddings in 
$\, \mathbb{T}^{3} \, $ are completely different from topological
point of view. Thus, the embedding of the trajectory shown at
Fig. \ref{ClosedAndPer}, a $\, $ is ``homologous'' to zero,
while the embedding of the trajectory shown at
Fig. \ref{ClosedAndPer}, b $\, $ represents a nonzero homology
class in $\, \mathbb{T}^{3}$. Certainly, the difference between the
trajectories shown at Fig. \ref{ClosedAndPer}, a $\, $ and
$\, $ Fig. \ref{ClosedAndPer}, b $\, $ is more evident 
in the covering  \linebreak
${\bf p}$ - space where they have different global geometry.

 Let us say here also, that both the situations shown at 
Fig. \ref{ClosedAndPer} a,b $\, $ represent the special case when
the conductivity tensor can be represented as a regular series
in powers of $\, (\omega_{B} \tau )^{-1} \, $ in the limit
$\, \omega_{B} \tau \, \rightarrow \, \infty $. Thus, we can actually
write here  
\begin{multline}
\label{SigmaRegExpClosed}
\sigma^{kl} (B) \,\,\,\, \simeq \,\,\,\,
{n e^{2} \tau \over m^{*}} \, \left(
\begin{array}{ccc}
0  &  0  &  0  \cr
0  &  0  &  0  \cr
0  &  0  &  *
\end{array}  \right) \,\,\, +  \\
+ \,\, {n e^{2} \tau \over m^{*}} \, \left(
\begin{array}{ccc}
0  &  *  &  *  \cr
*  &  0  &  *  \cr
*  &  *  &  0
\end{array}  \right) \,\, ( \omega_{B} \tau )^{-1} \,\,\,\, +  \\
+ \,\, {n e^{2} \tau \over m^{*}} \, \left(
\begin{array}{ccc}
*  &  *  &  *  \cr
*  &  *  &  *  \cr
*  &  *  &  *
\end{array}  \right) \,\, ( \omega_{B} \tau )^{-2}
\,\,\,\, + \,\,\,\, \dots  
\end{multline}
(closed trajectories),
\begin{multline}
\label{SigmaRegExpPeriodic}
\sigma^{kl} (B) \,\,\,\, \simeq \,\,\,\,
{n e^{2} \tau \over m^{*}} \, \left(
\begin{array}{ccc}
0  &  0  &  0  \cr
0  &  *  &  *  \cr
0  &  *  &  *
\end{array}  \right) \,\,\, +  \\
+ \,\, {n e^{2} \tau \over m^{*}} \, \left(
\begin{array}{ccc}
0  &  *  &  *  \cr
*  &  0  &  *  \cr
*  &  *  &  0
\end{array}  \right) \,\, ( \omega_{B} \tau )^{-1} \,\,\,\, +  \\
+ \,\, {n e^{2} \tau \over m^{*}} \, \left(
\begin{array}{ccc}
*  &  *  &  *  \cr
*  &  *  &  *  \cr
*  &  *  &  *
\end{array}  \right) \,\, ( \omega_{B} \tau )^{-2}
\,\,\,\, + \,\,\,\, \dots  
\end{multline}
(open periodic trajectories), where the even powers of 
$\, \omega_{B} \tau \, $ correspond to the symmetric part of 
$\, \sigma^{kl} (B) \, $, while the odd powers of 
$\, \omega_{B} \tau \, $ represent the anti-symmetric part in 
accordance with the Onsager relations.

 As we will see below, this situation does not take place for 
general open electron trajectories on the Fermi surface.

\vspace{1mm}

 In papers \cite{lifpes1,lifpes2} different examples of complicated
Fermi surfaces and more general types of open electron trajectories
in strong magnetic fields were considered. The trajectories
considered in \cite{lifpes1,lifpes2} are in general not periodic 
and are non-closed both in $\, \mathbb{T}^{3} \, $ and the 
${\bf p}$ - space and represent the first examples of the stable
open electron trajectories on a complicated Fermi surface.
The form of the trajectories found in  \cite{lifpes1,lifpes2} 
demonstrates also strongly anisotropic properties, so the strong 
anisotropy of the conductivity tensor
in the limit $\, \omega_{B} \tau \, \rightarrow \, \infty \, $
is also expected in this case. As was pointed out in 
\cite{KaganovPeschansky}, the analytical behavior of conductivity
in these examples can in fact be different from (\ref{3DimPeriodicTr})
and demonstrate a slower approach to its limiting form in the interval
of not extremely strong magnetic fields. Here we will try to study
this question in general case using the topological description of
the stable open trajectories on the Fermi surface.

\vspace{1mm}

 In the survey articles \cite{lifkag1,lifkag2,lifkag3} and also in 
the book \cite{etm} a remarkable review of both the theoretical and 
experimental investigations of this branch of electron theory of metals 
made at that time can be found. Let us give here also a reference on 
a survey article \cite{KaganovPeschansky} where the same topics were 
revisited after forty years and which contains also the aspects 
appeared in the later period.

\vspace{1mm}

 The general problem of classification of all possible types of
trajectories of dynamical system (\ref{QuasiclassicalEvolution})
with arbitrary (periodic) dispersion law $\, \epsilon ({\bf p}) \, $
was set by S.P. Novikov (\cite{MultValAnMorseTheory}) and was
intensively investigated in his topological school (S.P. Novikov,
A.V. Zorich, S.P. Tsarev, I.A. Dynnikov). The topological investigation
of system (\ref{QuasiclassicalEvolution}) has led to rather detailed
understanding of the geometry of the trajectories of different types
and gave finally the full classification of all possible types
of trajectories of (\ref{QuasiclassicalEvolution}). One of the most
important parts of the mathematical theory of the trajectories of
(\ref{QuasiclassicalEvolution}), which will be also of great
importance in the present paper, is connected with the detailed
description of the stable open trajectories of system
(\ref{QuasiclassicalEvolution}). Let us say here that the 
most important breakthroughs in this problem were made in papers
\cite{zorich1,dynn1} where rather deep topological theorems
about non-closed trajectories of system (\ref{QuasiclassicalEvolution})
were proved.

\vspace{1mm}

 Using the topological description of the stable open trajectories
of system (\ref{QuasiclassicalEvolution}) it was possible to introduce
important topological characteristics of the stable nontrivial regimes
of the magneto-conductivity behavior in the limit
$\, \omega_{B} \tau \, \rightarrow \, \infty \, $, which were 
introduced in paper \cite{PismaZhETF}. In general, the characteristics
introduced in \cite{PismaZhETF} can be described in the following way:

 Let us exclude now the ``trivial'' cases when we have just closed
electron trajectories on the Fermi surface in   \linebreak
${\bf p}$ - space and 
consider the situation when open trajectories (in ${\bf p}$ - space)
are present on the Fermi level. Besides that, let us require that
the open trajectories are stable with respect to small rotations of
$\, {\bf B} \, $ such that we have a ``Stability Zone'' in the space 
of directions of $\, {\bf B}$, where similar open trajectories
exist on the Fermi level.

\begin{figure}[t]
\begin{center}
\includegraphics[width=0.9\linewidth]{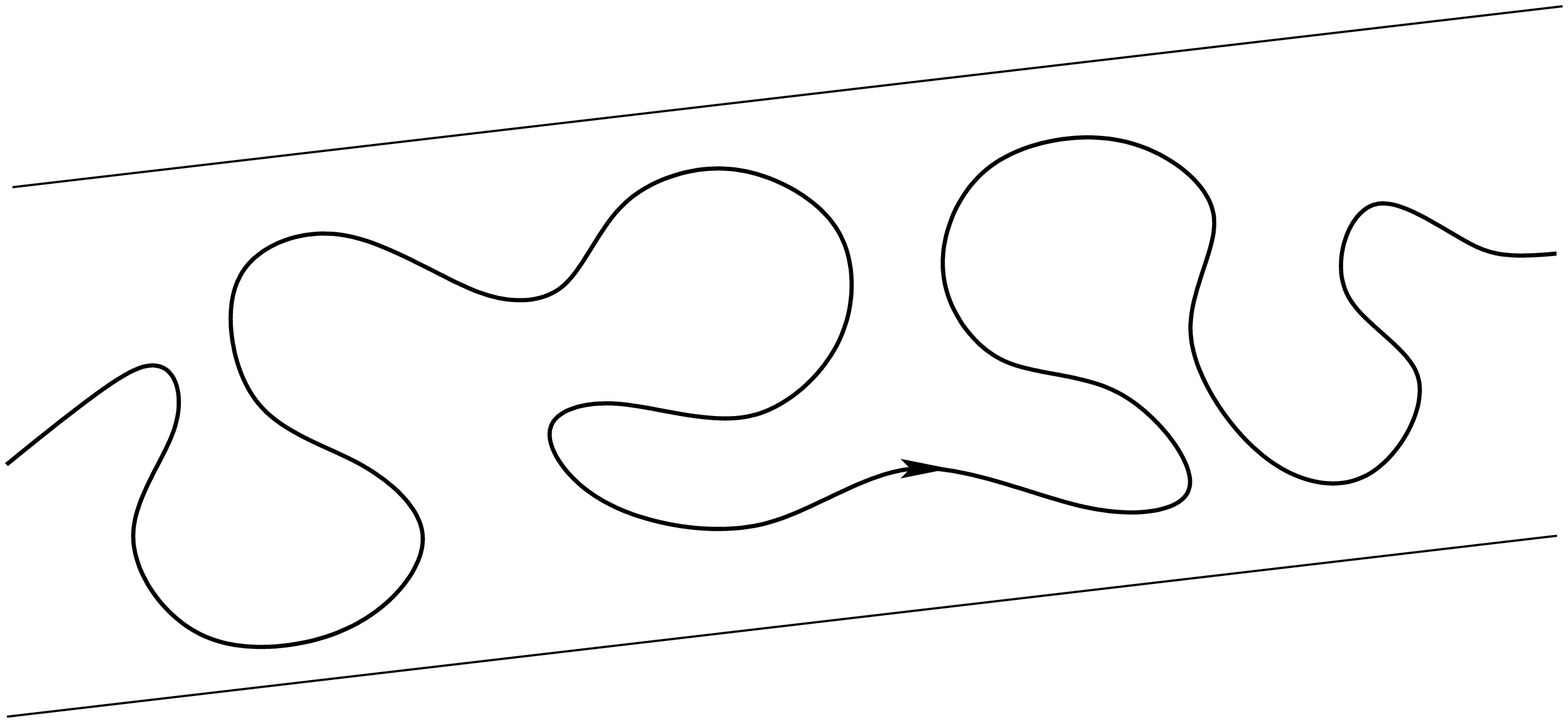}
\end{center}
\caption{A stable (quasiperiodic) open trajectory of system
(\ref{QuasiclassicalEvolution}) passing through a straight strip of 
a finite width in the plane orthogonal to $\, {\bf B}$.}
\label{StableOpenTr}
\end{figure}

 As can be extracted from the topological description of the stable 
open trajectories of system (\ref{QuasiclassicalEvolution}), the 
trajectories of this type demonstrate the following remarkable 
properties:

1) Every stable open (in ${\bf p}$ - space) trajectory of
(\ref{QuasiclassicalEvolution}) lies in a straight strip of a finite
width in the plane orthogonal to $\, {\bf B}$, passing through it
from $\, - \infty \, $ to $\, + \infty \, $ 
(Fig. \ref{StableOpenTr});

2) The mean direction of all open stable trajectories in 
${\bf p}$ - space is given by the intersection of the plane orthogonal 
to $\, {\bf B} \, $ and some integral (generated by two reciprocal
lattice vectors) plane $\, \Gamma $, which is the same for a given
``Stability Zone'' in the space of directions of $\, {\bf B}$.

\vspace{1mm}

 Let us note here that the statement (1) was first formulated by
S.P. Novikov in the form of a conjecture, which was proved later 
for the stable open trajectories of (\ref{QuasiclassicalEvolution})
(\cite{zorich1,dynn1}).

\vspace{1mm}

 As was pointed out in \cite{PismaZhETF}, the integral planes
$\, \Gamma_{\alpha} \, $ represent experimentally observable objects
view the remarkable geometric properties of the open stable
trajectories, described above. Thus, measuring the conductivity  
in the limit $\, \omega_{B} \tau \, \rightarrow \, \infty $, 
we can observe strongly anisotropic behavior of conductivity 
in the planes orthogonal to $\, {\bf B} \, $ for all
$\, {\bf B} / B \, \in \, \Omega_{\alpha} \, $ and extract the
integral plane $\, \Gamma_{\alpha} $, which is swept by the directions 
of the highest decreasing of conductivity in the ${\bf x}$ - space.

 More precisely, let us introduce the values
$$\sigma^{kl}_{\infty} \, ({\bf B}/B) \,\,\,\,\, = \,\,\,\,\,
\lim_{\omega_{B} \tau \rightarrow \infty} \,\,
\sigma^{kl} ({\bf B}) \quad ,  \quad \quad k, l \, = \, 1, 2, 3 $$

 Then, we can chose a basis in ${\bf x}$ - space
$$\Big( \, {\bf e}_{1} ({\bf B}) \, ,
\,\,\, {\bf e}_{2} ({\bf B}) \, , \,\,\,
{\bf e}_{3} \, = \, {\bf B}/B \, \Big) \quad , $$   
smoothly depending on $\, {\bf B} \, $ in $\, \Omega_{\alpha}$,
where the values $\, \sigma^{kl}_{\infty} \, $ will have the form
\begin{equation}
\label{3DimAnisotropicLimit}
\sigma^{kl}_{\infty} \,\,\,\,  =  \,\,\,\,
{n e^{2} \tau \over m^{*}} \,\, \left(
\begin{array}{ccc}
 0  &  0  &  0  \cr
 0  &  *  &  *  \cr   
 0  &  *  &  *
\end{array}  \right) 
\end{equation}

 As was said above, the direction of the vector 
$\, {\bf e}_{1} ({\bf B}) \, $ is given by the
intersection of the plane orthogonal to $\, {\bf B} \, $ with an
integral plane, i.e. the plane generated by two reciprocal lattice
vectors $\, {\bf q}_{1} $, $\, {\bf q}_{2} $, which is the same 
for the Zone $\, \Omega_{\alpha}$.

 The integral plane $\, \Gamma_{\alpha} \, $ can be represented by an
indivisible triple of integer numbers
$\, (M^{1}_{\alpha}, \, M^{2}_{\alpha}, \, M^{3}_{\alpha}) \, $
according to the equation
\begin{equation}
\label{IntPlaneEquation}
M^{1}_{\alpha} \, \left( {\bf x} \, , \, {\bf l}_{1} \right)
\,\,\, + \,\,\, 
M^{2}_{\alpha} \, \left( {\bf x} \, , \, {\bf l}_{2} \right)
\,\,\, + \,\,\,
M^{3}_{\alpha} \, \left( {\bf x} \, , \, {\bf l}_{3} \right)
\,\,\,\, = \,\,\,\, 0 
\end{equation}
where $\, ({\bf l}_{1}, \, {\bf l}_{2}, \, {\bf l}_{3}) \, $
are the basis direct lattice vectors. The numbers
$\, (M^{1}_{\alpha}, \, M^{2}_{\alpha}, \, M^{3}_{\alpha}) \, $
were called in \cite{PismaZhETF} the topological quantum numbers
observable in conductivity of normal metals. Let us say here that
the triples 
$\, (M^{1}_{\alpha}, \, M^{2}_{\alpha}, \, M^{3}_{\alpha}) \, $
can be rather nontrivial for complicated Fermi surfaces.

 The complete Zone $\, \Omega_{\alpha} \, $ 
can be called here the mathematical Stability Zone, 
corresponding to a given Fermi energy. We have to say 
here, however, that in experimental study of magneto-conductivity it 
is natural to introduce extended 
``experimentally observable Stability Zone''
$\,\, \hat{\Omega}_{\alpha} \,\, $ 
$(\Omega_{\alpha} \subset \hat{\Omega}_{\alpha}) \,\, $
due to specific behavior of trajectories of system 
(\ref{QuasiclassicalEvolution}) near the boundary of 
$\, \Omega_{\alpha}$. Everywhere in $\, \hat{\Omega}_{\alpha} \, $
the behavior of conductivity is experimentally indistinguishable
from that in $\, \Omega_{\alpha} \, $ even under good formal
implementation of the condition $\, \omega_{B} \tau \, \gg \, 1 \, $
until the magnitude $\, B \, $ of magnetic field becomes extremely
high. The conventional form of $\, \hat{\Omega}_{\alpha} \, $
depends actually on the experimental conditions and is determined
mainly by the maximal values of $\, B \, $ in experiment. The
difference between the Zones $\, \hat{\Omega}_{\alpha} \, $ and
$\, \Omega_{\alpha}$, which can be detected only in extremely strong
magnetic fields, represents in fact one of essential features of
analytical behavior of conductivity, considered here.

\vspace{1mm}

 Let us note now that the stable open quasiclassical trajectories 
do not represent all possible types of non-closed trajectories
of system (\ref{QuasiclassicalEvolution}) and trajectories with more
complicated geometry can arise on complicated Fermi surfaces for
some special directions of $\, {\bf B} $. Thus, the first example
of trajectory which can not be restricted by any straight strip of
a finite width was constructed by S.P. Tsarev (\cite{Tsarev}).
The trajectory of Tsarev demonstrates explicit chaotic behavior 
on the Fermi surface and is essentially different from the stable
open trajectories from this point of view. On the other hand, the
Tsarev trajectory is also strongly anisotropic in the 
${\bf p}$ - space and has an asymptotic direction in the plane
orthogonal to $\, {\bf B} $. So, from experimental point of view
the contribution of Tsarev trajectories to the magneto-conductivity
is also strongly anisotropic in the limit
$\, \omega_{B} \tau \, \rightarrow \, \infty \, $ and corresponds
to the regime (\ref{3DimAnisotropicLimit}). Let us say,
however, that the Tsarev chaotic trajectories are completely
unstable under small rotations of $\, {\bf B} \, $ and are not
connected with topologically stable characteristics unlike
the stable open trajectories discussed above. We have to say also,
that the trajectories of Tsarev type (if they present) can appear 
just for the set of zero measure in the space of directions of 
$\, {\bf B} $.

 Another example of non-closed trajectory of
(\ref{QuasiclassicalEvolution}) which demonstrates strongly chaotic
behavior both in $\, \mathbb{T}^{3} \, $ and in ${\bf p}$ - space
was constructed by I.A. Dynnikov in \cite{dynn2}. The trajectories 
of Dynnikov are rather different from the Tsarev chaotic trajectories
and require the maximal irrationality of the direction of 
$\, {\bf B} \, $ to be observed. The behavior of the conductivity
tensor in the presence of the chaotic trajectories constructed in
\cite{dynn2} was investigated in \cite{ZhETF1997} and appeared to be
rather different from both the regimes given by (\ref{3DimClosedTr})
and (\ref{3DimPeriodicTr}). The most interesting feature of the
contribution to the conductivity tensor of the trajectories of this
kind is that the presence of such trajectories suppresses the 
conductivity in all directions, including the direction of 
$\, {\bf B} $, in the limit
$\, \omega_{B} \tau \, \rightarrow \, \infty $.

\vspace{1mm}

 In general, the Tsarev and Dynnikov chaotic trajectories demonstrate
two different general types of possible chaotic behavior of trajectories
of system (\ref{QuasiclassicalEvolution}), so in this sense all chaotic
trajectories of (\ref{QuasiclassicalEvolution}) can be attributed to
the Tsarev or Dynnikov type. Let us say here that different properties
of chaotic trajectories of system (\ref{QuasiclassicalEvolution}) are
also intensively investigated in modern mathematical literature
(see \cite{zorich2,dynn3,DeLeo1,DeLeo2,DeLeo3,DeLeoDynnikov1,
dynn4,DeLeoDynnikov2,Skripchenko1,Skripchenko2,DynnSkrip}).

\vspace{1mm}

 The directions of $\, {\bf B} \, $, corresponding to the chaotic 
open trajectories, can appear only outside the ``Stability Zones'',
corresponding to the stable open trajectories of system
(\ref{QuasiclassicalEvolution}). In the same way, the chaotic open
trajectories of different types (Tsarev or Dynnikov type) can not
appear at the same direction of $\, {\bf B}$. Let us say, however,
that in general the contribution of non-closed trajectories to the
conductivity tensor should be added with the contribution 
(\ref{3DimClosedTr}) of closed trajectories of system 
(\ref{QuasiclassicalEvolution}), which usually present together with 
open trajectories (of any kind) at the same Fermi surface.

 For more rigorous description of possible situations in the behavior
of trajectories of (\ref{QuasiclassicalEvolution}) with arbitrary
periodic dispersion law let us give here a reference to a detailed
mathematical survey \cite{dynn3}, devoted to this question. The
detailed consideration of different physical results based on the 
results of topological investigations of system 
(\ref{QuasiclassicalEvolution}) can be found in
\cite{UFN,BullBrazMathSoc,JournStatPhys}.

\vspace{2mm}

 Our main goal in this paper will be more detailed investigation
of the conductivity behavior in the  ``Stability Zones'', based on the 
detailed topological description of the stable open electron trajectories 
arising for the corresponding directions of $\, {\bf B} $. As we said
already, the corresponding stable open trajectories demonstrate
rather regular geometric properties in the planes orthogonal to 
$\, {\bf B} \, $. Thus, all the stable open trajectories have the form 
shown at Fig. \ref{StableOpenTr} and the same 
mean direction in  \linebreak
$\, {\bf p}$ - space, given by the intersection of 
the plane orthogonal to $\, {\bf B} \, $
with an integral plane $\, \Gamma_{\alpha}$, which is fixed for the
corresponding Stability Zone $\, \Omega_{\alpha}$. As a result, the
conductivity tensor has the asymptotic form (\ref{3DimAnisotropicLimit})
in the limit $\, \omega_{B} \tau \, \rightarrow \, \infty \, $ and
demonstrates the specific geometric properties described above. These 
geometric (or topological) properties of the conductivity tensor give 
the most stable (topological) characteristics of the conductivity for a 
given ``Stability Zone'' and are connected with the ``purely geometric'' 
limit $\, \omega_{B} \tau \, \rightarrow \, \infty $. At the same time, 
we will see below, that the conductivity tensor demonstrates in general 
rather nontrivial analytical dependence on $\, {\bf B} \, $ in different 
parts of ``experimentally observable'' Stability Zones, which is caused 
by specific features of different trajectories of system 
(\ref{QuasiclassicalEvolution}) in these regions. Here we will 
investigate the analytical properties of conductivity for the case of 
rather big (but finite) values of the parameter 
$\, \omega_{B} \tau \, $ using topological description of 
``carriers of open trajectories'' on the Fermi surface, 
defined for the corresponding directions of $\, {\bf B} $. 
As we will see below, both the dependence on the parameter 
$\, \omega_{B} \tau \, $ and on the direction of $\, {\bf B} \, $ 
demonstrate here rather nontrivial properties.

 In the next section we will give a description of ``experimentally
observable'' Stability Zone and try to describe the main features of 
the analytical behavior of conductivity in its different parts, which
gives from our point of view the main points of the picture arising
in general case. In Sections III and IV we will present more detailed
consideration of the contribution of the stable open trajectories
to the magneto-conductivity, based on the topological description
of such trajectories on complicated Fermi surfaces. Sections
III and IV will have more mathematical character in comparison with 
Section II.

\vspace{5mm}

\section{The analytical behavior of conductivity in the
experimentally observable Stability Zone.}
\setcounter{equation}{0}

 To describe briefly the general picture, let us point out now the 
main points, which will play the most
essential role in our considerations below. This section will be
mostly descriptive, while more detailed analysis will be presented
in Sections III, IV.

\vspace{1mm}

 As we said already, the stable open trajectories 
demonstrate quasiperiodic properties for generic
directions of $\, {\bf B} \,\, $
(${\bf B}/B \, \in \, \Omega_{\alpha}$). However, they become purely 
periodic for some special directions of the magnetic field. It's not
difficult to see that the last situation arises every time when the
plane orthogonal to $\, {\bf B} \, $ intersects the plane
$\, \Gamma_{\alpha} \, $ along an integer vector in ${\bf p}$ - space:
\begin{equation}
\label{SpecDirBEq}
{\bf a} \,\,\, = \,\,\, k_{1} \, {\bf q}_{1} \, + \, 
k_{2} \, {\bf q}_{2} \,\,\, = \,\,\, m_{1} \, {\bf a}_{1} \, + \,
m_{2} \, {\bf a}_{2} \, + \, m_{3} \, {\bf a}_{3} 
\end{equation}
($ k_{1}, \, k_{2}, \, m_{1}, \, m_{2}, \, m_{3} 
\,\, \in \,\, \mathbb{Z} $).

 Both the cases of the quasiperiodic and the periodic trajectories
lead in this situation to the conductivity tensor 
(\ref{3DimAnisotropicLimit}) in the limit 
$\, \omega_{B} \tau \, \rightarrow \, \infty $, having the same
stable geometric properties for a given ``Stability Zone''.
However, as we will see below, the limiting values of the
components $\, \sigma^{22}$, $\, \sigma^{23}$, $\, \sigma^{32}$,
$\, \sigma^{33}$, $\, $ i.e. the values
$$\sigma^{kl}_{\infty} \, ({\bf B}/B) \,\,\,\,\, = \,\,\,\,\,
\lim_{\omega_{B} \tau \rightarrow \infty} \,\,
\sigma^{kl} ({\bf B}) \quad , \quad \quad \quad k, l \, = \, 2, 3 $$
(${\bf B}/B \, $ is fixed), will have here sharp ``jumps'' with
respect to the same values, defined for close generic directions
of $\, {\bf B} $.

 More precisely, we can introduce the functions
$\,\, {\bar \sigma}^{kl}_{\infty} \, ({\bf B}/B) \, $,
$\,\, (k, l \, = \, 2, 3) \, , \, $ which represent continuous 
functions of 
$\,\, {\bf n} \, = \, {\bf B}/B \, $ in $\, \Omega_{\alpha} \, $ 
and coincide with the corresponding values 
$\, \sigma^{kl}_{\infty} \, ({\bf B}/B) \, $ for generic directions
$\, {\bf B}/B \, \in \, \Omega_{\alpha} $. However, for special
directions of $\, {\bf B} \, $ described above we will have
$\,\, \sigma^{kl}_{\infty} \, ({\bf B}/B) \, \neq \,
{\bar \sigma}^{kl}_{\infty} \, ({\bf B}/B) \, , \,\, $ which is caused
by special features of statistical averaging over the open trajectories
of (\ref{QuasiclassicalEvolution}) near the Fermi surface. It's not
difficult to show also that we will always have the inequalities
$$\sigma^{22}_{\infty} \, ({\bf B}/B) \,\, - \,\,
{\bar \sigma}^{22}_{\infty} \, ({\bf B}/B) \,\,\,\,\, > \,\,\,\,\, 0 
\quad ,  $$
$$\sigma^{33}_{\infty} \, ({\bf B}/B) \,\, - \,\,
{\bar \sigma}^{33}_{\infty} \, ({\bf B}/B) 
\,\,\,\,\, > \,\,\,\,\, 0   \quad  \, $$
for these special directions of $\, {\bf B} $, while the signs of
$\,\,\, \sigma^{kl}_{\infty} \, ({\bf B}/B) \,\, - \,\,
{\bar \sigma}^{kl}_{\infty} \, ({\bf B}/B) \,\, , \,\,\, 
k \, \neq l \, , \,\,\, $
are in general indefinite.

\begin{figure}[t]
\begin{center}
\includegraphics[width=1\linewidth]{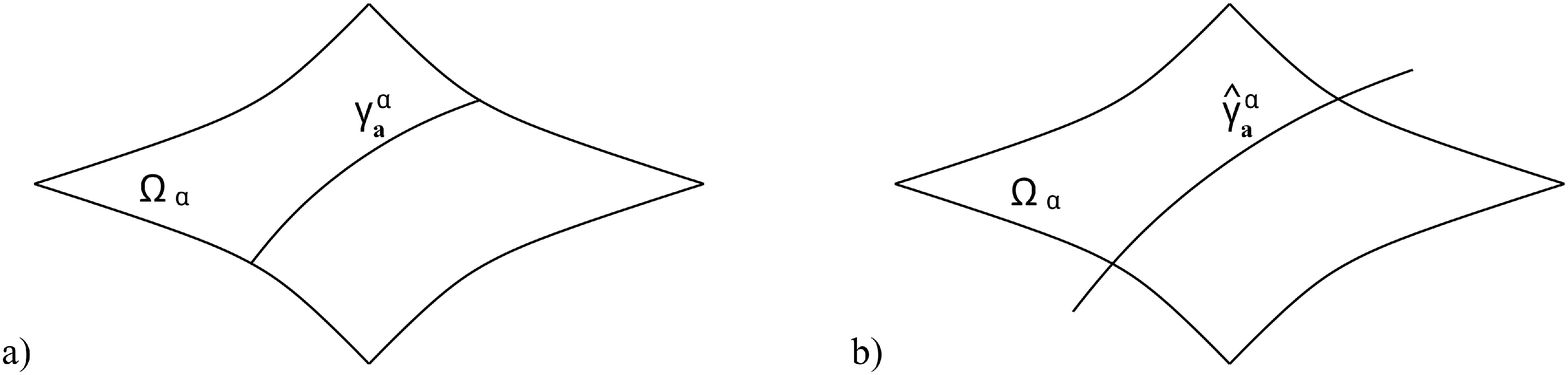}
\end{center}
\caption{(a) A schematic representation of the family 
$\, \gamma^{\alpha}_{\bf a} \, $ of special
directions of $\, {\bf B} $, corresponding to a rational mean
direction $\, {\bf a} \, $ of open trajectories, within a Stability
Zone $\, \Omega_{\alpha}$. $\,\, $ (b) The full one-dimensional set
$\, {\hat \gamma}^{\alpha}_{\bf a} \, $
of directions of $\, {\bf B} $, corresponding to existence of periodic 
trajectories with mean direction $\, {\bf a} $, intersecting the 
Stability Zone $\, \Omega_{\alpha}$.}
\label{SpecDirMagnField}
\end{figure}

\begin{figure}[t]
\begin{center}
\includegraphics[width=0.9\linewidth]{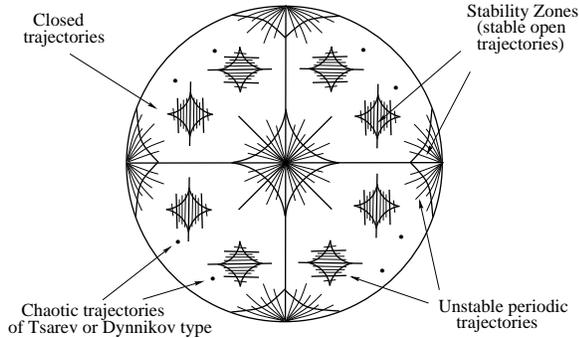}
\end{center}
\caption{A schematic angle diagram showing different situations
for different directions of $\, {\bf B} \, $ on the unit sphere.}
\label{AngleDiagramm}
\end{figure}

 The directions $\, {\bf B}/B \, \in \, \Omega_{\alpha} $, 
corresponding to a fixed rational mean direction
$\, {\bf a} \, $ of open trajectories, represent a one-dimensional
curve $\, \gamma^{\alpha}_{\bf a} \, $ (on $\, \mathbb{S}^{2}$) given 
by the intersection of the big circle, orthogonal to $\, {\bf a} $,
and the Stability Zone $\, \Omega_{\alpha} \, $
(Fig. \ref{SpecDirMagnField} a). We have to say, however,
that according to a more detailed consideration  
the periodic trajectories exist actually on some bigger curve
$\, {\hat \gamma}^{\alpha}_{\bf a} \, $ representing extension
of the curve $\, \gamma^{\alpha}_{\bf a} \, $ outside the
Stability Zone $\, \Omega_{\alpha} \, $
(Fig. \ref{SpecDirMagnField} b). All the open trajectories have
the same mean direction $\, {\bf a} \, $  for all
$\, {\bf B} / B \, \in \, {\hat \gamma}^{\alpha}_{\bf a} \, $ and
the area covered by the periodic trajectories on the Fermi surface
vanishes at the endpoints of $\, {\hat \gamma}^{\alpha}_{\bf a} $.
The open periodic trajectories demonstrate stability properties inside
the Stability Zone $\, \Omega_{\alpha} \, $ and become unstable
outside $\, \Omega_{\alpha} \, $ on the curve 
$\, {\hat \gamma}^{\alpha}_{\bf a} $. Let us note also here that 
the measure of generic (quasiperiodic) open trajectories on the Fermi 
surface remains finite up to the boundary of a Stability Zone. In general, 
the schematic angle diagram corresponding to a complicated Fermi surface 
can be represented by Fig. \ref{AngleDiagramm}.\footnote{We would like
to emphasize here that for generic Fermi level $\, \epsilon_{F} \, $
the measure of the stable open trajectories on the Fermi surface remains
finite up to the boundary of the ``mathematical Stability Zone''
$\, \Omega_{\alpha} $. However, it is often assumed in the 
literature that this measure vanishes at the boundary
of a Stability Zone on the angle diagram, which seems to be in
accordance with experimental data. As we will see below, the last
circumstance is caused by the existence of an extended
``experimentally observable'' Stability Zone 
$\, \hat{\Omega}_{\alpha} \, $ including an additional region 
$\, \Lambda_{\alpha} $, adjacent to the boundary of $\, \Omega_{\alpha} $,
with very special behavior of trajectories of system 
(\ref{QuasiclassicalEvolution}). It will be also shown, that the
presence of additional Zone $\, \Lambda_{\alpha} \, $ is responsible
also for arising of rather nontrivial analytical regimes of
conductivity behavior in the extended Zone $\, \hat{\Omega}_{\alpha} $.}

\begin{figure}[t]
\begin{center}
\includegraphics[width=1\linewidth]{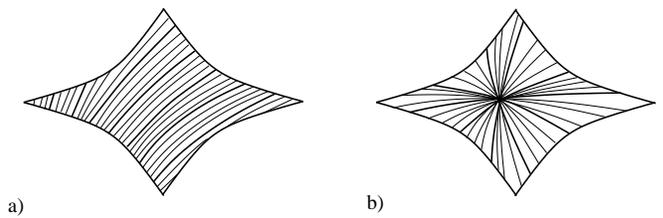}
\end{center}
\caption{The dense nets of directions of $\, {\bf B} \, $ inside
$\,\Omega_{\alpha} $, corresponding to arising of periodic
trajectories, in the cases when the direction orthogonal to
$\, \Gamma_{\alpha} \, $ does not belong to
$\, \Omega_{\alpha} \, $ (a) and when this direction belongs to
$\, \Omega_{\alpha} \, $ (b).}
\label{DenseNet}
\end{figure}

 We can see then that every Stability Zone $\, \Omega_{\alpha} \, $
is actually covered by an everywhere dense net representing the
``special directions'' of $\, {\bf B} \, $ corresponding to different
rational mean directions $\, {\bf a} \, $ of the stable open 
trajectories (Fig. \ref{DenseNet}). The set of all possible 
rational mean directions $\, {\bf a} \, $  of the open trajectories 
is given by two conditions:

1) $\, {\bf a} \,\, \in \,\, \Gamma_{\alpha} \,\, $,

2) $\, C_{\bf a} \, \cap \, \Omega_{\alpha} \,\,\, \neq \,\,\, 
\O \,\, $,

\noindent
where $\, C_{\bf a} \, $ is the big circle orthogonal to the direction
$\, {\bf a}$.

 Everywhere on the net we have the situation
$$\sigma^{kl}_{\infty} \, ({\bf B}/B) \,\,\,\, \neq \,\,\,\,
{\bar \sigma}^{kl}_{\infty} \, ({\bf B}/B) \,\,\,\,\, , $$  
however, the difference
$$|\sigma^{kl}_{\infty} \, ({\bf B}/B) \, - \,
{\bar \sigma}^{kl}_{\infty} \, ({\bf B}/B)| $$ 
decreases with the growth of the numbers 
$\, (m_{1}, \, m_{2}, \, m_{3}) \, $ in (\ref{SpecDirBEq}).
As we will see below, the approximate evaluation for this difference
for complicated Fermi surfaces can be written in the form
\begin{equation}
\label{sigmasigmabardiff}
|\sigma^{kl}_{\infty} \, ({\bf B}/B) \, - \,
{\bar \sigma}^{kl}_{\infty} \, ({\bf B}/B)| \,\,\, \sim \,\,\,
{{\rm ln}^{2} (|m_{1}| + |m_{2}| + |m_{3}|)  \over
\left( |m_{1}| + |m_{2}| + |m_{3}| \right)^{2}}  \,\,  , 
\end{equation}
($k,l = 2,3$).

 In the physical situation, when the values of $\, {\bf B} \, $ are
finite but rather big, the ``jumps'' on the curves 
$\, \gamma^{\alpha}_{\bf a} \, $ become very sharp ``peaks'' on these
curves, which can be called the ``rational peaks'' in the values of
magneto-conductivity according to rationality of the mean direction
of the corresponding open trajectories in ${\bf p}$ - space.

\vspace{1mm}

\begin{figure}[t]
\begin{center}
\includegraphics[width=1\linewidth]{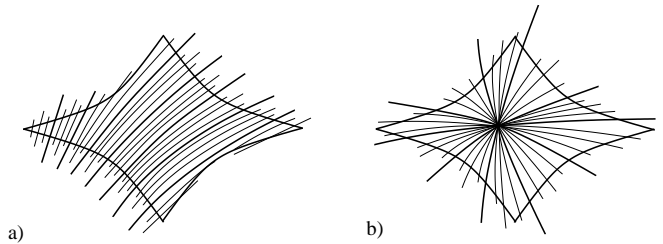}
\end{center}
\caption{The full sets of directions of $\, {\bf B} $
near $\,\Omega_{\alpha} $, corresponding to arising of periodic
trajectories on the Fermi surface.}
\label{DirOutStabZone}
\end{figure}

 The second important point in our picture is connected with the 
behavior of trajectories of system (\ref{QuasiclassicalEvolution}) 
near the boundaries of the Stability Zones $\, \Omega_{\alpha} \, $
outside the Stability Zones. As we have already said, the 
periodic trajectories of system (\ref{QuasiclassicalEvolution}) on 
the Fermi level exist in fact for directions of $\, {\bf B} \, $ 
belonging to the extended curves
$\, {\hat \gamma}^{\alpha}_{\bf a} $, so only at the endpoints
of $\, {\hat \gamma}^{\alpha}_{\bf a} \, $ the measure of the
trajectories with the period $\, {\bf a} \, $ on the Fermi surface 
is equal to zero. As a result, the full set of directions of 
$\, {\bf B} \, $ near $\, \Omega_{\alpha} \, $, corresponding to 
the presence of open trajectories on the Fermi level, represents the
Stability Zone $\, \Omega_{\alpha} \, $ with a set of additional
segments near $\, \Omega_{\alpha} $, which is dense on the
boundary of $\, \Omega_{\alpha} \, $ (Fig. \ref{DirOutStabZone}).
\footnote{Let us say here that the presence of additional segments
adjacent to the Stability Zone on the angle diagram and corresponding
to the presence of periodic open trajectories on the Fermi surface
was discussed in the literature (see e.g. 
\cite{lifpes1,lifpes2,lifkag1,lifkag2,lifkag3,etm,KaganovPeschansky}). 
It is often assumed, however, that this set is given by a finite 
number of segments, corresponding to some special (main) rational 
directions $\, {\bf a} $. As we could find, the first indication 
of the existence of an everywhere dense set of additional directions of 
$\, {\bf B} \, $ near the boundary of the zone of stable open trajectories, 
corresponding to the appearance of periodic trajectories of the system
(\ref{QuasiclassicalEvolution}), was given in the work \cite{GurzhyKop}. 
Let us say here, that the example in the work \cite{GurzhyKop} is connected 
with the system (\ref{QuasiclassicalEvolution}) on a special Fermi surface 
of genus 2, which are not considered as complex according to our terminology. 
However, an analogous property is also true in the general case for 
the Stability Zones on arbitrarily complex Fermi surfaces. This fact 
is connected actually with a special structure of the system 
(\ref{QuasiclassicalEvolution}) (and the existence of topological 
quantum numbers) at $\, {\bf B}/B \in \Omega_{\alpha} \, $,
which is a consequence of rather nontrivial topological theorems 
(\cite{zorich1,dynn1}) in the general case. We would like to emphasize 
here that the additional segments arise for every curve 
$\, \gamma^{\alpha}_{\bf a} \subset \Omega_{\alpha} \, $ 
connected with a rational mean direction of open trajectories
of (\ref{QuasiclassicalEvolution}). As a result, these 
segments form a dense set near the boundary of
$\, \Omega_{\alpha} $, which plays an important role in the formation 
of additional ``experimentally observable'' Zone 
$\, \Lambda_{\alpha} \, $ around the exact mathematical Stability Zone
$\, \Omega_{\alpha} $.}

\begin{figure}[t]
\begin{center}
\includegraphics[width=1\linewidth]{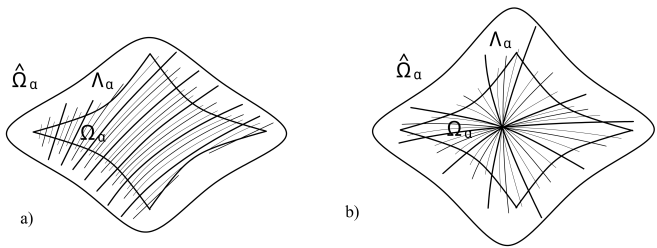}   
\end{center}
\caption{The ``extended Zones'' $\, \hat{\Omega}_{\alpha} \, $ 
observable in experiments.}
\label{ExtStabZone}
\end{figure}

 At the same time, the measure of open trajectories on the Fermi
surface for generic directions 
$\,\, {\bf B}/B \, \notin \, {\hat \gamma}^{\alpha}_{\bf a} \, $
remains finite up to the boundary of the Stability Zone and jumps
abruptly to zero outside $\, \Omega_{\alpha} $. The transformation
of the open trajectories on the boundary of $\, \Omega_{\alpha} \, $
leads in this case to appearance of very long closed trajectories
near the boundary of $\, \Omega_{\alpha} \, $ which are very similar
to open trajectories from experimental point of view up to
very (extremely) high values of $\, B $. We can state then that
in experimental observation of conductivity we observe actually some
``extended Stability Zone'' $\, \hat{\Omega}_{\alpha} $, including the
exact Stability Zone as a subset. The boundaries of the 
``extended Zone'' $\, \hat{\Omega}_{\alpha} \, $ are in a sense
conditional and depend on the maximal values of $\, B \, $ used in
the experiment. We can claim, however, that the experimentally
observable Stability Zone exceeds the exact Stability Zone by a set
of a finite measure for any arbitrary big maximal value of magnetic 
field $\, B_{max} \, $ (see Fig. \ref{ExtStabZone}).

 The set
$\,\, \Lambda_{\alpha} \, = \, \hat{\Omega}_{\alpha}/\Omega_{\alpha} \, $
can be called a ``supplement'' to the Stability Zone
$\,\Omega_{\alpha} \, $ arising in the experimental observation
of conductivity in strong magnetic fields.

 It is possible to introduce also the values
$\,\, \sigma^{kl}_{\infty} \, ({\bf B}/B) \, $  and
$\,\, {\bar \sigma}^{kl}_{\infty} \, ({\bf B}/B) \, , \, $
in the regions $\, \Lambda_{\alpha} $, which will have the same meaning
as in the regions $\,\Omega_{\alpha} $. In these regions we will have 
identically: 
$\, {\bar \sigma}^{22}_{\infty} \, \equiv \,
{\bar \sigma}^{23}_{\infty} \, \equiv \,
{\bar \sigma}^{32}_{\infty} \, \equiv \, 0 \, . \, $ 
Like in $\, \Omega_{\alpha} $, we will have here
$\,\, \sigma^{kl}_{\infty} \, ({\bf B}/B) \, \equiv \,
\,\, {\bar \sigma}^{kl}_{\infty} \, ({\bf B}/B) \, $
for generic directions of $\, {\bf B} \, $ and in general
$\,\, \sigma^{kl}_{\infty} \, ({\bf B}/B) \, \neq \,
\,\, {\bar \sigma}^{kl}_{\infty} \, ({\bf B}/B) \, $     
for $\,\, {\bf B}/B \, \in \, {\hat \gamma}^{\alpha}_{\bf a} \, . \, $

 Both inside the regions $\, \Omega_{\alpha} \, $ and
$\, \Lambda_{\alpha} \, $ the approach of the values of 
$\, \sigma^{kl} \, $ to their limiting values 
$\, \sigma^{kl}_{\infty} \, $ is rather irregular 
(but with some general trend), which makes
the conductivity behavior rather nontrivial from the experimental
point of view even for the values of $\, B $, satisfying the   
formal criterion $\, \omega_{B} \tau \, \gg \, 1 $. 

 First, as we have already noted, the values 
$\, \sigma^{kl}_{\infty} \, ({\bf B}/B) \, $ 
are different from the values
$\,\, {\bar \sigma}^{kl}_{\infty} \, ({\bf B}/B) \, $ 
on the curves $\, {\hat \gamma}^{\alpha}_{\bf a} \, $ both in the 
regions $\, \Omega_{\alpha} \, $ and $\, \Lambda_{\alpha} $. 
These values are continuous along every curve
$\, {\hat \gamma}^{\alpha}_{\bf a} \, $ up to the boundary of the
Zone $\, \hat{\Omega}_{\alpha} \, $ (more precisely, up to the
endpoints of $\, {\hat \gamma}^{\alpha}_{\bf a} $), at the same time,
they are discontinuous in the transverse directions to
$\, {\hat \gamma}^{\alpha}_{\bf a} $. 

 Formally speaking, all the open trajectories in the Zone
$\, \Lambda_{\alpha} \, $ are periodic, so their contribution to 
magneto-conductivity can be represented in the regular form 
(\ref{3DimPeriodicTr}). However, as can be easily seen, the period of 
the trajectories given by formula (\ref{SpecDirBEq}) can be rather large 
for big numbers $\, (m_{1}, m_{2}, m_{3} )$. As a result, the features 
of the regular expansion can be observed actually just on the scales
$$\omega_{B} \tau  \,\,\,\, \gg \,\,\,\, |m_{1}| + |m_{2}| + |m_{3}| 
\,\,\, . $$
At the same time, on the scales of the values of $\, B \, $, 
satisfying the condition
\begin{equation}
\label{BIntermediate}
1 \,\,\,\, < \,\,\,\, \omega_{B} \tau \,\,\,\, < \,\,\,\, 
|m_{1}| + |m_{2}| + |m_{3}| \,\,\, , 
\end{equation}
the contribution of the periodic trajectories reveals the properties,
analogous to those corresponding to generic open trajectories. 
We can state then, that for big values of
$\, |m_{1}| + |m_{2}| + |m_{3}| \, $ the periodic open trajectories
are indistinguishable from the generic ones if the values of $\, B \, $
satisfy restriction (\ref{BIntermediate}). Let us say here also, that 
under the same restrictions on $\, B \, $ the periodic 
open trajectories are also indistinguishable from the long closed 
trajectories in the vicinity of the curve
$\, {\hat \gamma}^{\alpha}_{\bf a} \, $ 
in the Zone $\, \Lambda_{\alpha} $.

 Let us now make some remarks concerning the experimental study of
the magneto-conductivity in (monocrystalline) metals having complicated 
Fermi surfaces. Let us note first that the measuring the 
magneto-conductivity in presence of the open trajectories very often 
is being made for directions of $\, {\bf B} \, $ belonging to the largest
``Stability Zones'', characterized by the biggest area and the highest
symmetry on the angle diagram. Rather often it is convenient also
to fix just a small set of directions of $\, {\bf B} \, $ inside the
Stability Zone and to measure the magneto-conductivity changing
the magnitude $\, B \, $ of the magnetic field. According to the
design of the research facility it appears rather often that the
corresponding set of directions of $\, {\bf B} \, $ represents in
fact some symmetric or ``rational'' points on the angle diagram,
such that all (or a part) the directions belong actually to some 
curves 
$\, {\hat \gamma}^{\alpha}_{\bf a} \, \subset \, 
\hat{\Omega}_{\alpha} \, $ corresponding to rather simple periods
$\, {\bf a} \, $ of the open trajectories. This is exactly the case
when the magneto-conductivity demonstrates the regular behavior
(\ref{SigmaRegExpPeriodic}) under the simple condition
$\, \omega_{B} \tau \, \gg \, 1 $.

 The corresponding magneto-resistance tensor can be written here in 
the form
\begin{multline}
\label{ResistPeriodTr}
\rho_{kl} \,\,\,\, \simeq \,\,\,\,
{m^{*} \over n e^{2} \tau } \, \left(
\begin{array}{ccc}
( \omega_{B} \tau )^{2}  &   \omega_{B} \tau   &
\omega_{B} \tau   \cr
\omega_{B} \tau   &  *  &  *  \cr
\omega_{B} \tau   &  *  &  *
\end{array}  \right)  \,\,\, +  \\
+ \,\, {m^{*} \over n e^{2} \tau } \, \left( 
\begin{array}{ccc}
0  &  *  &  *  \cr
*  &  0  &  *  \cr
*  &  *  &  0
\end{array}  \right) \,\, ( \omega_{B} \tau )^{-1} \,\,\,\, +  \\
+ \,\, {m^{*} \over n e^{2} \tau } \, \left( 
\begin{array}{ccc}
*  &  *  &  *  \cr
*  &  *  &  *  \cr
*  &  *  &  *     
\end{array}  \right) \,\, ( \omega_{B} \tau )^{-2}
\,\,\,\, + \,\,\,\, \dots 
\end{multline}

 In particular, the behavior of the transverse resistivity 
$\, ({\bf j} \perp {\bf B}) \, $ demonstrates a regular quadratic law:
\begin{equation}
\label{TransResPerTr}
\rho_{\perp} \quad \sim \quad B^{2} \, {\rm cos}^{2} \varphi
\,\, + \,\, {\rm const} \quad , \quad \quad 
\omega_{B} \tau \, \gg \, 1 \,\,\, , 
\end{equation}
where $\, \varphi \, $ is the angle with the $x$-axis in our 
coordinate system.

 The situation is rather different in the case when the open trajectories
have an irrational mean direction
\begin{equation}
\label{dbarkappa}
{\hat d} \,\,\,\, \sim \,\,\,\, {\bf q}_{1} \,\, + \,\,
\kappa \, {\bf q}_{2}  \quad  ,  \quad  \quad  \quad
\kappa \, \notin \, \mathbb{Q} 
\end{equation}
in the plane $\, \Gamma_{\alpha} $. From one point of view, the
limiting values of the conductivity tensor are also given here by the
relations (\ref{3DimAnisotropicLimit}) in the limit
$\, \omega_{B} \tau \, \rightarrow \, \infty $. At the same time,
the regular series (\ref{SigmaRegExpPeriodic}), (\ref{ResistPeriodTr})
can not be written in general for generic mean direction of the open 
trajectories. It can be shown also that the asymptotic behavior of 
the values $\, \sigma^{kl} (B) \, $ 
demonstrates here a slower (irregular) approach
to their limiting values 
$\, {\bar \sigma}^{kl}_{\infty} ({\bf B}/B) $. In general, the
asymptotic behavior of $\, \sigma^{kl} (B) \, $ inside 
$\, \Omega_{\alpha} \, $ for irrational mean direction of the open
trajectories can be represented in the form:
\begin{equation}
\label{SigmaOmega}
\sigma^{kl} (B) \,\, = \,\, \left(
\begin{array}{ccc}
a^{2} (B)  &  b (B)  &  c (B)  \cr
- \, b (B) \, &  \, {\bar \sigma}^{22}_{\infty} + q^{2} (B) \,\, &  
{\bar \sigma}^{23}_{\infty} + r (B)  \cr
- \, c (B)  &  {\bar \sigma}^{23}_{\infty} - r (B)  &
{\bar \sigma}^{33}_{\infty} + p^{2} (B)  
\end{array}  \right) \,\, ,  
\end{equation}
where the functions $\, a (B) $, $\, b (B) $, $\, c (B) $
have the following asymptotic behavior
\begin{equation}
\label{AsymptBehavior}
a (B) \,\,\, \sim \,\,\, b (B) \,\, \sim \,\,\, 
c (B) \,\,\, \sim \,\,\, \left( \omega_{B} \tau \right)^{-1} 
\end{equation}

 At the same time, the functions
$\, q (B) $, $\, p (B) \, $ have the following
``general trend''
\begin{equation}
\label{Trend}
q (B) \,\,\, \sim \,\,\, p (B) \,\,\, \sim \,\,\,
{{\rm ln} \,\, \omega_{B} \tau  \over \omega_{B} \tau}  
\end{equation}

\begin{figure}[t]
\begin{center}
\includegraphics[width=1\linewidth]{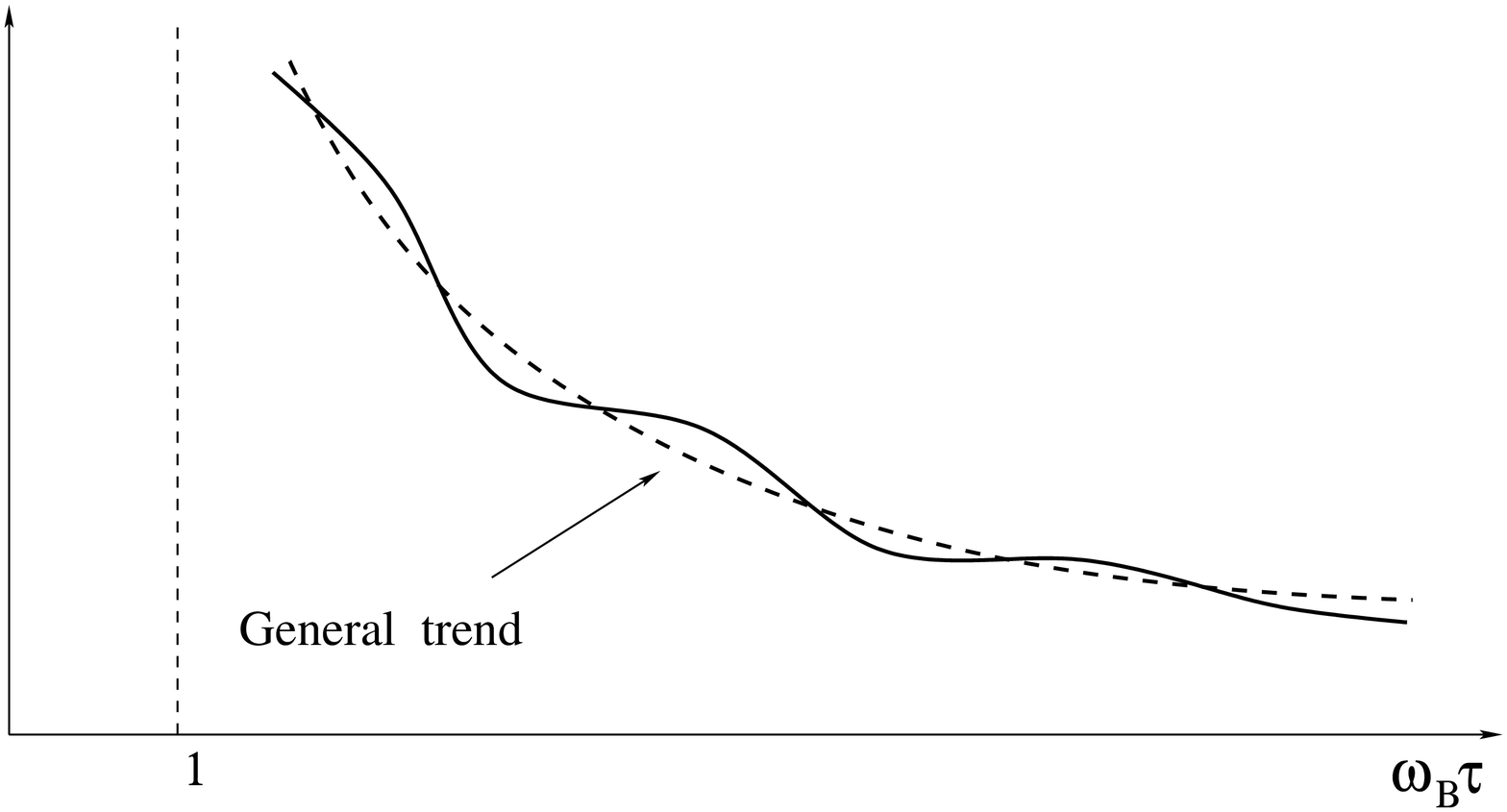}
\end{center}
\caption{The schematic sketch of the behavior of the functions
$\, q (B) \, $ and  $\, p (B) \, $
in the region $\, \omega_{B} \tau \gg 1 $.}
\label{Steps}
\end{figure}

 Let us emphasize here that the relation $\, \sim \, $ in (\ref{Trend})
means just some general trend in the behavior of the corresponding
functions which admits additional irregular corrections on the finite
scales. In general, these corrections can be characterized as
``cascades of step-like perturbations'' with the structure defined
by the properties of the number $\, \kappa \, $ (Fig. \ref{Steps}).
Let us note here that the behavior of $\, q (B) $, $\, p (B) \, $
can demonstrate noticeable local deviations from the trend for special 
irrational $\, \kappa \, $ which can be approximated by rational numbers 
with a high precision. In particular, the last circumstance can be 
important for the directions of $\, {\bf B} \, $ close to  the special 
directions $\, {\bf B}/B \, \in \, \gamma^{\alpha}_{\bf a} \, $ which 
were discussed above.

 The behavior of the function $\, r (B) \, $ is even more complicated
compared with that of $\, q (B) \, $ and $\, p (B) \, $. Here we
would like to suggest just the following ``reliable'' restriction
\begin{equation}
\label{rBTrend}
\left| r (B) \right| \,\,\,\,\, \leq \,\,\,\,\,
{{\rm ln} \,\, \omega_{B} \tau  \over \omega_{B} \tau}
\end{equation}
on the $\, B $ - dependence of the function $\, r (B) \, $ which
can demonstrate rather irregular behavior at different values of 
$\, B \, $.

 Another essential feature of the relations 
(\ref{Trend}) - (\ref{rBTrend}) is that
the simple condition $\, \omega_{B} \tau \, \gg \, 1 \, $ is not
sufficient here to get the values $\, \sigma^{kl}_{\infty} \, $
with a good precision. The latter circumstance can be understood from
the fact that the values $\, \sigma^{kl}_{\infty} \, $ are given now 
by the averaging of certain dynamical values over a complicated part
of the Fermi surface but not a one-dimensional circle.
As a result, relations (\ref{Trend}) -	(\ref{rBTrend}) can contain
in fact a large dimensionless coefficient, determined by the geometry
of the Fermi surface. We should expect then, that in 
the interval of not extremely high values of $\, B \, $ the geometric 
characteristics of the carriers of open trajectories will play important 
role in the behavior of $\, \sigma^{kl} (B) \, $ and can give noticeable 
deviations from the asymptotic regime (\ref{Trend}). As a consequence, 
we should expect also that in the interval of not extremely strong 
magnetic fields the interplay between the limiting values of 
$\, \sigma^{kl} (B) \, $ and the corrections depending on $\, B \, $ 
will play an essential role in conductivity behavior.

 In general, the picture described above demonstrates the main
features of the behavior of $\, \sigma^{kl} (B) \, $ inside the
mathematical Stability Zones $\, \Omega_{\alpha} \, $.

\vspace{1mm}

 In the Zone $\, \Lambda_{\alpha} \, $ we have to introduce the
function $\, \lambda ({\bf B}/B) \, $,
$\,\,\, 1 \leq \lambda < \infty $,
characterizing the mean size of long closed trajectories for generic
directions of $\, {\bf B} $. In common, the function
$\, \lambda ({\bf B}/B) \, $ can be defined as the ratio of the length
of such trajectories in the $\, {\bf p}$ - space to the size of the
Brillouen zone. Let us note that the definition of the function
$\, \lambda ({\bf B}/B) \, $ has actually rather approximate
character. It is natural then to introduce the
``intermediate'' stable values of conductivity
$\, {\bar \sigma}^{kl}_{int} ({\bf B}/B) \, $, 
$\,\,\, (k, l \, = \, 2, 3) $,
which are proportional to the measure of the long closed trajectories
on the Fermi surface. For the behavior of conductivity for not
extremely strong magnetic fields
$\,\, 1 \ll \omega_{B} \tau < \lambda ({\bf B}/B) \, $
and generic directions of $\, {\bf B} \, $ in $\, \Lambda_{\alpha} \, $
we can then use the following relations
\begin{equation}
\label{SigmaLowBTau}
\sigma^{kl} (B) \,\, = \,\, \left(
\begin{array}{ccc}
a^{2} (B)  &  b (B)  &  c (B)  \cr
- \, b (B) \, &  \, {\bar \sigma}^{22}_{int} + q^{2} (B) \,\, &
{\bar \sigma}^{23}_{int} + r (B)  \cr
- \, c (B)  &  {\bar \sigma}^{23}_{int} - r (B)  &
{\bar \sigma}^{33}_{int} + p^{2} (B)
\end{array}  \right)  
\end{equation}
($ 1 \ll \omega_{B} \tau < \lambda $), with the same remarks about the 
functions  $\, a (B) $, $\, b (B) $, $\, c (B) $,
$\, q (B) $, $\, r (B) $, $\, p (B) $.

\vspace{1mm}

 To describe the asymptotic behavior of conductivity in the region of 
extremely strong magnetic fields  \linebreak
$\omega_{B} \tau  \gg  \lambda ({\bf B}/B) \, $
for the same directions of $\, {\bf B} \, $ it is convenient to consider
separately the symmetric $\, s^{kl} (B) \, $ and the anti-symmetric part
$\, a^{kl} (B) \, $ of the tensor $\, \sigma^{kl} (B) \, $. As will be
shown below, the corresponding parts of $\, \sigma^{kl} (B) \, $ can
be approximated here by the following expressions
\begin{multline}
\label{SigmaHighBTauSymm}
s^{kl} (B) \,\,\,\,\, \simeq \,\,\,\,\,   \left(
\begin{array}{ccc}
0 \,\,\,\,\, &  0  &  \,\, 0   \cr
0  \,\,\,\,\, &  0  &  \,\, 0   \cr
0  \,\,\,\,\, &  0  &  \sigma^{\prime 33}
\end{array}
\right)  \,\,\, +   \cr
+ \,\,\, {n e^{2} \tau \over m^{*}} \left(
\begin{array}{ccc}
(\omega_{B} \tau)^{-2}  &  \lambda (\omega_{B} \tau)^{-2}  &
\lambda (\omega_{B} \tau)^{-2}   \cr
\lambda (\omega_{B} \tau)^{-2}  &
\lambda^{2} (\omega_{B} \tau)^{-2}  &
\lambda^{2} (\omega_{B} \tau)^{-2}  \cr
\lambda (\omega_{B} \tau)^{-2}  &
\lambda^{2} (\omega_{B} \tau)^{-2}  &
\lambda^{2} (\omega_{B} \tau)^{-2}
\end{array}
\right)  \,\,\,\,\, ,
\end{multline}
where
\begin{equation}
\label{SigmaPrime33}
\sigma^{\prime 33} \,\,\,\, < \,\,\,\, {\bar \sigma}^{33}_{int}  
\quad  ,
\end{equation}
and
\begin{equation}
\label{SigmaHighBTauAntiSym}
a^{kl} (B) \,\,\,\,\, \simeq \,\,\,\,\,
{n e^{2} \tau \over m^{*}} \left(
\begin{array}{ccc}
0  &  (\omega_{B} \tau )^{-1}  &  (\omega_{B} \tau )^{-1}  \cr
(\omega_{B} \tau )^{-1}  &  0  &  (\omega_{B} \tau )^{-1}  \cr
(\omega_{B} \tau )^{-1}  &  (\omega_{B} \tau )^{-1}  &  0
\end{array}
\right) \,\,\,\, ,
\end{equation}
respectively.

 Let us note, that the interplay between the symmetric and the 
anti-symmetric parts of $\, \sigma^{kl} (B) \, $ has here rather
nontrivial character. Thus, we can see that the values
$\, \sigma^{23} (B) \, $ and $\, \sigma^{32} (B) \, $ are defined
mostly by the symmetric part in the interval
$$\lambda ({\bf B}/B) \,\,\,\, \leq \,\,\,\, \omega_{B} \tau
\,\,\,\, \leq \,\,\,\, \lambda^{2} ({\bf B}/B)  \quad  , $$
and by the anti-symmetric part for 
$\,\, \omega_{B} \tau \,\, > \,\, \lambda^{2} ({\bf B}/B) \, $.
Another important feature in the behavior of conductivity is given here
by the relation (\ref{SigmaPrime33}), which indicates the suppression
of the conductivity along the direction of $\, {\bf B} \, $ in the 
region 
$\,\, \omega_{B} \tau \,\,\, \gg \,\,\, \lambda ({\bf B}/B)
\,\,\, \gg \,\,\,  1  \, $.

\vspace{1mm}

 The values of $\, \lambda ({\bf B}/B) \, $ are equal to $\, 1 \, $
on the exterior boundary of the region $\, \Lambda_{\alpha} $. At the
same time, we should put $\, \lambda ({\bf B}/B) = \infty \, $ on the
curves $\, {\hat \gamma}^{\alpha}_{\bf a} \, $ in 
$\, \Lambda_{\alpha} \, $ and on the boundary between the regions
$\, \Omega_{\alpha} \, $ and $\, \Lambda_{\alpha} $. It's not
difficult to see that the functions 
$\, {\bar \sigma}^{kl}_{int} ({\bf B}/B) \, $ lose their sense
and are not well defined on the exterior boundary of 
$\, \Lambda_{\alpha} $. At the same time, we can write
$\, {\bar \sigma}^{kl}_{int} ({\bf B}/B) \, = \, 
{\bar \sigma}^{kl}_{\infty +} ({\bf B}/B) $, where
$\, {\bar \sigma}^{kl}_{\infty +} ({\bf B}/B) \, $ are the boundary
values of $\, {\bar \sigma}^{kl}_{\infty} ({\bf B}/B) \, $ inside
$\, \Omega_{\alpha} $, on the boundary between 
$\, \Omega_{\alpha} \, $ and $\, \Lambda_{\alpha} $.

 According to the above remarks, we can see actually, that the
exact boundary between the regions 
$\, \Omega_{\alpha} \, $ and $\, \Lambda_{\alpha} \, $ 
is not usually observable in experiments.

 We can see that the picture described above makes the analytic
structure of tensor $\, \sigma^{kl} ({\bf B}) \, $ in the
Zone $\, \hat{\Omega}_{\alpha} \, $ rather complicated. 
The most complicated situation can arise in that part of the 
Zone $\, \Lambda_{\alpha} \, $ where we have the relation
$\,\, \omega_{B} \tau \, \simeq \, \lambda ({\bf B}/B) \,\, $ 
for typical values of $\, B \, $ in the experiment. In the last 
case the behavior of $\, \sigma^{kl} (B) \, $ can demonstrate 
the most complicated trends, which are intermediate between  
(\ref{SigmaLowBTau}) and 
(\ref{SigmaHighBTauSymm}) - (\ref{SigmaHighBTauAntiSym}). 
In experimental study of the behavior of conductivity it can be 
convenient to approximate the corresponding regimes with the aid 
of intermediate powers of $\, \omega_{B} \tau \, $ and write
$$\sigma^{kl} (B) \,\, \simeq \,\,
{n e^{2} \tau \over m^{*}} \left(
\begin{array}{ccc}
( \omega_{B} \tau )^{-2}  &  ( \omega_{B} \tau )^{-1}  &
( \omega_{B} \tau )^{-1}  \cr
( \omega_{B} \tau )^{-1}  &  ( \omega_{B} \tau )^{-2\mu}  &
( \omega_{B} \tau )^{-\nu}  \cr
( \omega_{B} \tau )^{-1}  &  ( \omega_{B} \tau )^{-\nu}  &  *
\end{array}   \!  \right) $$
($0 \, \leq \, \mu, \nu \, \leq \, 1\, , \,\,\, \mu \, \simeq \, \nu $)
in the intervals of rather high values of $\, B \, $.

 As an example, putting approximately $\, \mu \, = \, \nu \, $, 
we can write the relation for the resistivity behavior in strong 
magnetic fields:
$$\rho_{kl} (B) \,\, \simeq \,\,
{m^{*} \over n e^{2} \tau} \left(
\begin{array}{ccc}
( \omega_{B} \tau )^{2 - 2\mu}  & \omega_{B} \tau  &
(\omega_{B} \tau )^{1 - \mu}  \cr
\omega_{B} \tau  &  *  &  *  \cr
( \omega_{B} \tau )^{1 - \mu}  &  *  &  *
\end{array}  \right) $$
 
 In particular, we get the following approximation for the resistivity 
in the plane orthogonal to $\, {\bf B} \, $:
$$\rho_{\perp} \quad \sim \quad B^{2 - 2\mu} \,\, 
{\rm cos}^{2} \varphi \,\,\, + \,\,\, {\rm const} \quad , \quad \quad
\omega_{B} \tau \, \gg \, 1 \,\,\, , $$
where $\, \varphi \, $ is the angle with the $x$-axis in our
coordinate system.

\begin{figure}[t]
\begin{center}
\includegraphics[width=1\linewidth]{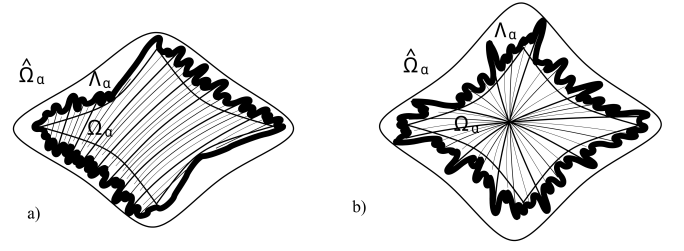}
\end{center}
\caption{The Zone of the most complicated behavior of conductivity
tensor (black) in the ``experimentally observable Stability Zone''
$\, \hat{\Omega}_{\alpha}$.} 
\label{ComplBehaviorZone}
\end{figure}

 We should say, however, that the powers $\, \mu \, $ and $\, \nu \, $
play here just the role of local approximating parameters and are 
unstable both with respect to rotations of $\, {\bf B} \, $ and big 
changes of its absolute value. Let us note also that the corresponding 
part of $\, \Lambda_{\alpha} \, $ has in general rather complicated 
structure view the complicated behavior of the function 
$\, \lambda ({\bf B}/B) \, $ (Fig. \ref{ComplBehaviorZone}) .

 We can see then that the experimental study of galvano-magnetic
phenomena in the presence of open trajectories can reveal rather 
non-trivial behavior of magneto-conductivity, which is caused both
by the complicated dependence of conductivity on the value of $\, B \, $
and possible small variations of directions of $\, {\bf B} \, $ in
experiment. The general analytical picture described above can be used
for description of the dependence of the magneto-conductivity both
on the value of $\, B \, $ and the direction of $\, {\bf B} \, $
inside the ``experimentally observable'' Stability Zone 
$\, \hat{\Omega}_{\alpha} $. Let us say again that the picture
described above is based just on geometric consideration of the
trajectories of system (\ref{QuasiclassicalEvolution}) and does
not include many other essential effects which can be important in
real metals.

 At the end of this section, we would like to note that the picture
described above is based on the assumption that the Fermi surface of
a metal has a complicated form. This assumption means, in particular,
that the carriers of the stable open trajectories have the form defined 
by general topological theorems for generic dispersion law. This form
will be described in the next section and has some special features
for generic Fermi surfaces $\, S_{F} $. However, for special simple
Fermi surfaces the carriers of open trajectories can have much simpler
form, which will result also in simpler behavior of conductivity in
comparison with the described above.

 As an example of the simple Fermi surface, we can consider just
a pair of slightly deformed integral planes in the $\, {\bf p}$ - space
(like for some classes of organic metals). It is easy to see that the
exact mathematical Stability Zone $\, \Omega \, $ can be identified
here with the whole sphere $\, \mathbb{S}^{2} \, $ and coincides with
the ``experimentally observable'' Stability Zone $\, \hat{\Omega} $.
As a result, we can not observe here the effects, connected with the
reconstruction of the open trajectories, typical for the Zones
$\, \Lambda_{\alpha} \, $. The ``rational peaks'' in the conductivity
can be observable here for the surfaces having essential deformations
(see e.g. \cite{OsKagMiura}), however, they decrease exponentially
with the growth of the numbers $\, (m_{1}, m_{2}, m_{3}) $.
As a result, we can expect here the experimental evidence of just a
finite number of the lines $\, \gamma_{\bf a} \, $ instead of the
complicated angle dependence of conductivity, observable in the Zones
$\, \Omega_{\alpha} \, $. It is easy to see also, that for the case of
vanishing amplitude of deformation of the planes the conductivity tensor
demonstrates just a very fast approach to its limiting form given by
formula (\ref{3DimAnisotropicLimit}).

 In the next sections we will make more detailed consideration of 
the effects described above using the kinetic approach to the 
transport phenomena. In general, all the statements above can be 
obtained within the $\, \tau$ - approximation for the kinetic 
equation for the electron gas.

\vspace{5mm}

\section{The geometry of the stable open trajectories and the 
conductivity tensor.}
\setcounter{equation}{0}

 Let us start with the description of generic picture arising 
on the Fermi surface in the case of presence of the stable open
trajectories at $\, \epsilon \, = \, \epsilon_{F} \, $ (\cite{dynn3}).
Everywhere below we will assume that the Fermi surface represents
a smooth non-selfintersecting 3-periodic surface in 
$\, {\bf p}$ - space. 

 Consider the general picture of intersection of a 3-periodic surface
in $\, {\bf p}$ - space by the planes orthogonal to $\, {\bf B} \, $
(Fig. \ref{GeneralSurface}). First, let us look at the closed
trajectories in $\, {\bf p}$ - space if they present on the Fermi 
surface.

\begin{figure}[t]
\begin{center}
\includegraphics[width=0.65\linewidth]{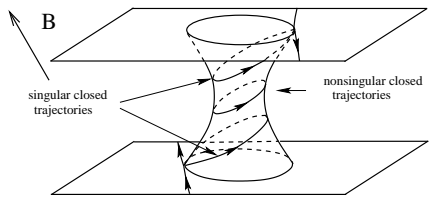}
\end{center}
\caption{A part of the Fermi surface representing a cylinder of closed 
trajectories bounded by singular trajectories.}
\label{CylinderClosedTr}

\vspace{5mm}

\end{figure}

 The non-singular closed trajectories in $\, {\bf p}$ - space are
always locally stable and can be combined into connected cylinders
consisting of trajectories of this type. In the case when a connected
component of the Fermi surface does not consist of the closed
trajectories only, these cylinders have finite heights and are
bounded by singular closed trajectories (Fig. \ref{CylinderClosedTr}).

\begin{figure}[t]
\begin{center}
\includegraphics[width=1\linewidth]{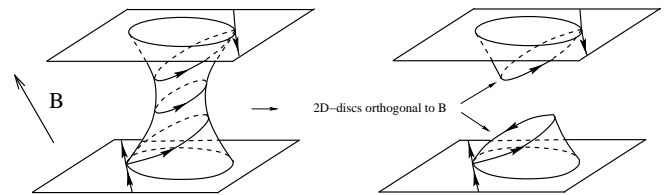}
\end{center}
\caption{The reconstruction of the Fermi surface given by elimination
of nonsingular closed trajectories and filling the holes by the discs
orthogonal to $\, {\bf B} $.}
\label{Reconstruction}
\end{figure}

 Let us make now the following reconstruction of the Fermi surface:

 We remove all the cylinders of closed trajectories from the Fermi
surface and fill the created holes by 2-dimensional discs orthogonal
to $\, {\bf B} \, $ (Fig. \ref{Reconstruction}).

 It is easy to see that the reconstructed Fermi surface carries the
same open electron trajectories as the initial one, so we can investigate
the behavior of open trajectories on the reconstructed surface.

 The reconstructed Fermi surface represents a 3-periodic surface
which contains in general infinitely many connected components in
$\, {\bf p}$ - space. One important theorem can be formulated
about the connected components of the reconstructed Fermi surface
in the case of presence of stable open trajectories
(\cite{zorich1,dynn1,dynn3}):

\vspace{1mm}

 Every connected component  of the reconstructed Fermi surface
carrying open electron trajectories represents a periodic
deformation of an integral plane in $\, {\bf p}$ - space.

\vspace{1mm}

 Let us remind here that the plane $\, \Gamma \, $ is called integral
if it is generated by any two reciprocal lattice vectors. So, every 
connected component  of the reconstructed Fermi surface has two 
independent periods represented by two vectors of the reciprocal lattice. 
Let us note also that the picture described above is evidently
stable with respect to all small rotations of $\, {\bf B} \, $ view
the local stability of nonsingular closed trajectories on the Fermi
surface.

 All the connected components of the reconstructed Fermi surface are
parallel to each other which leads to the coinciding of the mean 
directions of all stable open trajectories in $\, {\bf p}$ - space.
The last property was called in \cite{BullBrazMathSoc,JournStatPhys} 
the Topological Resonance and plays very important role in the
experimental observation of the anisotropic regimes of conductivity
behavior and of the Topological Numbers corresponding to a given
Stability Zone.

 In this paper we will use the picture described above as the general 
picture arising in the case of presence of stable non-closed electron 
trajectories on the Fermi surface. Let us note here that this
picture can have some additional features for some special directions
of $\, {\bf B} \, $ within the Stability Zone (see \cite{dynn3}) 
which can be also detected from the experimental point of view
(\cite{BullBrazMathSoc,JournStatPhys}). These phenomena, however,
have a non-generic character and will not be considered here.

 After the factorization over the vectors of the reciprocal lattice
every connected component of the reconstructed Fermi surface becomes
a two-dimensional torus $\, \mathbb{T}^{2} \, $ embedded in
$\, \mathbb{T}^{3} $. The number of non-equivalent components in 
$\, \mathbb{T}^{3} \, $ is always finite and represents an even number.

\begin{figure}[t]
\begin{center}
\includegraphics[width=1\linewidth]{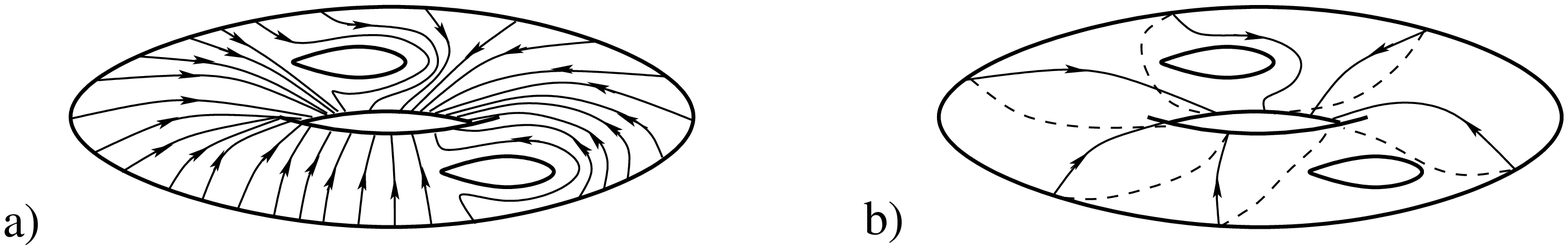}
\end{center}
\caption{Schematic topological description of the dense (non-periodic)
trajectory on $\, \mathbb{T}^{2} \, \setminus \cup \, D^{2}_{i} \, $
and closed (periodic in $\, {\bf p} $ - space) trajectory on the same
manifold after the factorization over the reciprocal lattice vectors.}
\label{TrajectOnT2}
\end{figure}

 The open (in $\, {\bf p} $ - space) trajectories of system
(\ref{QuasiclassicalEvolution}) on the Fermi level are given by the
intersections of the components of the reconstructed Fermi surface
with the planes orthogonal to $\, {\bf B} \, $ and represent
irrational coverings of the corresponding tori $\, \mathbb{T}^{2} \, $
(except the discs $\, D^{2} $) in generic situation
(Fig. \ref{TrajectOnT2} a). However, for
special directions of $\, {\bf B} \, $ the intersection of the plane
orthogonal to $\, {\bf B} \, $ with the integral plane 
$\, \Gamma_{\alpha} \, $ (${\bf B} \, \in \, \Omega_{\alpha}$)
can have rational (integral) direction, which means in fact that the 
corresponding trajectories of
(\ref{QuasiclassicalEvolution}) become periodic in 
$\, {\bf p} $ - space. The corresponding trajectories in 
$\, \mathbb{T}^{3} \, $ become now closed curves on the tori
$\, \mathbb{T}^{2} \, $ and are not dense anymore on the sets
$\, \mathbb{T}^{2} \, \setminus \cup \, D^{2}_{i} $ ,
which can be called the ``carriers of open trajectories'' on the
Fermi surface (Fig. \ref{TrajectOnT2} b).

 As we can see, the generic and the periodic open trajectories 
demonstrate an obvious difference in their statistical properties
on the ``carriers of open trajectories''. As will see below, this 
circumstance will play important role for the behavior of the 
conductivity tensor $\, \sigma^{kl} ({\bf B}) \, $
in the limit $\, \omega_{B} \tau \, \rightarrow \, \infty \, $.

 We can see here that the periodic trajectories can not appear for
completely irrational directions of $\, {\bf B} \, $ since the plane
orthogonal to $\, {\bf B} \, $ should contain at least one vector of 
the reciprocal lattice, giving the mean direction of such
trajectories. Let us note that the class 
of rationality of a direction of magnetic field $\, {\bf B} \, $ 
is defined as the number of linearly
independent reciprocal lattice vectors lying in the plane orthogonal
to $\, {\bf B} $. In particular, the direction 
$\, \hat{\bf B}^{0}_{\alpha} $, orthogonal to the plane
$\Gamma_{\alpha} $, represents a purely rational direction according
to this definition. 

 In general, it is not difficult to see that the directions of
$\, {\bf B} \, $  in a Stability Zone $\, \Omega_{\alpha} $,
corresponding to periodic open trajectories, form a dense net of
one-dimensional curves, which do not intersect each other in 
$\, \Omega_{\alpha} \, $ in the case when the direction
$\, \hat{\bf B}^{0}_{\alpha} \perp \Gamma_{\alpha} \, $ does
not belong to $\, \Omega_{\alpha} \, $ (Fig. \ref{DenseNet} a)
or all intersect at the point 
$\, \hat{\bf B}^{0}_{\alpha} \,\, $
if $\,\, \hat{\bf B}^{0}_{\alpha}  \in  \Omega_{\alpha} \, $
(Fig. \ref{DenseNet} b).

 As we already said above, all the one-dimensional curves
$\, \gamma^{\alpha}_{\bf a} \, $ should be actually extended outside
the Stability Zone $\, \Omega_{\alpha} \, $ since the periodic open
trajectories exist in fact for a wider set of directions of 
$\, {\bf B} \, $ surrounding $\, \Omega_{\alpha} $. The periodic open
trajectories become unstable with respect to small rotations of
$\, {\bf B} \, $ outside $\, \Omega_{\alpha} $. We have to note also
that the difference between the curves $\, \gamma^{\alpha}_{\bf a} \, $
and $\, {\hat \gamma}^{\alpha}_{\bf a} \, $ decreases with increasing
of the ``denominator'' of a rational mean direction of the periodic
trajectories and vanishes in the limit 
$\, | {\bf a} | \, \rightarrow \, \infty \, $
(Fig. \ref{DirOutStabZone}).

\vspace{1mm}

 Let us consider now the carriers of open trajectories in more detail.

\vspace{1mm}

\begin{figure}[t]
\begin{center}
\includegraphics[width=1\linewidth]{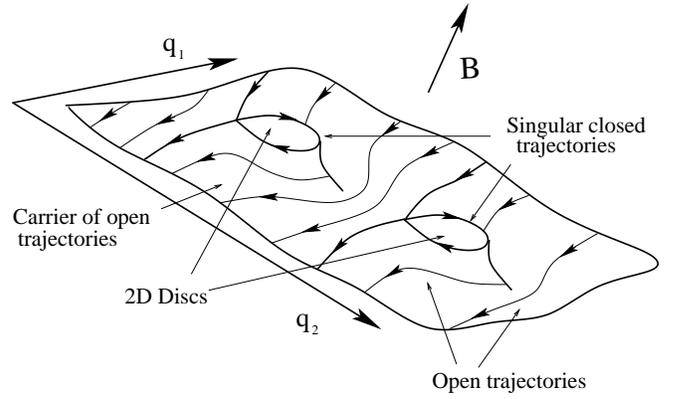}
\end{center}
\caption{The connected component of the reconstructed Fermi surface
representing a periodically deformed integral plane in 
$\, {\bf p} $ - space with periods 
$\, {\bf q}_{1}, \, {\bf q}_{2} \, \in \, L^{*}$.} 
\label{PeriodicPlane}
\end{figure}

 The general form of connected components of the reconstructed Fermi 
surface is shown at Fig. \ref{PeriodicPlane} and represents a periodically  
deformed integral plane in $\, {\bf p} $ - space with two periods
$\, {\bf q}_{1} $, $\, {\bf q}_{2} $, $\, $ given by two independent 
reciprocal lattice vectors.

 The carriers of open trajectories in $\, {\bf p} $ - space
represent the same periodically deformed integral planes without
``artificial'' two-dimensional discs $\, D^{2} $, attached after
the removal of closed trajectories. Let us exclude here the
non-generic case when the magnetic field $\, {\bf B} \, $ is
orthogonal to the plane $\, \Gamma_{\alpha} \, $
and consider the generic case
\begin{equation}
\label{NonParallel}
{\bf B} \,\,\, \nparallel \,\,\, 
\left[ {\bf q}_{1} \times {\bf q}_{2} \right] 
\end{equation}

 We will assume in our considerations that the $z$ - axis is always
chosen along the direction of $\, {\bf B} \, $ while the
$x$ - $\, $ and $y$ - axis are orthogonal to $\, {\bf B} $. In the
generic situation described above the values of $\, p_{z} \, $ 
separate all different trajectories at a given component and 
$\, p_{z} \, $ can be chosen as a ``coordinate'' on the corresponding 
carrier of open trajectories.

 The second coordinate on the carriers of open trajectories is
naturally given by the parameter $\, t $, measured along the
trajectories according to system (\ref{QuasiclassicalEvolution}).

 In general, a natural system of local coordinates in 
$\, {\bf p} $ - space (or $\, \mathbb{T}^{3} $) is given by the
triple $\, (p_{z}, \, t , \, \epsilon ) \, $ if we consider system 
(\ref{QuasiclassicalEvolution}) for all values of $\, \epsilon \, $.
The coordinates $\, (p_{z}, \, t , \, \epsilon ) \, $ are naturally
connected with system (\ref{QuasiclassicalEvolution}) and possess also
one more remarkable property:
\begin{equation}
\label{dpzdtdl}
d p_{z} \, d t \, d \epsilon \,\,\,\, \equiv \,\,\,\,
{c \over e B} \,\, d p_{x} \, d p_{y} \, d p_{z} 
\end{equation}

 The last property is caused in fact by the Hamiltonian properties
of system (\ref{QuasiclassicalEvolution}), which can be considered
as a Hamiltonian system with the Hamiltonian $\, \epsilon ({\bf p}) \, $
and the Poisson bracket
$$\{ p_{1} \, , \, p_{2} \} \,\,\, = \,\,\, {e \over c} \,\, B 
\,\,\, , \quad \{ p_{2} \, , \, p_{3} \} \,\, = \,\, 0
\,\,\, , \quad \{ p_{3} \, , \, p_{1} \} \,\, = \,\, 0 $$
(${\bf B} \, \parallel \, \nabla z $).

 It can be seen also that the integration over the layer restricted by
the energy values $\, \epsilon_{0} \, $ and 
$\, \epsilon_{0} \, + \, d \epsilon \, $ can be represented by the
integration over the energy surface 
$\, \epsilon ({\bf p}) \, = \, \epsilon_{0} \, $ according to the
formula
\begin{multline}   
\label{LayerIntegration}
\iiint_{\epsilon_{0} < \epsilon ({\bf p}) < \epsilon_{0} + d \epsilon}
\,\, f \left( p_{x}, \, p_{y}, \, p_{z} \right) \,\, 
d p_{x} \, d p_{y} \, d p_{z}   \quad  =   \\
= \quad
{e B \, d \epsilon \over c}  \iint_{\epsilon ({\bf p}) = \epsilon_{0}}
\,\, f \left( p_{z}, \, t , \, \epsilon_{0} \right) \,\,
d p_{z} \, d t
\end{multline}

 Formula (\ref{LayerIntegration}) gives in fact very convenient way
of statistical averaging of any value over the Fermi distribution in
presence of magnetic field.

\vspace{1mm}

\begin{figure}[t]
\begin{center}
\includegraphics[width=1\linewidth]{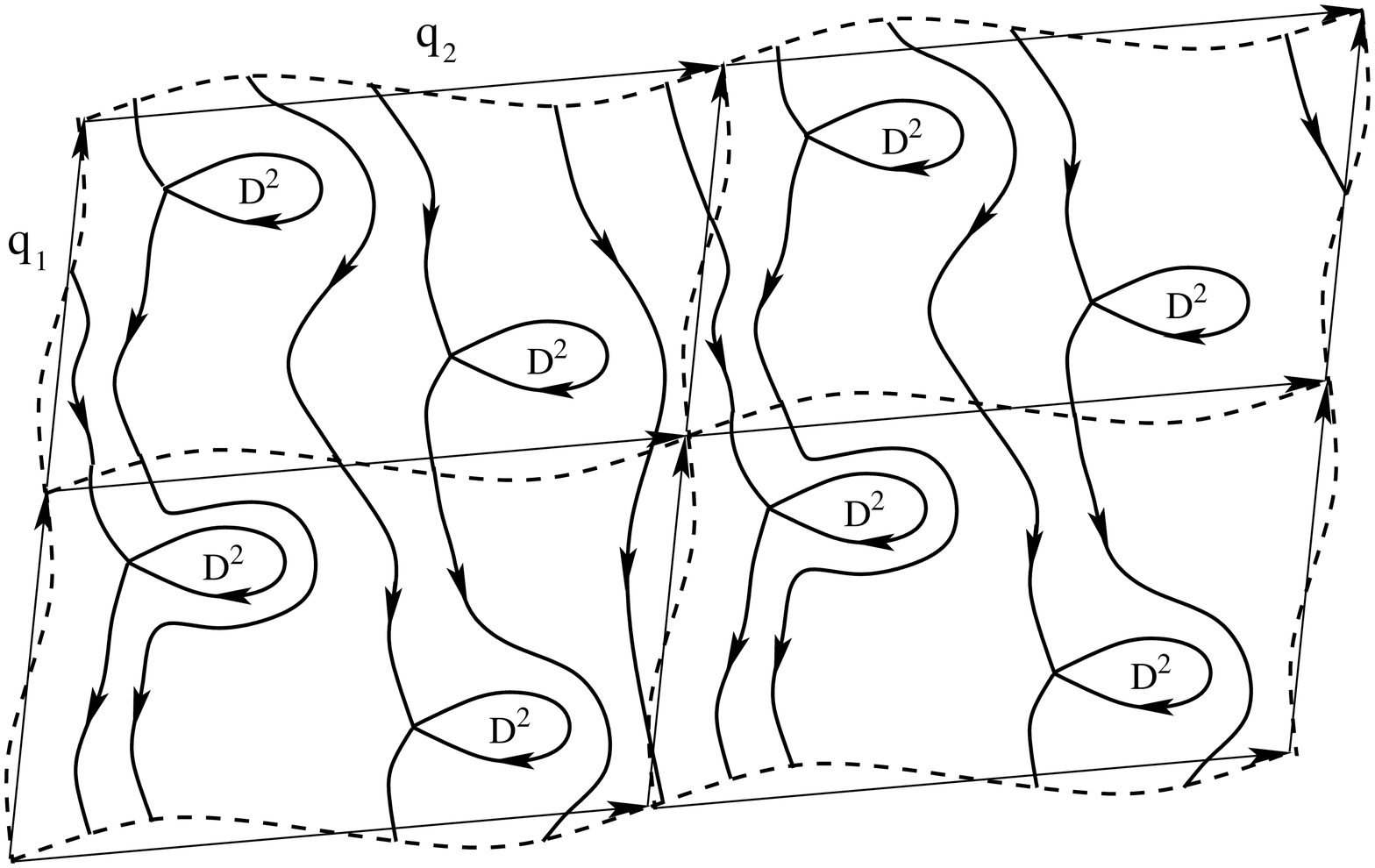}
\end{center}
\caption{The periodic set of the fundamental domains on a connected
component of the reconstructed Fermi surface.}
\label{PeriodicSet}
\end{figure}

 It's not difficult to see that every connected component of the
reconstructed Fermi surface can be divided into a periodic set of
the fundamental (minimal) domains identical to each other
(Fig. \ref{PeriodicSet}).

 It can be seen also that the fundamental domains can be chosen in the
form of ``curvilinear'' parallelograms with the sides, corresponding 
to the vectors $\, {\bf q}_{1} \, $ and $\, {\bf q}_{2} $. The division
of the connected components of the reconstructed Fermi surface gives
also a natural division of the corresponding carriers of open 
trajectories into fundamental domains.

 All the fundamental domains are physically equivalent to each other,
so every domain represents in fact the full set of the physical states
represented by the whole component. The two-dimensional tori
$\, \mathbb{T}^{2} \subset \mathbb{T}^{3} \, $, which we
discussed above, can be considered also as any of the fundamental 
domains with the identified opposite sides. As we said already, every 
nonsingular open trajectory becomes then a smooth curve on the torus 
$\, \mathbb{T}^{2} \, $ which can be open or closed (periodic in 
$\, {\bf p}$ - space) depending on the direction of $\, {\bf B} \, $
(Fig. \ref{TrajectOnT2}). 

 The difference between the closed and the non-closed trajectories on 
the tori $\, \mathbb{T}^{2} \, $ causes an evident difference in the 
averaging of any of the physical quantities over time on the 
trajectories of this kind. Thus, we can claim that any time average 
over generic (non-closed on $\, \mathbb{T}^{2} $) trajectory coincides 
with the  averaging of the corresponding quantity over the carrier of 
open trajectories, which depends smoothly on the direction of 
$\, {\bf B} \, $ inside the ``Stability Zone''. On the other hand, 
the time average over a periodic (closed on $\, \mathbb{T}^{2} $) 
trajectory is in general different from the mean value of the same 
quantity on the carrier of open trajectories. As can be seen, 
the last feature can lead to irregular dependence of the conductivity 
tensor on the direction of $\, {\bf B} \, $ inside 
$\, \Omega_{\alpha} $.

\vspace{1mm}

 Let us consider now the standard kinetic approach 
(see e.g. \cite{Abrikosov,etm,Ziman}) to the magneto-transport 
phenomena, using the geometric picture described above.

\vspace{1mm}

 The full dynamical system for the adiabatic evolution of the electron
states both in presence of magnetic and electric fields can be written
as
\begin{equation}
\label{FullDynSyst}
{\dot {\bf p}} \,\,\,\, = \,\,\,\, 
{e \over c} \,\, \left[ \nabla \epsilon ({\bf p})
\, \times \, {\bf B} \right] \,\,\, + \,\,\, e \, {\bf E} 
\end{equation}

 In system (\ref{FullDynSyst}) the electric field $\, {\bf E} \, $
will be considered as a small value while the value of magnetic field is 
supposed to satisfy the strong magnetic field limit condition
$\,\, \omega_{B} \, \tau \, \gg \, 1 \, $.

 In the kinetic approach we have to introduce the
single-particle distribution function $\, f({\bf p}, \, t) \, $
satisfying the Boltzmann equation
\begin{multline}
\label{KinEq}
f_{t} \,\,\, + \,\,\, {e \over c} \, \sum_{l=1}^{3} \,
\left[ \nabla \epsilon ({\bf p}) \times {\bf B} \right]^{l} \,\,
{\partial f \over \partial p^{l}} \,\,\, + \,\,\, e \, \sum_{l=1}^{3} \,
E^{l} \, {\partial f \over \partial p^{l}} \,\,\,\, =  \\
= \,\,\,\, I [f] ({\bf p}, \, t) \quad , 
\end{multline}
where $\, I [f] \, $ represents the collision integral. Here we will
be interested in the stationary solutions of the equation (\ref{KinEq}),
so we put now $\, f({\bf p}, \, t) \, = \, f({\bf p}) $.

 In the absence of electric field the distribution function
$\, f ({\bf p}) \, $ is given by its equilibrium values
$$f_{0} ({\bf p}) \,\,\, = \,\,\, 
{1 \over {\rm exp} \, [(\epsilon ({\bf p}) - \epsilon_{F}) / T ] 
\, + \, 1} $$
and we have identically $\, I [f] ({\bf p}) \, \equiv \, 0 $.
The conductivity tensor $\, \sigma^{kl} ({\bf B}) \, $ is defined by
the linear in $\, {\bf E} \, $ correction to the function
$\, f_{0} ({\bf p}) $, satisfying the equation
\begin{multline}
\label{LinKinEq}
{e \over c} \, \sum_{l=1}^{3} \,
\left[ \nabla \epsilon ({\bf p}) \times {\bf B} \right]^{l} \,\,
{\partial f_{(1)} \over \partial p^{l}} \,\,\, + \,\,\, 
e \, \sum_{l=1}^{3} \, E^{l} \, 
{\partial f_{0} \over \partial p^{l}} \,\,\,\, =   \\
=  \,\,\,\,
\left[ {\hat L}_{[f_{0}]} \, \cdot \, f_{(1)} \right] ({\bf p}) \quad ,
\end{multline}
where $\, {\hat L}_{[f_{0}]} \, $ is the linearization of the
functional $\, I [f] ({\bf p}) \, $ on the corresponding function
$\, f_{0} $. The value $\, {\hat L}_{[f_{0}]} \, \cdot \, f_{(1)} \, $ 
can be considered as a term, connected with the relaxation of
non-equilibrium perturbations to the equilibrium state. In the
so-called $\, \tau $ - approximation the right-hand part of system
(\ref{LinKinEq}) can be replaced by the value 
$\, - f_{1} ({\bf p}) / \tau \, $, where $\, \tau \, $ plays the role
of the characteristic relaxation time.  It will be convenient here to
use the $\, \tau $ - approximation for the equation (\ref{LinKinEq}),
which is enough to catch all the main features of our consideration.
The characteristic relaxation time $\, \tau \, $ is usually identified
with the mean free electron motion time.

 Let us introduce now the variable $\, s \, = \, t e B / c \, $, where
$\, t \, $ is the parameter along the trajectories of system
(\ref{QuasiclassicalEvolution}), introduced above, and consider the
coordinate system $\, (p_{z}, \, s, \, \epsilon ) \, $ in the
$\, {\bf p} $ - space. The parameter $\, s \, $ has a purely geometrical
meaning in $\, {\bf p} $ - space and does not depend on the value of
$\, B $. According to (\ref{dpzdtdl}) - (\ref{LayerIntegration}),
we can naturally write
$$d p_{z} \, d s \, d \epsilon \,\,\,\, \equiv \,\,\,\,
d p_{x} \, d p_{y} \, d p_{z} \quad , $$
\begin{multline*}
\iiint_{\epsilon_{0} < \epsilon ({\bf p}) < \epsilon_{0} + d \epsilon}
\,\, f \left( p_{x}, \, p_{y}, \, p_{z} \right) \,\,
d p_{x} \, d p_{y} \, d p_{z}   \quad  =   \\
= \quad  d \epsilon  \, \iint_{\epsilon ({\bf p}) = \epsilon_{0}}
\,\, f \left( p_{z}, \, s , \, \epsilon_{0} \right) \,\,
d p_{z} \, d s 
\end{multline*}

 It's not difficult to see also, that under our assumptions above the
equation on the function $\, f_{(1)} (p_{z}, \, s , \, \epsilon ) \, $
can be represented as
$${e B \over c} \,\, {\partial f_{(1)} \over \partial s} \,\,\, + \,\,\,
e \, \left( {\bf E} \cdot {\bf v}_{gr} \right) \, 
{\partial f_{0} \over \partial \epsilon } \,\,\,\, = \,\,\,\,
- \, f_{(1)} / \tau $$
in the coordinate system $\, (p_{z}, \, s, \, \epsilon ) $.

 After the substitution
\begin{equation}
\label{f1Subst}
f_{(1)} (p_{z}, \, s , \, \epsilon ) \,\,\, =  \,\,\, - \,
{\partial f_{0} (\epsilon) \over \partial \epsilon } \,\,\, 
\sum_{l=1}^{3} \,\,  E^{l} \, g^{l} (p_{z}, \, s , \, \epsilon ) 
\end{equation}
we get the linear systems
\begin{equation}
\label{gvSyst}
{e B \over c} \,\, {\partial g^{l} \over \partial s} \,\,\, - \,\,\,
e \, v^{l}_{gr} (p_{z}, \, s , \, \epsilon ) \,\,\, = \,\,\, - \,\,
{1 \over \tau} \,\, g^{l}
\end{equation}
for the functions $\, g^{l} (p_{z}, \, s , \, \epsilon ) $.

 The solutions of systems (\ref{gvSyst}) which we need can be written
in the form
\begin{equation}
\label{Gengl}
g^{l} (p_{z}, \, s , \, \epsilon ) \,\,\,\, = \,\,\,\,
{c \over B} \,\, \int_{-\infty}^{s} \,\, v^{l}_{gr} \left(
p_{z}, \, s^{\prime} , \, \epsilon \right) \,\,
e^{{c (s^{\prime} - s) \over e B \tau}} \,\, d s^{\prime} \quad , 
\end{equation}
where the integration is taken along the whole part of a trajectory
preceding the point $\, (p_{z}, \, s, \, \epsilon ) $. 

 The mean value of the density of the electric current is given 
by the formula
$$j^{k} \,\,\,\, = \,\,\,\, e \,\, \int
v^{k}_{gr} \left( p_{z}, \, s , \, \epsilon \right) \,\,
f_{(1)} \left( p_{z}, \, s , \, \epsilon \right) \,\,   
{d p_{z} \, d s \, d \epsilon \over (2 \pi \hbar)^{3}} $$
(in the linear approximation in $\, E $).

 The form (\ref{f1Subst}) of the correction $\, f_{(1)} \, $ gives  
in fact very strong concentration of
$\, f_{(1)} (p_{z}, \, s, \, \epsilon ) \, $ near the Fermi surface
for most of the normal metals, so the term
$\, - \, \partial f_{0} / \partial \epsilon \, $ can be actually
replaced by the delta-function $\, \delta (\epsilon - \epsilon_{F}) \, $
in the integration over $\, \epsilon $. Finally, we come to the following
simple expression for the value of the electric current density:
\begin{multline*}
j^{k} \,\,\,\,\, = \,\,\,\,\, {e c  \over  B}  \, \iint_{S_{F}}
\, {d p_{z} \, d s \over (2 \pi \hbar)^{3}} \,\,\,
v^{k}_{gr} (p_{z}, s) \,\,\,  \times   \\ 
\times \, \sum_{l=1}^{3} \, \int_{-\infty}^{s} \,
v^{l}_{gr} (p_{z}, s^{\prime}) \,\,\,
e^{{c (s^{\prime} - s) \over e B \tau}} \,\, d s^{\prime} \,\,\, E^{l}
\end{multline*}

 The general form of the conductivity tensor $\, \sigma^{kl} \, $
can be represented as
\begin{multline}
\label{GenSigmakl}
\sigma^{kl} (B) \,\,\,\,\, = \,\,\,\,\, 
{e c  \over  B}  \, \iint_{S_{F}}
\, {d p_{z} \, d s \over (2 \pi \hbar)^{3}} \,\,\,
v^{k}_{gr} (p_{z}, s) \,\,\,  \times   \\
\times \, \int_{-\infty}^{s} \,
v^{l}_{gr} (p_{z}, s^{\prime}) \,\,\,
e^{{c (s^{\prime} - s) \over e B \tau}} \,\, d s^{\prime} 
\end{multline}

 At the same time, the contribution of the open trajectories 
$\, \Delta \sigma^{kl} \, $ is given by the restriction of the integral
in (\ref{GenSigmakl}) to the set of the carriers of open trajectories
$\, {\hat S}_{F} \, $ instead of the whole Fermi surface, so we can 
write:
\begin{multline}
\label{DeltaSigmaB}
\Delta \, \sigma^{kl} (B) \,\,\,\,\, = \,\,\,\,\, 
{e c  \over  B}  \, \iint_{\hat{S}_{F}}
\, {d p_{z} \, d s \over (2 \pi \hbar)^{3}} \,\,\,
v^{k}_{gr} (p_{z}, s) \,\,\,  \times   \\
\times \, \int_{-\infty}^{s} \,
v^{l}_{gr} (p_{z}, s^{\prime}) \,\,\,
e^{{c (s^{\prime} - s) \over e B \tau}} \,\, d s^{\prime} 
\end{multline}

 Let us say, that the irregular behavior of $\, \sigma^{kl} ({\bf B}) \, $
in $\, \Omega_{\alpha} \, $ is completely determined by the functions
$\, \Delta \sigma^{kl} ({\bf B}) \, $, so we will be interested here 
mostly in the behavior of $\, \Delta \sigma^{kl} ({\bf B}) \, $ in the 
limit $\, B \, \rightarrow \, \infty $ .

 It's not difficult to see that for all types of trajectories in our case 
we can write in the limit $\, B \, \rightarrow \, \infty \, $ for the 
conductivity tensor
$$\sigma^{kl}_{\infty} \,\,\, = \,\,\, e^{2} \tau
\iint_{S_{F}}  v^{k}_{gr} (p_{z}, s)  \,\,\,
\langle v^{l}_{gr} \rangle_{tr} \, (p_{z}, s) \,\,\,
{d p_{z} d s \over (2 \pi \hbar)^{3}} \,\,\, , $$
where 
$\, \langle v^{l}_{gr} \rangle_{tr} \, (p_{z}, s) \, $
is the mean value of the group velocity component on the trajectory
passing through the point $\, (p_{z}, s) \, $:
$$\langle v^{l}_{gr} \rangle_{tr} \, (p_{z}, s)
\,\,\,\, = \,\,\,\, \lim_{s_{0} \rightarrow \infty} \,\,\,
{1 \over s_{0}} \, \int_{s-s_{0}}^{s} \, 
v^{l}_{gr} \left( p_{z}, \, s^{\prime} \right) \,\,
d s^{\prime}  $$

 To get the contribution of the open trajectories we have now to 
restrict the integration to the set of the carriers of open trajectories
$\, {\hat S}_{F} \, $ on the Fermi surface, so we can write for the
contribution of the stable open trajectories to the conductivity in 
the limit $\, B \, \rightarrow \, \infty \, $:
$$\Delta \, \sigma^{kl}_{\infty} \,\,\, = \,\,\, e^{2} \, \tau \,
\iint_{{\hat S}_{F}} \,\, v^{k}_{gr} (p_{z}, s) \,\,\,
\langle v^{l}_{gr} \rangle_{tr} \, (p_{z}, s)
\,\,\, {d p_{z} d s \over (2 \pi \hbar)^{3}} $$

 Let us note here that according to our choice of coordinate system
$\, (x, \, y, \, z) \, $ we will always have
$\, \langle v^{x}_{gr} \rangle_{tr} \, = \, 0 \, $ and
$\, \langle v^{y}_{gr} \rangle_{tr} \, \neq \, 0 \, $,
$\, \langle v^{z}_{gr} \rangle_{tr} \, \neq \, 0 \, $ for the stable
open trajectories.

 We can see also, that according to the definition above, the values 
$\, \langle v^{k}_{gr} \rangle_{tr} \, (p_{z}, s) \, $
do not depend in fact on the variable $\, s $ for the trajectories of 
our type. As a corollary, the first multiplier can be also replaced by 
its mean value on every trajectory of system 
(\ref{QuasiclassicalEvolution}). Finally, we can
write for the contribution of the stable open trajectories to the
magneto-conductivity for $\, B \, \rightarrow \, \infty \, $:
\begin{equation}
\label{GenFormSigmakl}
\Delta \, \sigma^{kl}_{\infty} \,\,\, = \,\,\, e^{2} \, \tau 
\iint_{{\hat S}_{F}} \,
\langle v^{k}_{gr} \rangle_{tr} \, (p_{z}) \,\,\,
\langle v^{l}_{gr} \rangle_{tr} \, (p_{z}) 
\,\,\, {d p_{z} d s \over (2 \pi \hbar)^{3}} 
\end{equation}

 Let us note that the tensor $\, \sigma^{kl}_{\infty} \, $ is purely
symmetric according to (\ref{GenFormSigmakl}), which is not the case
for finite values of $\, B $. 

\vspace{1mm}

 Let us give now qualitative derivation of the regimes of the
conductivity behavior described in the previous section.

\vspace{1mm}

 As it is not difficult to see, in the case of generic direction of
$\, {\bf B} \, $ (non-periodic open trajectories) the mean values
$\, \langle v^{k}_{gr} \rangle_{tr} \, (p_{z}) \, $
can be identified with the mean values of the same quantities on the
corresponding (connected) carrier of open trajectories 
$\, {\hat S}^{\gamma}_{F} \, $:
$$\langle v^{k}_{gr} \rangle_{{\hat S}^{\gamma}_{F}} 
\,\,\, = \,\,\, \iint_{{\hat S}^{\gamma}_{F}} \, v^{k}_{gr} 
(p_{z}, \, s) \,\, d p_{z} \, d s \,\,\, \Big/ 
\iint_{{\hat S}^{\gamma}_{F}}  d p_{z} \, d s $$
and are constant on each carrier of the stable open trajectories.

 In general, the mean values
$\, \langle v^{k}_{gr} \rangle_{{\hat S}^{\gamma}_{F}} \, $
can be also expressed through the values
$\, \langle v^{k}_{gr} \rangle_{tr} \, (p_{z}) \, $
by the formula
\begin{equation}
\label{CarrierTrajAvConn}
\langle v^{k}_{gr} \rangle_{{\hat S}^{\gamma}_{F}} \,\, = \,\,
\iint_{{\hat S}^{\gamma}_{F}} \,
\langle v^{k}_{gr} \rangle_{tr} \, (p_{z})
\,\,\, d p_{z} \, d s \,\,\, \Big/ 
\iint_{{\hat S}^{\gamma}_{F}} d p_{z} \, d s 
\end{equation}

 Let us introduce now the functions
$$\Delta \, {\bar \sigma}^{kl}_{\infty} ({\bf B} / B) \,\,\, = \,\,\,
e^{2} \, \tau \, \sum_{\gamma} \, \iint_{{\hat S}^{\gamma}_{F}} \,
\langle v^{k}_{gr} \rangle_{{\hat S}^{\gamma}_{F}} \,\,
\langle v^{l}_{gr} \rangle_{{\hat S}^{\gamma}_{F}} 
\,\,\, {d p_{z} \, d s \over (2 \pi \hbar)^{3}} $$
where the summation is taken over all the connected carriers 
$\, {\hat S}^{\gamma}_{F} \, $ of the stable open trajectories.

 The functions $\, \Delta {\bar \sigma}^{kl}_{\infty} ({\bf B} / B) \, $
are smooth functions of the direction of $\, {\bf B} \, $ inside any
Stability Zone $\, \Omega_{\alpha} $. According to (\ref{GenFormSigmakl}),
the functions $\, \Delta {\bar \sigma}^{kl}_{\infty} ({\bf B} / B) \, $ 
coincide with the limiting values 
$\, \Delta \sigma^{kl}_{\infty} ({\bf B} / B) \, $
for generic directions of $\, {\bf B} \, $, 
corresponding to non-periodic form of the open trajectories.

 At the same time, for special directions of $\, {\bf B} \, $, 
corresponding to the periodic form of the stable open trajectories,
the values (\ref{GenFormSigmakl}) do not coincide with
$\, \Delta {\bar \sigma}^{kl}_{\infty} ({\bf B} / B) \, $ 
due to the dependence of the values 
$\, \langle v^{k}_{gr} \rangle_{tr} \, (p_{z}) \, $ on $\, p_{z} $. 
Just from the Schwartz inequality we can write in this case
$$\Delta \, \sigma^{22}_{\infty} \, ({\bf B}/B) \,\,\, - \,\,\,
\Delta \, {\bar \sigma}^{22}_{\infty} \, ({\bf B}/B) 
\,\,\,\, > \,\,\,\, 0 \quad ,  $$
$$\Delta \, \sigma^{33}_{\infty} \, ({\bf B}/B) \,\,\, - \,\,\,
\Delta \, {\bar \sigma}^{33}_{\infty} \, 
({\bf B}/B) \,\,\,\, > \,\,\,\, 0  \quad  \, $$
as it was written above. The values
$$\Delta \, \sigma^{kl}_{\infty} \, ({\bf B}/B) \,\,\, - \,\,\,
\Delta \, {\bar \sigma}^{kl}_{\infty} \, ({\bf B}/B) \,\,\, , \quad
k \, \neq l \,\, , $$ 
can have in general arbitrary signs. We have also
$\, \Delta \sigma^{kl}_{\infty} \, ({\bf B}/B) \,\, = \,\,
\Delta {\bar \sigma}^{kl}_{\infty} \, ({\bf B}/B) \,\, = \,\, 0 \,\, , \, $
if $k$ or $l$ is equal to $1$, in our coordinate system.

\begin{figure}[t]
\begin{center}
\includegraphics[width=0.9\linewidth]{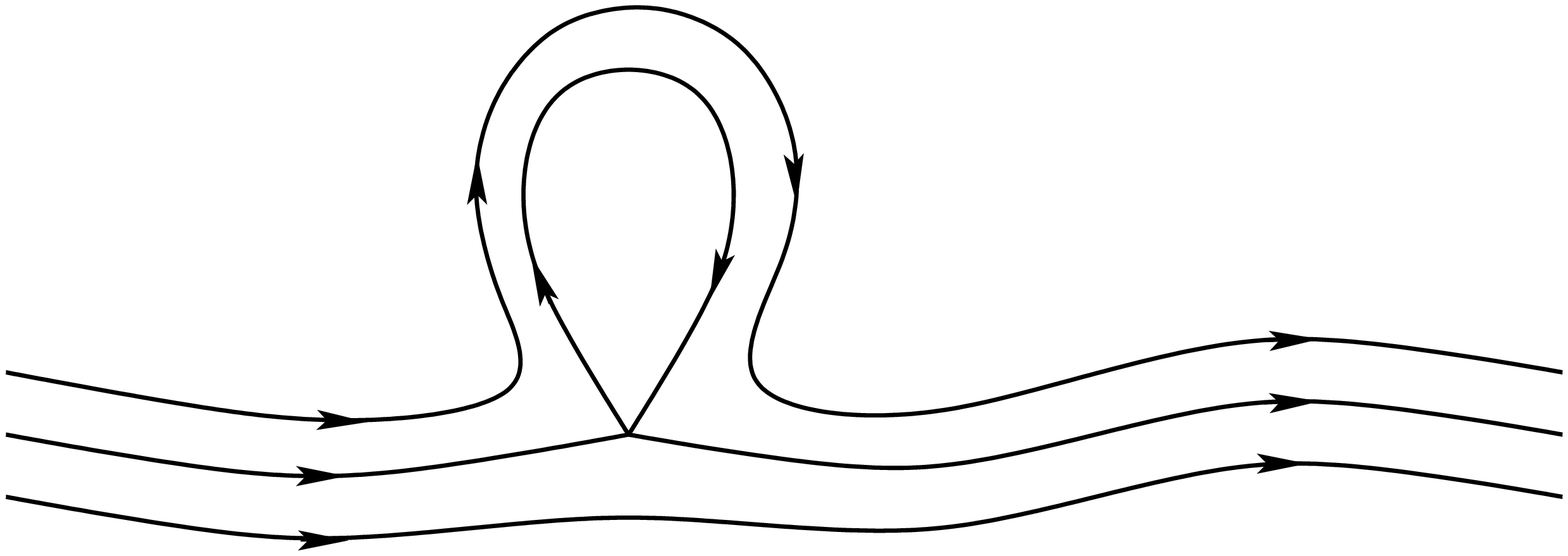}
\end{center}
\caption{Open periodic trajectories with rather different geometry   
near the singular trajectory on the Fermi surface.}  
\label{DiffTraject}
\end{figure}

 Let us point out an important role of the geometry of the carriers of 
open trajectories on the complicated Fermi surfaces for the effect which 
we discuss here. As it is not difficult to see, the difference
$\, \Delta \sigma^{kl}_{\infty} \, ({\bf B}/B) \,\, - \,\,
\Delta {\bar \sigma}^{kl}_{\infty} \, ({\bf B}/B) \,\, $ is defined 
actually by the ``inconstancy'' of the functions
$\, \langle v^{k}_{gr} \rangle_{tr} \, (p_{z}) \, $
on the carriers of the stable open trajectories. As can be seen at
Fig. \ref{DiffTraject}, the presence of the singular trajectories 
on the carriers of open trajectories makes the behavior of
$\, \langle v^{k}_{gr} \rangle_{tr} \, (p_{z}) \, $ rather irregular 
in the case of periodicity of the open trajectories on the
Fermi surface. Thus, the periodic trajectories adjacent to the singular 
trajectory from different sides have rather different geometry in
$\, {\bf p}$ - space. As a result, we can expect that the mean
variations of the values
$\, \langle v^{k}_{gr} \rangle_{tr} \, (p_{z}) \, $ 
on the carrier of open trajectories can be
comparable with the mean values of the same quantities and the same
can be stated also about the difference
$\, \Delta \sigma^{kl}_{\infty} \, ({\bf B}/B) \, - \,
\Delta {\bar \sigma}^{kl}_{\infty} \, ({\bf B}/B) \,\, , \,\,$ 
$(k, l = 2,3) \, $. It should be also noted here, however,
that this effect decreases with increasing of ``denominators'' of
the rational mean direction of the open trajectories due to the 
increasing of the density of the periodic open trajectory on 
a carrier of open trajectories. As we also said already, we have the 
identities $ \, \langle v^{x}_{gr} \rangle_{tr} \, \equiv \, 0 \, $
in our coordinate system.

\begin{figure}[t]
\begin{center}
\includegraphics[width=0.8\linewidth]{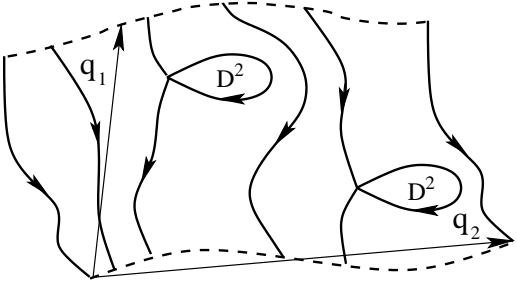}
\end{center}
\caption{The fundamental domain on the carrier of open trajectories
having the form of curvilinear parallelogram with two opposite sides,
represented by two equivalent parts of open electron trajectories.}
\label{NewFundDom}
\end{figure}

 To evaluate the dependence of 
$\, \Delta \sigma^{kl}_{\infty} \, ({\bf B}/B)  \, - \,
\Delta {\bar \sigma}^{kl}_{\infty} \, ({\bf B}/B) $ 
on the numbers $\, (k_{1}, k_{2}) \, $
it is convenient to choose the fundamental domain on the carrier
of open trajectories in the form of curvilinear parallelogram with
two opposite sides, represented by two equivalent parts of periodic
open trajectories (Fig. \ref{NewFundDom}).

 The values $\, p_{z} \, $ and $\, s \, $ give a good parametrization
of the fundamental domain under the requirement (\ref{NonParallel}).
We can assume without loss of generality that $\, {\bf B} \, $ is not
orthogonal to the vector $\, {\bf q}_{2} \, $ which will also mean
that the mean direction of the open trajectories does not coincide
with the direction of $\, {\bf q}_{2} $. Let us put also that the 
values of $\, p_{z} \, $ belong to the interval
$\, [0, p^{0}_{z}] \, $ for our fundamental domain, where
$\,\, p^{0}_{z} \, = \, |({\bf B}, {\bf q}_{2})| / B \, $.

 The parameter $\, s \, $ is not continuous in our fundamental domain
and has singularities on the levels of $\, p_{z} \, $, containing 
singular points within the domain. As we will see, this circumstance 
will not actually be important in our considerations. Let us put that
the values of $\, s \, $ are restricted by the values
$\, s_{1} (p_{z}) \, $ and $\, s_{2} (p_{z}) \, $ at a given value of
$\, p_{z} \, $ for our fundamental domain. The functions
$\, s_{1} (p_{z}) \, $ and $\, s_{2} (p_{z}) \, $ also have 
singularities at the same $\, p_{z} \, $, defined by the singular
behavior of $ \,\, s_{2} (p_{z}) \, - \, s_{1} (p_{z}) \, $:
$$s_{2} (p_{z}) \, - \, s_{1} (p_{z}) \,\,\,\,\, \sim \,\,\,\,\,
| \, {\rm ln} \, (|\Delta p_{z}| / p^{0}_{z}) \, | $$ 
near the singular levels.

 It's not difficult to see also that the singularities of
$ \, s_{2} (p_{z}) - s_{1} (p_{z}) \, $ have asymmetric character
since the corresponding trajectories approach the singular point
once or twice depending on the sign of $\, \Delta p_{z} $.
Let us note here that the last circumstance played important role
in the proof of the fact of mixing in the dynamical systems analogous
to those considered here. This property was proved in paper
\cite{SinaiKhanin} where systems of this kind were considered in
connection with a different problem of Hamiltonian dynamics
(\cite{Arnold}).

 Every trajectory with the period
$\,\, {\bf q} \, = \, k_{1} {\bf q}_{1} + k_{2} {\bf q}_{2} \, $
is represented in our domain by $\, |k_{1}| + |k_{2}| \, $ separate
lines, giving a closed curve after the identification of the 
equivalent points in the $\, {\bf p} $ - space. The values of the 
functions $\, \langle v^{l}_{gr} \rangle_{tr} \, $ are the same on 
every set of $\, |k_{1}| + |k_{2}| \, $ components representing one 
periodic trajectory. 

 The presence of singular points plays also important role in the
global motion of electrons along the non-singular trajectories since
they appear many times at rather small distances from a periodic
trajectory with a large period
$\, {\bf q} \, = \, k_{1} {\bf q}_{1} + k_{2} {\bf q}_{2} $.
As a result, the length $\, S \, $ of a periodic trajectory in the
parameter $\, s \, $ ($S \, \sim \, |k_{1}| + |k_{2}|$) has essential
fluctuations, caused by the presence of singular points in the
fundamental domain. Let us note, that according to system
(\ref{QuasiclassicalEvolution}) we can write in our coordinate
system
\begin{equation}
\label{VyGrAvTr}
\langle v^{y}_{gr} \rangle_{tr} \,\,\, = \,\,\,
L (k_{1}, k_{2}) \left/ S (p_{z}) \right.  \,\,\,  ,
\end{equation}  
where $\, L (k_{1}, k_{2}) \, $ is the length of the corresponding
topological cycle in the $\, {\bf p}$ - space, depending just on the 
numbers $\, (k_{1}, k_{2}) $. In general, we have
$\,  (|k_{1}| + |k_{2}|) \cdot N \, $ singular points
$\,\, p^{j}_{z} \, \in \, [0 , p^{0}_{z}] \, $ of the function
$\, S (p_{z}) \, $, where $\, N \, $ is the number of (non-equivalent) 
singular points of system (\ref{QuasiclassicalEvolution}) on the carrier 
of open trajectories.

 The function $\, S (p_{z}) \, $ represents a periodic function with
the period $\, P \, = \, p^{0}_{z} / (|k_{1}| + |k_{2}|) \, $ and
integrable (asymmetric) singularities
$$ S (p_{z}) \,\,\, \sim \,\,\, {\rm ln} \left( p_{z}^{0} /
|\Delta p_{z}| \right) $$
at the values of $\, p_{z} \, $ corresponding to singular electron
trajectories. In particular, in the interval 
$\, I^{j} \, = \, [p^{j}_{z} , p^{j+1}_{z}] $, restricted by two
subsequent ``singular'' values of $\, p_{z} \, $
($|p^{j+1}_{z} - p^{j}_{z}| \leq P $), we can write the contribution
of the singularities at the points $\, p^{j}_{z} \, $ and
$\, p^{j+1}_{z} \, $ to the value of $\, S (p_{z}) \, $ in the form
\begin{multline}
\label{SingularContr}
\delta_{1} S (p_{z}) \,\,\, \sim \,\,\, \alpha^{j}_{+} \,
{\rm ln} \left( p_{z}^{0} / (p_{z} - p^{j}_{z}) \right) \,\,\, +   \\
+ \,\,\, \alpha^{j+1}_{-} \, {\rm ln} \left( p_{z}^{0} / 
(p^{j+1}_{z} - p_{z}) \right) \,\, , 
\end{multline}
($\alpha^{j}_{+}, \alpha^{j+1}_{-} \, \sim \, 1$). 

 Besides that, the singularities at the points
$\, ( \dots , \, p^{j-2}_{z} , \, p^{j-1}_{z} ) \, $ and
$\, (p^{j+2}_{z} , \, \dots ) $, close to the interval $\, I^{j} $,
play also in fact important role for the variations of
$\, S (p_{z}) \, $ inside the interval $\, I^{j} \, $ for big values
of $\, k_{1} \, $ and $\, k_{2} \, $.

 The corresponding contribution to $\, S (p_{z}) \, $ can be written
as a sum of terms analogous to (\ref{SingularContr}), which number
is proportional to $\, |k_{1}| + |k_{2}| $. We actually will be
interested here in the fluctuations of $\, S (p_{z}) \, $ caused by
these singularities, so only the variation of this contribution with
the variable $\, p_{z} \, $ is important for us. It's not difficult to 
show also that the difference of this contribution at the points
$\, p^{j}_{z} \, $ and $\, p^{j+1}_{z} \, $ has the order
$\,\, \sim \, {\rm ln} (|k_{1}| + |k_{2}|) $. For our purposes here
(we omit here rigorous estimations) the corresponding contribution
$\, \delta_{2} S (p_{z}) \, $ to the function $\, S (p_{z}) \, $ 
can be actually approximated by a linear function
$$\delta_{2} S (p_{z}) \,\,\,\, \sim \,\,\,\, 
\Gamma^{j} \,\, (|k_{1}| + |k_{2}|) \,\,
{\rm ln} \left( |k_{1}| + |k_{2}| \right) \,\,\,
{p_{z} - p^{j}_{z} \over p^{0}_{z}} \,\,\, , $$
($|\Gamma^{j}| \sim 1$), on every interval $\, I^{j} $.

 The functions $\, \delta_{1} S (p_{z}) \, $ and 
$\, \delta_{2} S (p_{z}) \, $ represent the main corrections to the
``basic value'' $\, {\bar S} \, $ of the function $\, S (p_{z}) $,
which are responsible for the main terms in the dependence of 
$\, \Delta \sigma^{kl}_{\infty} ({\bf B} / B)  \, - \,
\Delta {\bar \sigma}^{kl}_{\infty} ({\bf B} / B) \, $ on the numbers
$\, (k_{1}, k_{2}) \, $ (or $(m_{1}, m_{2}, m_{3})$). The basic
value $\, {\bar S} \, $ of the function $\, S (p_{z}) \, $ is given
by a constant function on the whole interval $\, [0, p^{0}_{z}] \, $,
which in the main order is proportional to the length
$\, L (k_{1}, k_{2}) \, $ (or $L (m_{1}, m_{2}, m_{3})$) of the 
topological cycle $\, (m_{1}, m_{2}, m_{3}) \, $ in the
$\, {\bf p}$ - space. Easy to see that for any sequence of rational
directions converging to some fixed
(irrational) direction we can write in the main order
$\,  L (k_{1}, k_{2}) \, \sim \, |k_{1}| + |k_{2}| $.

\begin{figure}[t]
\begin{center}
\includegraphics[width=0.9\linewidth]{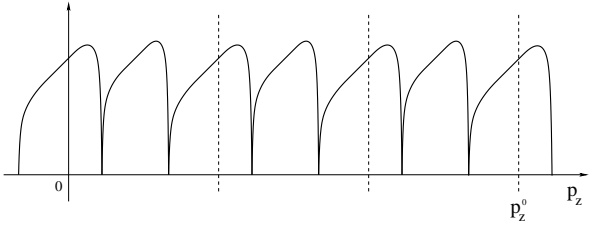}
\end{center}
\caption{The general form of the periodic function
$\, \langle v^{y}_{gr} \rangle_{tr} \, (p_{z}) \, $
on a carrier of open trajectories for the case
of rational mean direction
$\,\, {\bf q} \, = \, k_{1} {\bf q}_{1} + k_{2} {\bf q}_{2} \, $
of the trajectories.}
\label{graphvlgr}
\end{figure}

 The functions $\, \langle v^{y}_{gr} \rangle_{tr} \, $,
$\,\, \langle v^{z}_{gr} \rangle_{tr} \, $ are continuous
functions of $\, p_{z} $, however, they are not smooth on the curves,
representing singular trajectories. Thus, for the component
$\, v^{y}_{gr} \, $ we can write the following relations
$$ \langle v^{y}_{gr} \rangle_{tr} \,\,\,\,\, \sim \,\,\,\,\,
\left[ \, {\rm ln} \, (p^{0}_{z} / |\Delta p_{z}|) \, \right]^{-1} 
\,\,\, \rightarrow \,\,\, 0 $$
near these curves.  In general, the values 
$\, \langle v^{y}_{gr} \rangle_{tr} \, $
represent a periodic function of $\, p_{z} \, $ with the period
$\, P \, = \, p^{0}_{z} / (|k_{1}| + |k_{2}|) \, $ and very narrow
cusps at the values of $\, p_{z} \, $, corresponding to singular
electron trajectories (Fig. \ref{graphvlgr}). It is easy to see also
that the number of cusps on the period of the function
$\, \langle v^{y}_{gr} \rangle_{tr} \, $
is equal to the number $\, N \, $ of non-equivalent singular points
of system (\ref{QuasiclassicalEvolution}) on the corresponding 
carrier of open trajectories.

 In our situation we can put approximately
\begin{multline*}
S (p_{z}) \,\,\,\, \simeq \,\,\,\, {\bar S} \, + \, \delta_{1} S (p_{z}) 
\, + \, \delta_{2} S (p_{z}) \,\,\,\, \sim    \\
\sim \,\,\,\, |k_{1}| + |k_{2}| \,\,\, + \,\,\, \alpha^{j}_{+} \,
{\rm ln} \left( p_{z}^{0} / (p_{z} - p^{j}_{z}) \right) \,\,\, +   \\
+ \,\,\, \alpha^{j+1}_{-} \, {\rm ln} \left( p_{z}^{0} /
(p^{j+1}_{z} - p_{z}) \right) \,\,\, +  \\
+ \,\,\, \Gamma^{j} \,\, (|k_{1}| + |k_{2}|) \,\,
{\rm ln} \left( |k_{1}| + |k_{2}| \right) \,\,\,
{p_{z} - p^{j}_{z} \over p^{0}_{z}}   \quad  ,
\end{multline*}
and
$$\langle v^{y}_{gr} \rangle_{tr} \,\,\, \simeq \,\,\,
L (k_{1}, k_{2}) \left/ \left( {\bar S} + \delta_{1} S (p_{z}) +
\delta_{2} S (p_{z})  \right)  \right. $$
on every interval $\, I^{j} $. Let us say that the approximation
above should be used in our scheme also for the calculation of the
approximate values of 
$\, \langle v^{y}_{gr} \rangle_{\hat{S}^{\gamma}_{F}} \, $ according to
formula (\ref{CarrierTrajAvConn}). Using this approximation to compute
the values  
\begin{multline}
\label{SigmaklFormula}
\Delta \, \sigma^{kl}_{\infty} ({\bf B} / B)  \,\,\, - \,\,\, 
\Delta \, {\bar \sigma}^{kl}_{\infty} ({\bf B} / B) \,\,\,\,\, =   \\
= \,\,\, e^{2} \, \tau  \,\, \sum_{\gamma}  \,\, 
\iint_{{\hat S}^{\gamma}_{F}}  \langle v^{k}_{gr} \rangle_{tr} \,\,
\langle v^{l}_{gr} \rangle_{tr} \,\,\, 
{d p_{z} d s \over (2 \pi \hbar)^{3}} \,\,\, -  \\
- \,\,\, e^{2} \, \tau \,\, \sum_{\gamma} 
\,\, \iint_{{\hat S}^{\gamma}_{F}} \,
\langle v^{k}_{gr} \rangle_{{\hat S}^{\gamma}_{F}} \,\,
\langle v^{l}_{gr} \rangle_{{\hat S}^{\gamma}_{F}}
\,\,\, {d p_{z} \, d s \over (2 \pi \hbar)^{3}} 
\end{multline}
for $k, l = 2, \, $ we finally get the relations
$$\left| \Delta \, \sigma^{22}_{\infty} ({\bf B} / B)  \, - \,
\Delta \, {\bar \sigma}^{22}_{\infty} ({\bf B} / B) \right| 
\,\,\,\,\, \sim \,\,\,\,\, {{\rm ln}^{2} (|k_{1}| + |k_{2}|) \over
( |k_{1}| + |k_{2}| )^{2}} $$

 To give the necessary approximation for the function
$\, \langle v^{z}_{gr} \rangle_{tr} \, $ we have to evaluate also 
the integral
$$J (p_{z}) \,\,\, = \,\,\, 
\oint \, v^{z}_{gr} (p_{z}, s) \,\, ds $$
over a periodic trajectory.

 First, we should note, that we have to prescribe here also the
``basic value''
$$J (p_{z}) \,\,\,\,\, \simeq \,\,\,\,\, {\bar J} 
\,\,\,\,\, \simeq \,\,\,\,\,
{\bar v}^{z}_{gr} \, (|k_{1}| + |k_{2}|) $$
to this integral, which is constant on the interval
$\, [0, p^{0}_{z}] $. The essential corrections to the value
$\, {\bar J} \, $ on the intervals $\, I^{j} \, $ are caused again
by the presence of singular points and can be represented by two
main contributions:

1) The contribution of the endpoints of $\, I^{j} $, which can be
approximately written as
\begin{multline*}
\delta_{1} J (p_{z}) \,\,\,\, \simeq \,\,\,\, 
v^{z}_{j} \, \alpha^{j}_{+} \, {\rm ln} \left( p_{z}^{0} / (
p_{z} - p^{j}_{z}) \right) \,\,\,\, +   \\
+ \,\,\,\, v^{z}_{j+1} \, \alpha^{j+1}_{-} \,\, {\rm ln} 
\left( p_{z}^{0} / (p^{j+1}_{z} - p_{z}) \right) \,\, ,
\end{multline*}
where $\, v^{z}_{j} \, $ and $\, v^{z}_{j+1} \, $ are the values of
$\, v^{z}_{gr} \, $ at the singular points, adjacent to the 
corresponding singular trajectories;

2) The contribution of other singular points close to the interval
$\, I^{j} $, which can again be approximated by the linear function
$$\delta_{2} J (p_{z}) \,\,\,\, \simeq \,\,\,\, 
v^{z}_{*} \,\, \Gamma^{j} \,\, (|k_{1}| + |k_{2}|) \,\,
{\rm ln} \left( |k_{1}| + |k_{2}| \right) \,\,\,
{p_{z} - p^{j}_{z} \over p^{0}_{z}} \,\, , $$
where $\, v^{z}_{*} \, $ represents some weighted average of the
values $\, v^{z}_{j} \, $ over all the singular points in the
fundamental domain.

 We can use now the following approximation for the function
$\, \langle v^{z}_{gr} \rangle_{tr} (p_{z}) \, $:
\begin{multline}
\label{VzGrAvAppr}
\langle v^{z}_{gr} \rangle_{tr} (p_{z}) \,\,\, \simeq   \\
\simeq  \,\,
\left( {\bar J} + \delta_{1} J (p_{z}) +
\delta_{2} J (p_{z})  \right) \left/ 
\left( {\bar S} + \delta_{1} S (p_{z}) +
\delta_{2} S (p_{z})  \right)  \right.
\end{multline}

 We can see here that the function 
$\, \langle v^{z}_{gr} \rangle_{tr} (p_{z}) \, $ does not have cusps 
at the values $\, p^{j}_{z} \, $ and the values
$\, \langle v^{z}_{gr} \rangle_{tr} (p^{j}_{z}) \, $ are equal to
$\, v^{z}_{j} $. However, the structure of the function
$\, \langle v^{z}_{gr} \rangle_{tr} (p_{z}) \, $ still has many
common features with the structure of the function
$\, \langle v^{y}_{gr} \rangle_{tr} (p^{j}_{z}) $. In particular,
using the approximation (\ref{VzGrAvAppr}) in the formula
(\ref{SigmaklFormula}), we obtain actually the same relations 
\begin{equation}
\label{lnk1k2}
\left| \Delta \, \sigma^{kl}_{\infty} ({\bf B} / B)  \, - \,
\Delta \, {\bar \sigma}^{kl}_{\infty} ({\bf B} / B) \right|
\,\,\,\,\, \sim \,\,\,\,\, {{\rm ln}^{2} (|k_{1}| + |k_{2}|) \over
( |k_{1}| + |k_{2}| )^{2}} 
\end{equation}
for $k, l = 2, 3 $, which implies also relations
(\ref{sigmasigmabardiff}) for a given Stability Zone 
$\, \Omega_{\alpha} $.

\vspace{1mm}

 The arguments used above can be applied also to the estimation of
the functions $\, q (B) $, $\, r (B) $, $\, p (B) \, $ in relations 
(\ref{SigmaOmega}). Let us say at once, that we will not present here
rigorous calculations, however, the approximate considerations below 
capture all the essential features which are necessary for our 
purposes. Thus, for the estimation of the functions
(\ref{Gengl}) we can use the approximation when the integration
in (\ref{Gengl}) is taken just over a part of a trajectory having
length $\, \sim e B \tau / c \, $ in the parameter $s$. 
In the same way, we can use for the averaged values
\begin{equation}
\label{VAveragedB}
\langle v^{k}_{gr} \rangle_{B} \, (p_{z}, s)
\,\,\,\, \equiv \,\,\,\,
{c \over e B \tau} \, \int_{-\infty}^{s} \,
v^{k}_{gr} \,(p_{z}, s^{\prime}) \,\,\,
e^{{c (s^{\prime} - s) \over e B \tau}} \,\, d s^{\prime} 
\end{equation}
the approximation
$$\langle v^{k}_{gr} \rangle_{B} \, (p_{z}, s)
\,\,\,\, \simeq \,\,\,\, {c \over e B \tau} \,
\int_{s - e B \tau / c}^{s} \,\,\,
v^{k}_{gr} \,(p_{z}, s^{\prime}) \,\, d s^{\prime}  $$

 Let us take again the fundamental domain in the form shown at 
Fig. \ref{NewFundDom}. The open trajectories have now irrational mean 
direction and are not periodic in the $\, {\bf p}$ - space anymore.
The corresponding parts of open trajectories are represented by nets
of approximately $\, e B \tau / m^{*} c \, = \, \omega_{B} \tau \, $
curves in the fundamental domain with approximate distances 
$\, \sim p^{0}_{z} / \omega_{B} \tau \, $ between the curves.

 The functions $\, \langle v^{k}_{gr} \rangle_{B} \, (p_{z}, s) \, $ 
are almost constant along the curves 
$\, p_{z} = {\rm const} \, $ in the fundamental domain and
change for the most part along the coordinate $\, p_{z} $. For generic
mean direction of open trajectories the functions 
$\, \langle v^{2,3}_{gr} \rangle_{B} \, (p_{z}, s) \, $
do not have periods smaller than 
$\, p^{0}_{z} \, $ in the fundamental domain, however, in other 
essential features the behavior of 
$\, \langle v^{2,3}_{gr} \rangle_{B} \, (p_{z}, s) \, $
in our approximation is similar to the above behavior of the functions
$\, \langle v^{2,3}_{gr} \rangle_{tr} \, (p_{z}) \, $, 
considered for a rational mean direction of open trajectories with
$\,\, |k_{1}| + |k_{2}| \, \simeq \, \omega_{B} \tau $. In particular, 
we can also introduce here the values $\, p^{j}_{z} \, $ in the 
fundamental domain, defined by the requirement of presence of a 
singular point on the preceding part of the corresponding trajectory
at a distance not exceeding the value 
$\, \sim \omega_{B} \tau \, p_{F} \, $ in the $\, {\bf p}$ - space.
The number of the values $\, p^{j}_{z} \, $ in the fundamental domain
has the order $\, \sim \omega_{B} \tau \, $ for 
$\, B \rightarrow \infty \, $ and we can consider again the division
of the interval $\, [0, p^{0}_{z}] \, $ to the segments $\, I^{j} \, $
analogous to those considered above. The similar analysis shows also
that the behavior of the functions 
$\, \langle v^{2,3}_{gr} \rangle_{B} \, (p_{z}, s) \, $
demonstrates on the intervals $\, I^{j} \, $ the features analogous
to those observed in the behavior of 
$\, \langle v^{2,3}_{gr} \rangle_{tr} \,  (p_{z}) \, $ for 
a rational mean direction of the open trajectories with the numbers 
$\, k_{1} $, $\, k_{2} \, $ having the order $\, \omega_{B} \tau $.

 Let us introduce now the symmetric and the anti-symmetric parts of
the conductivity tensor
$$\sigma^{kl} \,\,\,\,\, = \,\,\,\,\, s^{kl} \,\,\, + \,\,\, a^{kl} $$

 According to the Onsager relations we have the identities 
$$s^{kl} (- {\bf B}) \,\,\, = \,\,\, s^{kl} ({\bf B})   \quad \, 
\quad \quad 
a^{kl} (- {\bf B}) \,\,\, = \,\,\, - \, a^{kl} ({\bf B}) $$

 After some elementary calculations we can get from the formula
(\ref{DeltaSigmaB}):
\begin{equation}
\label{SymmetricPart}
\Delta \, s^{kl}(B) \,\,\,\, = \,\,\,\,
e^{2} \, \tau \,\, \sum_{\gamma} \,\, \iint_{{\hat S}^{\gamma}_{F}} \,
\langle v^{k}_{gr} \rangle_{B} \,\,
\langle v^{l}_{gr} \rangle_{B}
\,\,\, {d p_{z} \, d s \over (2 \pi \hbar)^{3}}  \,\,\,\, , 
\end{equation}
\begin{multline}
\label{AntiSymPart}
\Delta \, a^{kl}(B) \,\,\,\, =   \\
= \,\,\, {e^{2} \, \tau \over 2} \,\,
\sum_{\gamma} \iint_{{\hat S}^{\gamma}_{F}}   \Big( v^{k}_{gr} \,\, 
\langle v^{l}_{gr} \rangle_{B} \,\,\, - \,\,\, 
\langle v^{k}_{gr} \rangle_{B} \,\, v^{l}_{gr} \Big) \,
{d p_{z} \, d s \over (2 \pi \hbar)^{3}} 
\end{multline}

 It is not difficult to see that for any irrational mean direction of 
the open trajectories the values 
$\, \langle v^{k}_{gr} \rangle_{B} \, $,
$\, \langle v^{l}_{gr} \rangle_{B} \, $ coincide with the values
$\, \langle v^{k}_{gr} \rangle_{{\hat S}^{\gamma}_{F}} \, $,
$\, \langle v^{l}_{gr} \rangle_{{\hat S}^{\gamma}_{F}} \, $
in the limit $\, B \rightarrow \infty $. At the same time, for large
but finite values of $\, B \, $ the behavior of the functions
$\, \langle v^{2,3}_{gr} \rangle_{B} (p_{z}, s) \, $ is analogous
to the behavior of $\, \langle v^{2,3}_{gr} \rangle_{tr} \, $ for
a rational mean direction of the open trajectories with 
$\, k_{1}, k_{2} \, \simeq \, \omega_{B} \tau \, $.
As a result, in full analogy with formula (\ref{SigmaklFormula}),
we can write the relations
\begin{equation}
\label{lnOmegaTau}
\Delta \, s^{kl} (B) \,\, - \,\, \Delta \, s^{kl}_{\infty} 
\quad \simeq \quad
{{\rm ln}^{2} (\omega_{B} \tau)  \over  (\omega_{B} \tau)^{2}} \quad ,
\quad \,\,\,\,\,  k, l \, = \, 2, 3 
\end{equation}
for generic directions of the open trajectories. 

 It is easy to see that the estimation (\ref{lnOmegaTau}) gives now 
the relations (\ref{Trend}) for the functions $\, q (B) \, $ and 
$\, p (B) \, $. At the same time, the relations (\ref{rBTrend}) for 
the function $\, r (B) \, $ are defined by the behavior of the 
anti-symmetric part of the tensor $\, \sigma^{kl} (B) \, $. 
Using the same approximations for the functions 
$\, \langle v^{2,3}_{gr} \rangle_{B} (p_{z}, s) \, $, we get now the
estimation
\begin{equation}
\label{SkewlnOmegaTau}
\left| \Delta \, a^{23} (B) \right|  \quad \leq \quad
{{\rm ln} \,\, \omega_{B} \tau  \over  \omega_{B} \tau} \quad ,
\end{equation}

 Using also the identity $\, a_{\infty}^{kl} \, \equiv \, 0 \, $
we then get relations (\ref{rBTrend}) for the function 
$\, r (B) \, $. 

 Let us say that a unified description of the function
$\, \Delta \, a^{23} (B) \, $ can not be actually obtained in 
general situation, since its the behavior admits really essential 
variations depending on a concrete form of the Fermi surface.

\vspace{1mm}

 As we said already, the relations (\ref{lnOmegaTau}) can be considered
just as a ``general trend'' in the behavior of conductivity in strong
magnetic fields. Let us give here a more detailed explanation of our
statement. 

 As we can see, the analogy between the relations (\ref{lnk1k2}) and
(\ref{lnOmegaTau}) requires in fact the assumption of the
``uniform covering'' of the carrier of open trajectories by parts
of the trajectories having length 
$\,\, \simeq \, \omega_{B} \tau \, p_{F} $. This assumption is
fulfilled ``in average'', however, it can be violated for concrete
values of $\, \omega_{B} \tau \, $ for special directions of the
open trajectories. It is not difficult to see, that this violation
definitely takes place when the direction of the open trajectories
can be approximated by a rational direction with a very high
precision ($\, \ll \, |k_{1}| + |k_{2}| \, $). As we can see, the
corresponding open trajectories will then be very close to the
periodic trajectories on the lengths
$$\omega_{B} \tau \, p_{F} \,\,\, \gg \,\,\, 
\left( |k_{1}| + |k_{2}| \right) \, p_{F} \,\,\, , $$
so the parameter $\, |k_{1}| + |k_{2}| \, $ should be actually used
instead of $\, \omega_{B} \tau \, $ in 
(\ref{lnOmegaTau}) up to the values 
$\, \omega_{B} \tau \, \gg \, |k_{1}| + |k_{2}| \, $.
As a result, the general trend (\ref{lnOmegaTau}) can have essential 
corrections.

 For better description of the behavior of $\, \sigma^{kl} (B) \, $ 
for generic mean directions of the stable open trajectories let us 
just say some words about the problem of approximation of irrational 
numbers by rational ones. In our case we have to investigate the 
possibility of rational approximations of the number $\, \kappa \, $ 
in relation (\ref{dbarkappa}), which correspond to rational 
approximations of the mean direction of the open trajectories 
in the plane $\, \Gamma_{\alpha} $. 

 According to classical (Dirichlet) theorem we can state that for
any number $\, N \in \mathbb{N} \, $ there exist two integer numbers
$\, k_{1} $, $\, k_{2} $:
$$1 \, \leq  \, k_{1} \, \leq \, N \quad  ,  \quad \quad 
k_{1}, k_{2} \, \in \, \mathbb{Z} \,\,\, , $$
such that
\begin{equation}
\label{DirRelation}
\left| \, \kappa \, - \, {k_{2} \over k_{1}} \, \right| 
\,\,\,\,\,  <  \,\,\,\,\, {1 \over k_{1} N} 
\end{equation}

 Let us note here that we can put without loss of generality
$\, | \kappa | < 1 $, so we will assume also 
$\, |k_{2}| \leq |k_{1}| \, $ for precise approximations of 
$\, \kappa $. 

 For the value $\, N \simeq \omega_{B} \tau \, $ we can rewrite the
relation (\ref{DirRelation}) in the form
\begin{equation}
\label{TrajAppr}
\omega_{B} \tau \, p_{F} \,
\left| \, \kappa \, - \, {k_{2} \over k_{1}} \, \right|
\,\,\,\,\,  <  \,\,\,\,\, {p_{F} \over k_{1}}  \,\,\,  ,
\end{equation}
which means that the deviation of the open trajectory from the 
corresponding periodic trajectory does not exceed the value
$\,\, \simeq \, p_{F} / k_{1} \, $ on the length 
$\,\, L \, \simeq \omega_{B} \tau \, p_{F} \, $. The relation
(\ref{TrajAppr}) can be considered as the first step in the
approximation of irrational directions of open trajectories by
rational ones since the mean distance between the curves representing
a periodic trajectory in the fundamental domain is of order of
$$p_{F} / (|k_{1}| + |k_{2}|) \,\,\,\, \simeq \,\,\,\, p_{F} / k_{1} 
\,\,\, . $$

 We have to say, however, that relation (\ref{TrajAppr}) does not give
a precise connection between the values $\, \omega_{B} \tau \, $ and
$\, k_{1} $, so this connection should be specially studied for every
concrete irrational number $\, \kappa $. Using the theorem of Dirichlet,
we can see that for any irrational number $\, \kappa \, $ there always
exists an infinite sequence of rational approximations
$$k^{(s)}_{2} / k^{(s)}_{1} \quad , \quad \quad s = 1, \dots , \infty 
\,\,\, , $$
such that the following relations
\begin{equation}
\label{QuadraticAppr}
\left| \, \kappa \, - \, {k^{(s)}_{2} \over k^{(s)}_{1}} \, \right|
\,\,\,\,\,  <  \,\,\,\,\, {1 \over (k^{(s)}_{1})^{2}} 
\end{equation}
take place for every $\, s $. If we consider now a sequence 
$\, B^{(s)} \, $ of the values of $\, B $, such that
$\, \omega_{B^{(s)}} \tau \, \simeq \, k^{(s)}_{1} \, $,
we get the relations
$$\omega_{B^{(s)}} \tau \, p_{F} \,
\left| \, \kappa \, - \, {k^{(s)}_{2} \over k^{(s)}_{1}} \, \right|
\,\,\,\,\,  <  \,\,\,\,\, {p_{F} \over k^{(s)}_{1}}  \,\,\,  , $$
analogous to (\ref{TrajAppr}), for this sequence.

 It can be seen, that the estimation above should not cause in general
crucial deviations from the general trend according to our picture. 

 Let us say, however, that the estimation (\ref{QuadraticAppr}) can be
actually improved if we consider ``generic'' irrational numbers 
$\, \kappa \, $, representing a set of the full measure in 
$\, \mathbb{R} \, $. For example, the following theorem can be 
formulated for generic $\, \kappa \, $:

 Theorem (see e.g. \cite{Hinchin}):

 Let $\, f (x) \, $ be a positive function of positive $\, x \, $,
such that the function $\, x f (x) \, $ is non-increasing. Then for
almost all irrational $\, \kappa \, $ the inequality
$$\left| \, \kappa \, - \, {k^{(s)}_{2} \over k^{(s)}_{1}} \, \right|
\,\,\,\,\,  <  \,\,\,\,\, {f (k^{(s)}_{1}) \over k^{(s)}_{1}} $$
has infinitely many solutions $\, (k^{(s)}_{1}, k^{(s)}_{2}) \, $
if for some positive $\, c \, $ the integral
$$\int_{c}^{\infty} \, f (x) \,\, d x $$
is divergent.

 The theorem above gives an improvement of the estimation 
(\ref{QuadraticAppr}) for ``almost all'' irrational numbers 
$\, \kappa \, $ in the sense of the standard measure in 
$\, \mathbb{R} $. Easy to see, that the set of the functions 
$\, f (x) \, $, which can be used for the improvement of the 
estimation (\ref{QuadraticAppr}), is rather big. For example, we can
put
$$f_{1} (x) \,\,\, = \,\,\, \left[ x \,\, {\rm ln} \, x \right]^{-1}
\quad , \quad \quad x \, > \, 1   \quad  , $$
$$f_{2} (x)	\,\,\, = \,\,\,	\left[ x \,\, {\rm ln} \, x 
\,\,\, {\rm ln} \, {\rm ln} \, x \right]^{-1}
\quad ,	\quad \quad x \, >	\, e \quad , $$
etc. In particular, using the function $\, f_{1} (x) \, $, we get for
the corresponding sequence of $\, B^{(s)} \, $ the estimation
\begin{equation}
\label{BetterEst}
\omega_{B^{(s)}} \tau \, p_{F} \,
\left| \, \kappa \, - \, {k^{(s)}_{2} \over k^{(s)}_{1}} \, \right|
\quad  <  \quad
{p_{F} \over k^{(s)}_{1} \, {\rm ln} \, k^{(s)}_{1}}  \quad , 
\end{equation}
($\omega_{B^{(s)}} \tau  \, \simeq  \, k^{(s)}_{1} $).

 The estimation above shows that the general trend (\ref{Trend}) for
the functions $\, q (B) $, $\, p (B) \, $ should have
in fact an additional structure represented by the presence of 
``plateaus'' near the values $\, B^{(s)} \, $.

 As we said already, the estimation (\ref{BetterEst}) can be used 
for ``almost all'' irrational numbers, characterizing the mean direction 
of open trajectories. For special irrational numbers these estimations 
can be much stronger and we should say in this case that the 
corresponding number has ``very good'' approximations by rational 
numbers.

 As we can see then, the general trend in the behavior of 
$\, s^{kl} (B) \, $  $\,\, (k, l = 2, 3) \, $ can be understood
as an implementation of the relations (\ref{Trend})
for the ``reference values'' of $\, B \, $, which represent 
an infinite sequence on the interval 
$\, m^{*} c / e \tau \, < B \, < \infty $. At the same time, near the
values $\, B^{(s)} \, $ the behavior of $\, s^{kl} (B) \, $
demonstrates some kind of ``plateaus'' where the local dependence on
$\, B \, $ is actually different from the general trend. As a result,
the global behavior of	$\, s^{kl} (B) \, $, 
$\,\, k, l = 2, 3 \, $ demonstrates a kind of ``cascade structure''
defined by the properties of the irrationality $\, \kappa \, $. Let
us note also that the cascade structure is more distinct for the numbers  
$\, \kappa \, $, having better rational approximations, and is smoothed
for generic irrational $\, \kappa \, $.

 According to the above remark, we can also see that the trend
(\ref{Trend}) has in some sense ``conventional'' character and can be
also considered just as a convenient choice among close possible
descriptions of the asymptotic behavior of conductivity.
We have to note also that in the special case when the carriers of open
trajectories do not contain holes with singular points the relations
(\ref{Trend}) should be actually changed to the simpler trend
$$q (B) \,\,\, \sim \,\,\, p (B) \,\,\, \sim \,\,\, 
\left( \omega_{B} \tau \right)^{-1} $$

 As it is not difficult to see, all the remarks above apply also to
the estimation (\ref{SkewlnOmegaTau}), which can be violated in the
same situations as (\ref{lnOmegaTau}). For this reason, we should
consider the relation (\ref{rBTrend}) just as a ``reliable'' 
restriction on the function $\, r (B) \, $, defined by relation
(\ref{SigmaOmega}).

\vspace{1mm}

 At the end of this section, let us consider the behavior of the
components $\, \sigma^{kl} (B) \, $, where $\, k \, $ or $\, l \, $
is equal to $1$. Let us note here that relations (\ref{SigmaOmega})
represent in this case the main terms in the behavior of these
components which determines the difference between the functions
$\, a (B) $, $\, b (B) $, $\, c (B) \, $ and
$\, q (B) $, $\, r (B) $, $\, p (B) \, $. According to system
(\ref{QuasiclassicalEvolution}) the integral
$$\int_{-\infty}^{s} \, v^{x}_{gr} (p_{z}, s^{\prime}) \,\,
d s^{\prime} $$
represents the value $\,\, - \, p_{y} (s) \, + \, {\rm const} \, $ 
and is a bounded function
in our coordinate system. An elementary estimation of the integral
(\ref{VAveragedB}) gives now the main term
\begin{equation}
\label{v1grRel}
\langle v^{1}_{gr} \rangle_{B} \, (p_{z}, s)
\,\,\,\, \sim  \,\,\,\, \left(\omega_{B} \tau \right)^{-1} 
\end{equation}
in the asymptotic behavior of the function
$\, \langle v^{1}_{gr} \rangle_{B} \, $, which can be used in the
formulae (\ref{SymmetricPart}) - (\ref{AntiSymPart}). In particular,
we easily get from (\ref{SymmetricPart}) the relation
$$\Delta \, \sigma^{11} (B) \,\,\,\, \sim \,\,\,\, 
\left( \omega_{B} \tau \right)^{-2}  \,\,\,  , $$
which, being added also with the corresponding term of
(\ref{SigmaRegExpClosed}), gives the first part of the relation
(\ref{AsymptBehavior}).

 Let us note now that the value 
$\, \langle v^{1}_{gr} \rangle_{B} \, $ demonstrates also the property
$$\iint_{\hat{S}^{\gamma}_{F}} \, \langle v^{1}_{gr} \rangle_{B}
\, (p_{z}, s) \,\,\, {d p_{z} \, d s \over (2\pi \hbar)^{3}}
\,\,\,\,\, = \,\,\,\,\, 0 $$
(view the same property of $\, v^{1}_{gr} (p_{z}, s)$). 
Using the representation
$$ \langle v^{l}_{gr} \rangle_{B} \,\,\,\,\, = \,\,\,\,\,
\langle v^{l}_{gr} \rangle_{\hat{S}^{\gamma}_{F}} \,\,\, + \,\,\,
o (1) \,\,\, , \quad  \quad   B \, \rightarrow \, \infty  $$
$(l = 2, 3) \, $ for generic mean direction of the open trajectories
in the formula (\ref{SymmetricPart}), we then get
the relations 
$$\Delta \, s^{12} (B)  \,\,\,\,\, \sim \,\,\,\,\,
\Delta \, s^{13} (B) \,\,\,\,\,	= \,\,\,\,\,
o \left(  (\omega_{B} \tau )^{-1} \right) $$
for the symmetric part of $\, \Delta \sigma^{kl} (B) \, $.

 We can see then, that the main terms in the asymptotic behavior
of the values $\, \sigma^{1l} (B) \, $, $\, \sigma^{k1} (B) \, $,
$\,\, (l, k = 2, 3) \, $ are defined by the anti-symmetric part of
the conductivity tensor. To estimate the corresponding terms let us
note now that the estimation (\ref{v1grRel}) is also valid for the
function
$$\langle v^{1}_{gr} \rangle_{- B} \, (p_{z}, s)
\,\,\,\, \equiv \,\,\,\,
{c \over e B \tau} \, \int_{s}^{+\infty} \,
v^{1}_{gr} \,(p_{z}, s^{\prime}) \,\,\,
e^{{c (s - s^{\prime}) \over e B \tau}} \,\, d s^{\prime} $$
for our type of open trajectories. Using now the change of the order 
of integration in formula (\ref{AntiSymPart}) we can write
\begin{multline*}
\Delta \, a^{1l} (B) \,\,\,\, =   \\
= \,\,\,\, {e^{2} \, \tau \over 2} \,
\sum_{\gamma} \iint_{\hat{S}^{\gamma}_{F}} \, \Big( 
\langle v^{1}_{gr} \rangle_{- B}  \, - \, 
\langle v^{1}_{gr} \rangle_{B} \Big) \,\, v^{l}_{gr}
\,\,\, {d p_{z} \, d s \over (2\pi \hbar)^{3}} 
\end{multline*}
$(l = 2, 3)$, which gives the necessary estimation for the functions
$\, \Delta a^{1l} (B) \, $. Adding also the anti-symmetric part of
the tensor (\ref{SigmaRegExpClosed}), we get finally the remaining
part of the relation (\ref{AsymptBehavior}).

 Let us note also that the higher corrections to the components
$\, \sigma^{1l} (B) \, $, $\, \sigma^{k1} (B) \, $ have more
complicated behavior for generic open trajectories in
$\, {\bf p} $ - space but usually are not considered in applications.

\vspace{1mm}

 Let us say here, that the structure of $\, \sigma^{kl} (B) \, $
described above has in some sense abstract theoretical character and
probably can not be observed in all features in most of the experiments.
However, we expect that the basic elements of the picture described
above still can be detected in special studies of the conductivity
behavior in strong magnetic fields. In general, we can expect that
in the most of the real experiments it can be just established that
the values $\, \sigma^{kl} (B) \, - \bar{\sigma}^{kl}_{\infty} \, $, 
$\,\, k, l = 2, 3 \, $
demonstrate some general decreasing with the increasing of $\, B \, $
with not uniquely specified decreasing law. Quite often it is convenient
to approximate the behavior of $\, \sigma^{kl} (B) \, $ by some
intermediate powers $\, (\omega_{B} \tau)^{\mu} \, $ of the
parameter $\, \omega_{B} \tau \, $, however, we should admit an
irregular dependence of the parameter $\, \mu \, $ on the direction
and the value of $\, {\bf B} \, $ in this case.
Let us say also that the right-hand parts in 
(\ref{lnOmegaTau}) - (\ref{SkewlnOmegaTau}) can
arise with big dimensionless coefficients having geometric origin, so
for complicated Fermi surfaces the interplay between the values
$\, \bar{\sigma}^{kl}_{\infty} \, $ and $\, q^{2} (B) \, $,
$\, r (B) \, $, $\, p^{2} (B) \, $ can be important even for rather
big values of $\, \omega_{B} \tau \, $.

 Let us say that the questions considered above are connected mostly 
with the geometry of the stable open trajectories. In the next section
we will consider the aspects of the conductivity behavior connected with
the reconstruction of the trajectories of this type.

\vspace{5mm}

\section{The reconstruction of the open trajectories and the conductivity
behavior in the ``experimentally observable Stability Zone''
$\, \hat{\Omega}_{\alpha} $ .} 
\setcounter{equation}{0}

\begin{figure}[t]
\begin{center}
\includegraphics[width=0.9\linewidth]{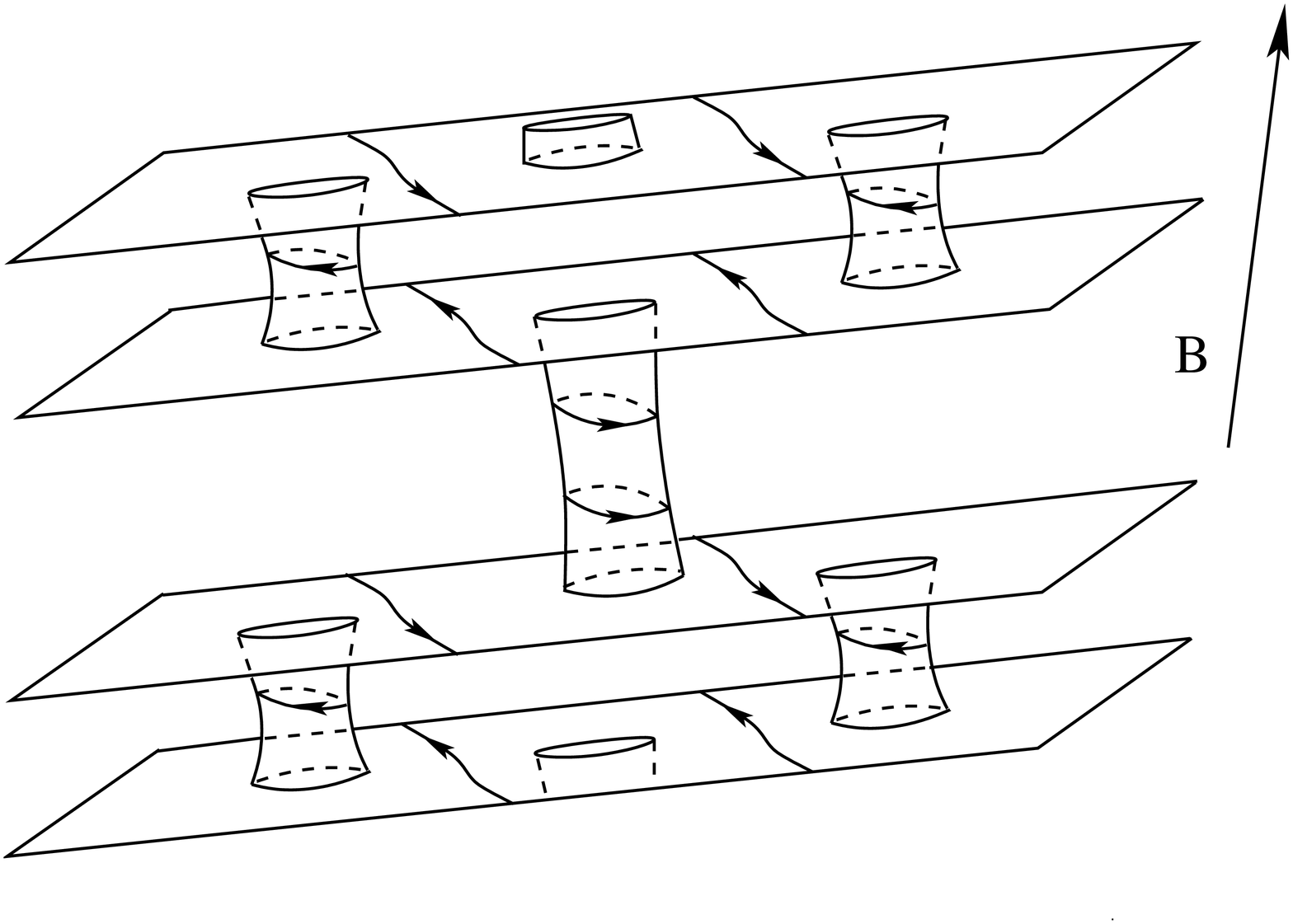}
\end{center}
\caption{The simplified (schematic) Fermi surface, used for
demonstration of the main features arising at the boundary of the
Stability Zone $\, \Omega_{\alpha} $.}
\label{SchemFermiSurf}
\end{figure}

 To explain the behavior of conductivity in the Zone 
$\, \Lambda_{\alpha} \, $ we have to consider the reconstruction of the
carriers of open trajectories after crossing the boundary of the
Zone $\, \Omega_{\alpha} \, $ on the unit sphere. In general, the most 
essential features of such reconstruction can be demonstrated by the
following simplified scheme:

 Let us imagine the Fermi surface in $\, \mathbb{R}^{3} \, $ in the 
form of an infinite set of integral planes (with holes) connected by
parallel cylinders of finite heights (Fig. \ref{SchemFermiSurf}). It 
is assumed that all the planes are divided into the ``even-numbered''
and the ``odd-numbered'' planes, which gives also the separation 
of the cylinders into two different classes. We will assume here,
that all the planes of a given class represent the same object after 
the factorization over the reciprocal lattice vectors and the same 
is assumed also about the cylinders of a given class. Thus, the 
physical Fermi surface represents here two parallel two-dimensional
tori (with holes) $\, \mathbb{T}^{2} \subset \mathbb{T}^{3} \, $
connected by two cylinders in $\, \mathbb{T}^{3} $.

 It is easy to see that for those $\, {\bf B} $, which are almost 
parallel to the axes of the cylinders, the cylinders constructed above 
represent cylinders of closed electron trajectories, while the planes 
with holes play the role of carriers of open trajectories. It can 
be seen also, that the open trajectories on the planes of different 
types represent the motion in two opposite directions, while two types 
of finite-sized cylinders carry closed trajectories of the
``electron type'' and the ``hole type'' respectively. Thus, we get
a Stability Zone $\, \Omega_{\alpha} \, $ around the direction 
$\, {\hat {\bf B}}^{\parallel} $, parallel to the axes of the 
cylinders connecting the integral planes.

 The cylinders of the closed trajectories almost coincide with the
cylinders, connecting the planes, for directions of $\, {\bf B} $,
close to $\, {\hat {\bf B}}^{\parallel} \, $, 
however, they become shorter and
represent just parts of these cylinders for $\, {\bf B} $, deviating
from the direction $\, {\hat {\bf B}}^{\parallel} \, $. The boundary of 
the Zone $\, \Omega_{\alpha} \, $ is defined by the condition that
the height of the cylinders of closed trajectories of one type
(say, electron type) becomes zero, while the height of the cylinders
of closed trajectories of another type remains finite. After crossing
the boundary of $\, \Omega_{\alpha} \, $
the individual integral planes do not represent 
carriers of open trajectories anymore, however, we can still claim 
that near the boundary of $\, \Omega_{\alpha} \, $
the cylinders of closed trajectories of the
second type cut our Fermi surface into pairs of connected
integral planes in $\, \mathbb{R}^{3} $. So, we can investigate here 
the behavior of trajectories of system 
(\ref{QuasiclassicalEvolution}) separately on these separated parts
of the Fermi surface, which gives actually rather simple description
of the trajectories in $\, \mathbb{R}^{3} $ :

 Let us denote here by $\, \Gamma_{\alpha} \, $ the integral plane,
giving the common integral direction of the planes considered above,
and by $\, \Pi ({\bf B}) \, $ the plane orthogonal to $\, {\bf B} $.
It is not difficult to see that in the picture described above all
the trajectories on our Fermi surface become closed if the intersection
of $\, \Gamma_{\alpha} \, $ and $\, \Pi ({\bf B}) \, $ has irrational 
direction. 

\begin{figure}[t]
\begin{center}
\includegraphics[width=1.0\linewidth]{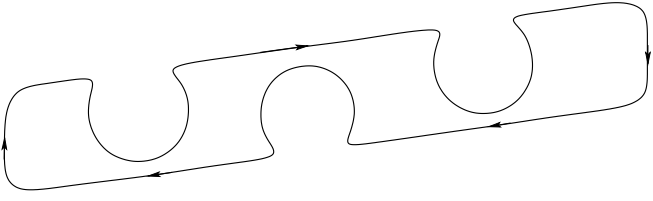}
\end{center}
\caption{The long closed trajectories arising as a result of a
reconstruction of a pair of open trajectories on a pair of connected
planes after crossing the boundary of the Zone 
$\, \Omega_{\alpha} \, $ on the angle diagram.}
\label{ReconstrTraject}
\end{figure}

 The appearance of the closed trajectories on every pair of connected
planes can be considered as a result of a reconstruction of the open
trajectories, which is caused by the ``jumps'' of the trajectories 
between two planes outside $\, \Omega_{\alpha} \, $
(Fig. \ref{ReconstrTraject}). It is easy to see
that these closed trajectories will have very big length in the 
immediate vicinity of the boundary of $\, \Omega_{\alpha} \, $
due to the low probability of the ``jumps'' in this region. 
At the same time, it can be seen also that for any fixed irrational
direction of intersection of $\, \Gamma_{\alpha} \, $ and 
$\, \Pi ({\bf B}) \, $ all the open trajectories will necessarily
pass through the reconstruction after the crossing the boundary of 
$\, \Omega_{\alpha} $.

 The situation is rather different when the intersection of
$\, \Gamma_{\alpha} \, $ and $\, \Pi ({\bf B}) \, $ has a rational
direction $\, {\bf a} $. In this case just a part of open trajectories
undergoes reconstruction after the crossing the boundary of
$\, \Omega_{\alpha} \, $ by the direction of the magnetic field,
while the other trajectories still remain open. So, in this
situation the long and the open periodic trajectories coexist on the 
integral planes, having different measure for different directions of
$\, {\bf B} $. It is not difficult to see also, that the open 
trajectories have the maximal measure near the boundary of
$\, \Omega_{\alpha} \, $ and disappear at the endpoints of the curve
$\, \hat{\gamma}^{\alpha}_{\bf a} $.

 From a simple estimation of the integrals in (\ref{GenSigmakl}) we
easily get that in the interval 
$\, 1 \, \ll \, \omega_{B} \tau \, \ll \, \lambda ({\bf B}/B) \, $
the contribution of the long closed trajectories to the conductivity
is similar to that of the stable open trajectories, so we get formula
(\ref{SigmaLowBTau}) in this interval. 

 To estimate the contribution of the trajectories shown at 
Fig. \ref{ReconstrTraject} in the limit 
$\, \omega_{B} \tau \, \gg \, \lambda ({\bf B}/B) \, $ let us just 
first note that we have in general the relations
$$\langle v_{gr}^{y} \rangle_{tr} \, (p_{z}) \,\, = \,\, 0
\quad , \quad \quad
\langle v_{gr}^{z} \rangle_{tr} \, (p_{z}) \,\, \neq \,\, 0 $$
for this type of trajectories. At the same time, we can see that the
trajectory shown at Fig. \ref{ReconstrTraject} can be considered as 
consisting of two parts belonging to different former carriers of open
trajectories with opposite values of
$\, \langle v_{gr}^{y} \rangle_{\hat{S}^{\gamma}_{F}} \, $ and
$\, \langle v_{gr}^{z} \rangle_{\hat{S}^{\gamma}_{F}} \, $.
As a result, the typical value of 
$\, \langle v_{gr}^{z} \rangle_{tr} \, $ on the closed trajectories
is rather small and is defined by the density of the trajectory on
the (former) carriers in the Brillouen zone. The exact dependence of
$\, \langle v_{gr}^{z} \rangle_{tr} \, (p_{z}) \, $ on the value
$\, \lambda ({\bf B}/B) \, $ can have in fact a complicated
character, it's not difficult to see, however, that for our pretty
rough definition of the function $\, \lambda ({\bf B}/B) \, $
the simple approximation
$$\langle v_{gr}^{z} \rangle_{tr} \, (p_{z}) \,\,\,\,\, \sim \,\,\,\,\, 
\lambda^{-1} ({\bf B}/B) $$
can be used. It can be seen also that with the same accuracy we can
also use the approximations
\begin{equation}
\label{vyzLCTest}
\langle v_{gr}^{y} \rangle_{B} \,\,\,\,\,\, \sim \,\,\,\,\,\,
\langle v_{gr}^{z} \rangle_{B} \, - \, 
\langle v_{gr}^{z} \rangle_{tr} \,\,\,\,\,\, \sim \,\,\,\,\,\,
\lambda ({\bf B}/B) \, / \, \omega_{B} \tau 
\end{equation}
for large but finite values of $\, B \, $. At the same time, for the
values $\, \langle v_{gr}^{x} \rangle_{B} \, $ we can use the
relation
\begin{equation}
\label{vxLCTest}
\langle v_{gr}^{x} \rangle_{B} \,\,\,\,\, \sim \,\,\,\,\,
\langle v_{gr}^{x} \rangle_{- B} \,\,\,\,\, \sim \,\,\,\,\,
\left( \omega_{B} \tau \right)^{-1} 
\end{equation}
for all $\omega_{B} \tau \, \gg \, 1 \, $.
 Using the above relations, we then easily get from 
(\ref{SymmetricPart}) the following estimations
\begin{multline*}
\Delta \, s^{kl} (B) \,\,\,\,\, \simeq \,\,\,\,\,
{n e^{2} \tau \over m^{*}} \left(
\begin{array}{ccc}
0 \,\,\,\,\, &  0  &  \,\, 0   \cr
0  \,\,\,\,\, &  0  &  \,\, 0   \cr
0  \,\,\,\,\, &  0  &  \,\,\,\, * \, \cdot \, \lambda^{-2}
\end{array}
\right)  \,\,\, +   \cr
+ \,\,\, {n e^{2} \tau \over m^{*}} \left(
\begin{array}{ccc}
(\omega_{B} \tau)^{-2}  &  \lambda (\omega_{B} \tau)^{-2}  &
\lambda (\omega_{B} \tau)^{-2}   \cr
\lambda (\omega_{B} \tau)^{-2}  &  
\lambda^{2} (\omega_{B} \tau)^{-2}  &
\lambda^{2} (\omega_{B} \tau)^{-2}  \cr
\lambda (\omega_{B} \tau)^{-2} 	&
\lambda^{2} (\omega_{B} \tau)^{-2}  &
\lambda^{2} (\omega_{B} \tau)^{-2}  
\end{array} 
\right)   
\end{multline*}
for the contribution of the long closed trajectories to the symmetric 
part of the conductivity tensor. Let us note also here that the
integration over a set of carriers of open trajectories 
$\,\, \hat{S}_{F} \, = \, \cup \, \hat{S}^{\gamma}_{F} \, $ 
in formula (\ref{SymmetricPart}) 
should be understood now as the integration over the part of the Fermi 
surface $\, S_{LCT} \, $ occupied by the long closed trajectories.

\vspace{1mm}

 Let us remind now, that the tensor $\, \Delta s^{kl} (B) \, $ should
be added also with the symmetric part of the contribution of the short 
closed trajectories extracted from the relations 
(\ref{SigmaRegExpClosed}). As a result, the value $\, s^{33} (B) \, $ 
will get in fact a finite contribution, which does not depend on the 
value of $\, \lambda ({\bf B}/B) \, $. We can claim, however, that the
conductivity behavior demonstrates here a partial suppression of the
conductivity along the direction of $\, {\bf B} \, $, which is
expressed by the formula (\ref{SigmaPrime33}).

\vspace{1mm}

 To evaluate the contribution of the long closed trajectories to the
anti-symmetric part of the conductivity tensor we need actually a more 
detailed consideration of the behavior of 
$\, v^{k}_{gr} (p_{z}, s) \, $ on the trajectories of this type.
Thus, using formula (\ref{AntiSymPart}) in the form
\begin{multline}
\label{LCTAnSymPart}
\Delta \, a^{kl} (B) \,\,\,\,\, =  \\
= \,\,\,\,\, {e^{2} \tau \over 2} \,
\iint_{S_{LCT}} \,\,  
\Big( \langle v^{k}_{gr} \rangle_{- B}  \, - \,
\langle v^{k}_{gr} \rangle_{B} \Big) \,\, v^{l}_{gr}
\,\,\, {d p_{z} \, d s \over (2\pi \hbar)^{3}}  \,\,\,\,  ,
\end{multline}
we can easily get the estimation
$$\Delta \, a^{12} (B) \,\,\,\,\, \sim \,\,\,\,\,
\Delta \, a^{13} (B) \,\,\,\,\, \sim \,\,\,\,\,
\left( \omega_{B} \tau \right)^{-1} $$
from the relation (\ref{vxLCTest}).

 At the same time, the estimation of the value 
$\, \Delta a^{23} (B) \, $ requires in fact a more detailed 
analysis of the formula (\ref{LCTAnSymPart}). Thus, the estimations
(\ref{vyzLCTest}) can not be simply used in formula (\ref{LCTAnSymPart})
to get the corresponding estimation of $\, \Delta a^{23} (B) \, $
because of the specific form of the trajectories shown at
Fig. \ref{ReconstrTraject}. Indeed, as can be easily seen, the values
$\, v^{y}_{gr} (s) \, $ and $\, v^{z}_{gr} (s) \, $ on the trajectories
of this form are strongly correlated on big scales
$\, ( \, \sim \, \lambda ({\bf B}/B)) $, so, their slow harmonics
will in fact annihilate each other in the anti-symmetric part of
$\, \sigma^{kl} (B) \, $. More precisely, let us represent the values
$\, v^{k}_{gr} (p_{z}, s) \, $ on the long closed trajectories in the
form
$$v^{k}_{gr} (p_{z}, s) \,\,\,\, = \,\,\,\,
\sum_{m = - \infty}^{+\infty} \,\, v^{k}_{gr \, (m)} (p_{z})
\,\,\,\, \exp \left(i m \, \omega_{0} (p_{z}) \, s \right) \,\,\, , $$
$(v^{k}_{gr \, (-m)} \, = \, \bar{v}^{k}_{gr \, (m)}), \,\, $ where
$$\omega_{0} (p_{z}) \,\,\,\, \simeq \,\,\,\, \left( m^{*} \, \cdot \,
\lambda ({\bf B}/B) \right)^{-1}  \quad  . $$

 It is not difficult to check then by direct calculation that the
contribution of the long closed trajectories to the anti-symmetric
part of the conductivity tensor can be written as
\begin{multline}
\label{AntiSymmFourie}
\Delta \, a^{kl} (B) \,\,\,\,\, = \,\,\,\,\, {e c \over 2 B} \,
\int \, {d p_{z}  \over  (2 \pi \hbar)^{3}} \,\,\, \times   \\
\times \,\,\, \sum_{m = - \infty}^{+\infty} \,\
{2 \pi i m \, \left( v^{k}_{gr \, (m)} 
\, \bar{v}^{l}_{gr \, (m)} \,\, - \,\, \bar{v}^{k}_{gr \, (m)} 
\, v^{l}_{gr \, (m)} \right)
\over \left( m \, \omega_{0} (p_{z}) \right)^{2} \,\,\, + \,\,\,
\left( c / e B \tau \right)^{2}}
\end{multline}
where the integration over $\, p_{z} \, $ is taken along the heights 
of the cylinders of long closed trajectories. Let us note here that
the net height of all non-equivalent cylinders of the long closed
trajectories has the order of $\, p_{F} / \lambda ({\bf B}/B) \, $
while the density of the frequencies
$\, \omega_{(m)} (p_{z}) \, = \, m \, \omega_{0} (p_{z}) \, $
is proportional to $\, \lambda ({\bf B}/B) \, $.

 The correlation of the functions 
$\, v^{y}_{gr} (p_{z}, s) \, $ and $\, v^{z}_{gr} (p_{z}, s) \, $ 
on the trajectory means now that the coefficients 
$\, v^{y}_{gr \, (m)} (p_{z}) \, $ and 
$\, v^{z}_{gr \, (m)} (p_{z}) \, $ are phase correlated, i.e. have
almost the same complex phases at small values of $\, m \, $. 
In general, we can use the estimation
$$\left| \, {\rm Arg} \,\, v^{y}_{gr \, (m)} (p_{z}) \,\,\, - \,\,\,
{\rm Arg} \,\, v^{z}_{gr \, (m)} (p_{z}) \, \right| 
\,\,\,\,\, \sim \,\,\,\,\, {|m| \over \lambda ({\bf B}/B)} $$
in the interval
$$ - \, \lambda ({\bf B}/B) \,\,\,\,\, \leq \,\,\,\,\, m 
\,\,\,\,\, \leq \,\,\,\,\, \lambda ({\bf B}/B)  \quad  . $$  

 At the same time, we can write for the same values of $\, m \, $
the relations
$$\left| \, v^{y}_{gr \, (m)} \, \right| 
\,\,\,\,\, \simeq \,\,\,\,\,
\left| \, v^{z}_{gr \, (m)} \, \right| 
\,\,\,\,\, \sim \,\,\,\,\, \lambda^{-1/2} ({\bf B}/B) $$
for the trajectories of our type.

 As a result, we can write approximately
$$\left| \, 2 \pi i m \left( v^{y}_{gr \, (m)}
\, \bar{v}^{z}_{gr \, (m)} \,\, - \,\, \bar{v}^{y}_{gr \, (m)}
\, v^{z}_{gr \, (m)} \right) \right|
\,\,\, \sim \,\,\, {m^{2} \over \lambda^{2} ({\bf B}/B)} $$
in the interval $\,\, |m| \,\, \leq \,\, \lambda ({\bf B}/B) \, $.

 Using the above estimations in the formula (\ref{AntiSymmFourie})
we can get now the regular expansion for the value 
$\, \Delta a^{23} (B) \, $ in the region
$\, \omega_{B} \tau \, \gg \, \lambda ({\bf B}/B) \, $,
having the form
\begin{multline*}
\Delta \, a^{23} (B) \,\,\,\,\, \simeq   \\
\simeq  \,\,\,\,\,  \left( \omega_{B} \tau \right)^{-1} \,  \left( 
\Delta  \, a^{23}_{(0)} \,\, + \,\, \sum_{k\geq 1} \,
\Delta  \, a^{23}_{(k)} \, \left( {\omega_{B} \tau \over \lambda} 
\right)^{-2k} \right)  
\end{multline*}
where all $\, \Delta  a^{23}_{(k)} \, $ have the order of
$\,\, n e^{2} \tau / m^{*} \, $. In particular, we can write in the 
main order
$$\Delta \, a^{23} (B) \,\,\,\,\,  \simeq  \,\,\,\,\,
{n e^{2} \tau \over m^{*}} \,\, \left( \omega_{B} \tau \right)^{-1} $$

 Finally, we can write the estimation
$$\Delta \, a^{kl} (B) \,\,\,\,\, \simeq \,\,\,\,\,
{n e^{2} \tau \over m^{*}} \left(
\begin{array}{ccc}
0  &  (\omega_{B} \tau )^{-1}  &  (\omega_{B} \tau )^{-1}  \cr
(\omega_{B} \tau )^{-1}  &  0  &  (\omega_{B} \tau )^{-1}  \cr
(\omega_{B} \tau )^{-1}  &  (\omega_{B} \tau )^{-1}  &  0
\end{array}
\right) $$
for the contribution of the long closed trajectories to the
anti-symmetric part of the conductivity tensor.

 Easy to see also that after the addition of the contribution of the
short closed trajectories we get finally the formula
(\ref{SigmaHighBTauAntiSym}) for the anti-symmetric part of the
conductivity tensor.

\vspace{1mm}

 For the values 
$\, \omega_{B} \tau \, \simeq \, \lambda ({\bf B}/B) \, $ formula
(\ref{GenSigmakl}) gives some intermediate regimes between 
(\ref{SigmaLowBTau}) and 
(\ref{SigmaHighBTauSymm}) - (\ref{SigmaHighBTauAntiSym})
which have in general some irregular form. 
We don't consider here these regimes in detail and 
just speak of the experimental approach to description of 
$\, \sigma^{kl} (B) \, $ using intermediate powers of 
$ \, \omega_{B} \tau \, $.  As we have already said, the corresponding
powers of $\, \omega_{B} \tau \, $ have in this case just a local
character and are not well defined as stable characteristics of the 
conductivity behavior.

\begin{figure}[t]
\begin{center}
\includegraphics[width=0.9\linewidth]{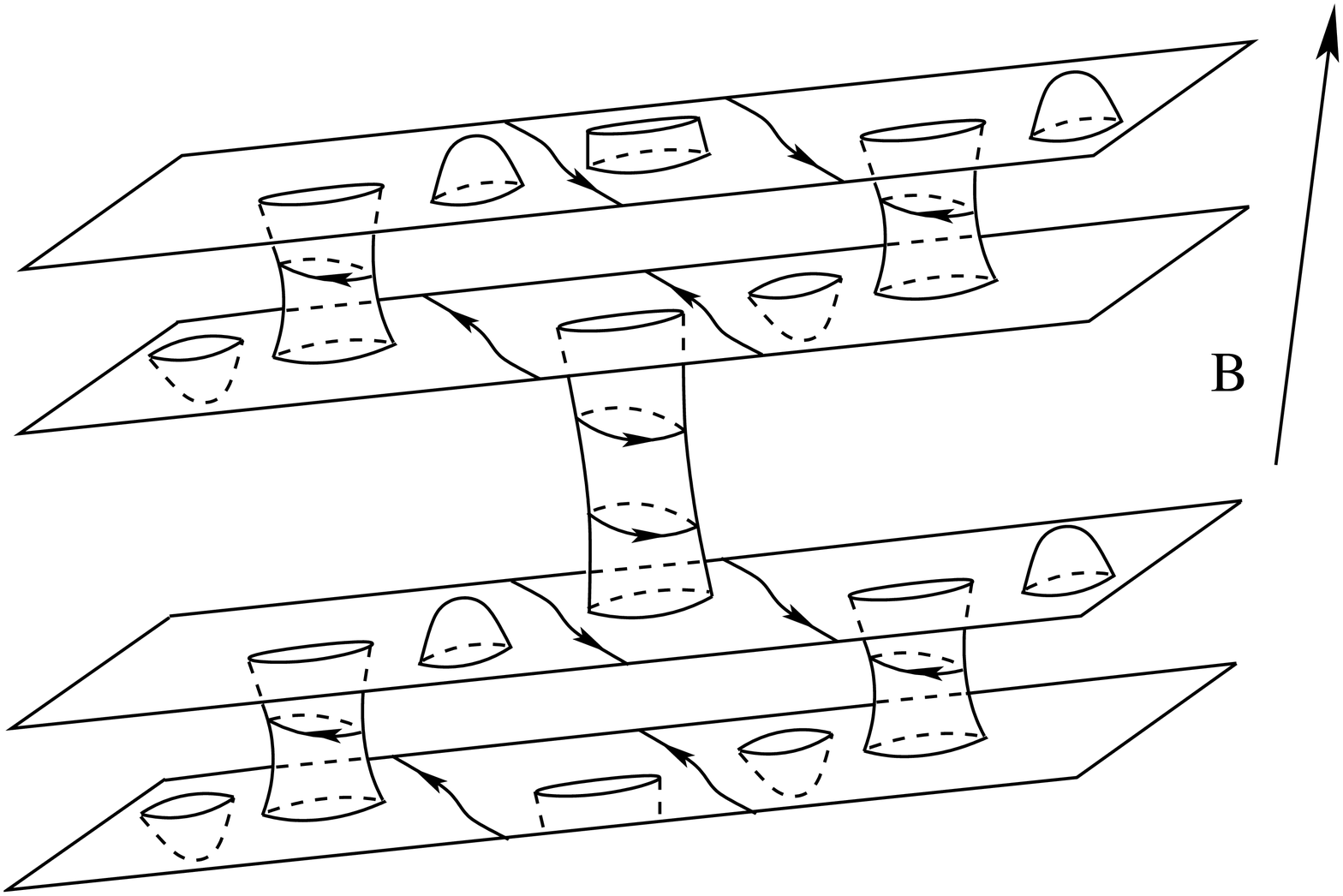}
\end{center}
\caption{More general schematic surface representing the structure
of trajectories on real Fermi surface for directions of 
$\, {\bf B} \, $ belonging to a fixed Stability Zone 
$\, \Omega_{\alpha} $.}
\label{EnrichedSchemSurf}
\end{figure}

 The schematic Fermi surface considered above looks rather special
from the general point of view. However, it demonstrates all the main
points of the behavior of open trajectories near the boundary of the
Stability Zone $\, \Omega_{\alpha} \, $ which are important for us.
In general, the schematic picture represented above can be enriched
by additional cylinders of closed trajectories, connecting the
integral planes or just having a point as a cylinder base
(Fig. \ref{EnrichedSchemSurf}). Theoretically, we can also consider 
the situation when the connected parts of the Fermi surface joining 
integral planes represent compound structures consisting of several 
cylinders of closed trajectories (in fact, this situation is extremely 
unlikely for real Fermi surfaces). Still, the main feature of the
reconstruction of the carriers of open trajectories after crossing the
boundary of $\, \Omega_{\alpha} \, $ for generic Fermi surface
is given by the vanishing of the height of one particular cylinder 
of closed trajectories and the division of the Fermi surface into the 
pairs of integral planes, separated by other cylinders of closed 
trajectories. The most general picture of the division of the Fermi 
surface for $\, \hat{\bf B} \in \Omega_{\alpha} \, $ is given actually 
by the scheme containing more than two non-equivalent integral planes
(with the same integral direction), connected by cylinders of closed
trajectories. In the last situation, which is theoretically possible
for Fermi surfaces of very high genus, the pairs of integral planes
and individual carriers of open trajectories can coexist after the
reconstruction of a part of the open trajectories. Mathematically
we will have in this case an overlapping of two Stability Zones
$\, \Omega_{\alpha} \, $ and $\, \Omega^{\prime}_{\alpha} \, $
with the same ``topological quantum numbers'', which will result
also in overlapping of the zones $\, \hat{\Omega}_{\alpha} \, $
and $\, \hat{\Omega}^{\prime}_{\alpha} \, $ in the experimental
study of conductivity. The full conductivity tensor in the 
overlappings
$$\Omega_{\alpha} \, \cap \, \Omega^{\prime}_{\alpha} 
\,\,\, , \quad
\Omega_{\alpha} \, \cap \, \Lambda^{\prime}_{\alpha}
\,\,\, , \quad
\Lambda_{\alpha} \, \cap \, \Omega^{\prime}_{\alpha}
\,\,\, , \quad
\Lambda_{\alpha} \, \cap \, \Lambda^{\prime}_{\alpha} $$
is given in this case by the sum of the corresponding asymptotic 
expressions above, so we have here a theoretical possibility to observe
very complicated analytical behavior of conductivity in the limit
$\, B \rightarrow \infty $.

 Let us point out here that the possibility of representation of the
Fermi surface in the special form described above in the case of
presence of stable open trajectories is based on rather deep
topological theorems playing the most important role in the 
considerations of papers \cite{zorich1,dynn1,dynn3}. So, the
picture described above represents in fact (under some generic 
requirements) the most general situation.

 At the same time, we have to note that the statement above should
be understood just as a topological characterization of system
(\ref{QuasiclassicalEvolution}) in the case of presence of stable
open trajectories. Thus, the topological structure shown at 
Fig. \ref{EnrichedSchemSurf} can have in fact much more complicated
(visual) geometric representation in the $\, {\bf p} $ - space.

\vspace{1mm}

 Finally, we can briefly describe the structure of the 
``experimentally observable Stability Zone'' 
$\,\, \hat{\Omega}_{\alpha} \, $, arising in the quasiclassical
approximation, as a union of the following main parts:

\vspace{1mm}

I) The central part of the Zone $\, \hat{\Omega}_{\alpha} \, $
(containing the ``mathematical Stability Zone''
$\, \Omega_{\alpha} $). The conductivity behavior in this part 
demonstrates the ``most regular'' dependence on the value of 
$\, B \, $, represented by the asymptotic regimes 
(\ref{SigmaOmega}) or (\ref{SigmaLowBTau}). At the same time,
the angular dependence of $\, \sigma^{kl} ({\bf B}) \, $
demonstrates here rather irregular character due to the presence of 
the ``rational peaks'' in the values of 
$\, \sigma^{kl} ({\bf B}) \, $ on a dense set of directions of 
$\, {\bf B} \, $ in this area (Fig. \ref{ComplBehaviorZone}).

\vspace{1mm}

II) The zone of rather complicated form around the central part of 
the Zone $\, \hat{\Omega}_{\alpha} \, $ (the black area at
Fig. \ref{ComplBehaviorZone}). The tensor 
$\, \sigma^{kl} ({\bf B}) \, $ demonstrates here the most complicated
dependence both on the value and the direction of $\, {\bf B} \, $,
corresponding to a gradual transition from the regime	
(\ref{SigmaLowBTau}) to 
(\ref{SigmaHighBTauSymm}) - (\ref{SigmaHighBTauAntiSym}).
For an approximation of the behavior of 
$\, \sigma^{kl} ({\bf B}) \, $ the intermediate powers
$\, (\omega_{B} \tau)^{\mu} \, $ can be locally used, in general the 
value of $\, \mu \, $ demonstrates unstable behavior. At the same
time, the behavior of $\, \sigma^{kl} ({\bf B}) \, $ demonstrates
a suppression of the conductivity along the direction of 
$\, {\bf B} \, $ ($\sigma^{33} ({\bf B})$) to some lower values
on the exterior boundary of this zone.

\vspace{1mm}

III) The ``boundary region'' of the Zone 
$\, \hat{\Omega}_{\alpha} \, $. The behavior of
$\, \sigma^{kl} ({\bf B}) \, $ demonstrates here a gradual 
transition from the regime 
(\ref{SigmaHighBTauSymm}) - (\ref{SigmaHighBTauAntiSym})
to the simpler behavior (\ref{3DimClosedTr}). The behavior of 
$\, \sigma^{33} ({\bf B}) \, $ demonstrates here a gradual increasing 
of the conductivity along the direction of $\, {\bf B} \, $ to its 
maximal value on the boundary of $\, \hat{\Omega}_{\alpha} \, $.

\vspace{1mm}

 At the end of the paper, we would like to make also one additional
substantial remark. Namely, it can be actually shown, that the 
topological structure of system (\ref{QuasiclassicalEvolution}) 
admits also a similar description being considered for the whole 
dispersion relation $\, \epsilon ({\bf p}) \, $ and not just for a 
fixed Fermi level
$\, \epsilon \, = \, \epsilon_{F} \, $ (see \cite{dynn3}). 
In particular, the ``Stability Zones'' (corresponding to the
presence of stable open trajectories at least at one energy level)
can be introduced for the whole spectrum $\, \epsilon ({\bf p}) \, $.
From this point of view, the ``experimentally observable Stability
Zones'' $\, \hat{\Omega}_{\alpha} \, $ have in fact a connection with
exact mathematical Stability Zones, defined for the whole
energy spectrum. In particular, every ``experimentally observable 
Stability Zone'' $\, \hat{\Omega}_{\alpha} \, $ always belongs
to a bigger mathematical Stability Zone 
$\, \Omega^{\prime}_{\alpha} \, $, defined for the whole spectrum
$\, \epsilon ({\bf p}) \, $.

\vspace{5mm}

\section{Conclusions.}
\setcounter{equation}{0}

 We investigate the analytical behavior of conductivity of normal
metals in strong magnetic fields in the case of presence of stable
open electron trajectories on the Fermi surface. As it is shown in
the paper, the behavior of conductivity can demonstrate rather
nontrivial analytical properties in ``experimentally observable
Stability Zones'' even under the condition 
$\, \omega_{B} \tau \, \gg \, 1 \, $. In particular, in different
parts of the experimentally observable Stability Zone the behavior
of conductivity can demonstrate different types of ``regular'' or
more complicated intermediate ``irregular'' regimes depending on the
value of the magnetic field. The results of the paper are based on
topological description of the carriers of open trajectories on 
complicated Fermi surfaces obtained in recent mathematical 
investigations of the corresponding geometric problem.

\end{document}